\documentclass[12pt,a4paper]{article}
\usepackage{amssymb, amsmath}
\usepackage{array}   
\usepackage{cite}

\usepackage[dvips]{graphicx}   
\graphicspath{{Figures/}}      

\newcommand{\sign}{\mathop{\rm sign}\nolimits}       

\begin{document}

\title{\Large \textbf{Competitive Adsorption of a Two-Component Gas on a Deformable Adsorbent}}

\medskip

\author{A.~S.~Usenko\footnote{E-mail: usenko@bitp.kiev.ua} \\
Bogolyubov Institute for Theoretical Physics, \\[-3pt]
Ukrainian National Academy of Sciences, Kyiv 03680, Ukraine}

\date{}

\maketitle


\begin{abstract}

We investigate the competitive adsorption of a two-component gas on the surface
of an adsorbent whose adsorption properties vary in adsorption due to the
adsorbent deformation. The essential difference of adsorption isotherms for a
deformable adsorbent both from the classical Langmuir adsorption isotherms of a
two-component gas and from the adsorption isotherms of a one-component gas
taking into account variations in adsorption properties of the adsorbent in
adsorption is obtained. We establish bistability and tristability of the system
caused by variations in adsorption properties of the adsorbent in competitive
adsorption of gas particles on it. Conditions under which adsorption isotherms
of a binary gas mixture have two stable asymptotes are derived. It is shown
that the specific features of the behavior of the system under study can be
described in terms of a potential of the known explicit form.

\end{abstract}




PACS numbers: 68.43.-h;  68.43.Mn;  68.43.Nr;  68.35.Rh












\section{Introduction}  \label{Introduction}

Problems of adsorption on the surface of different bodies belong to a wide
class of problems of physics, chemistry, and biology that are very important
both from the theoretical point of view and for various practical applications.
The results of numerous investigations show that adsorption of particles leads
to considerable changes in physical and chemical characteristics of adsorbents.
Detailed analysis of the changes in the properties of the adsorbent surface due
to adsorption is given, e.g., in
\cite{ref.Mor,ref.RobMc,ref.Nau_78,ref.JaP,ref.KiK,ref.Vol,ref.Zan,ref.Zhd,ref.LNP,ref.AdC,ref.Nau_03}.
\looseness=-1

Since processes of adsorption and desorption are obligatory stages of
heterogeneous-catalytic reactions, the results of the adsorption theory are
extremely important for investigation of various problems of heterogeneous
catalysis \cite{ref.Bar,ref.Rog,ref.Roz,ref.Bor,ref.KSh,ref.Kry}. \looseness=-1

Generalizations of the classical Langmuir adsorption theory aimed at a more
correct description of the adsorbent surface and adsorbed particles give the
qualitatively new  behavior of the amount of adsorbed substance and its
kinetics.  An extensive material obtained on the basis of different models and
applications to various problems of adsorption and catalysis are widely
presented in the literature (see, e.g.,
\cite{ref.Zhd,ref.Rog,ref.Roz,ref.Bor,ref.KSh,ref.Kry,ref.FrK,ref.Rut,ref.Tov,ref.Do,ref.KSt}).
In particular, due to lateral interactions between adparticles, adsorption
isotherms can have a hysteresis shape, and different structural changes in the
adsorbent surface occur (reviews of the theoretical and experimental results
are given, e.g., in
\cite{ref.Zhd,ref.LNP,ref.AdC,ref.Nau_03,ref.Rut,ref.Tov,ref.Do,ref.KSt}). In
turn, a qualitative change in the surface structure in adsorption leads to a
series of specific features of oscillatory surface reactions and formation of
different spatiotemporal patterns (for the oscillatory kinetics in
heterogeneous catalysis and related problems, see, e.g., the reviews
\cite{ref.IEr,ref.Imb_1,ref.Imb_2} and the monograph \cite{ref.Ert}).
\looseness=-2

It is established in \cite{ref.Usenko} that, parallel with lateral interactions
between adparticles, there is another factor (the adsorption-induced
deformation of an adsorbent) leading to hysteresis-shaped isotherms of
localized adsorption of a one-component gas on the flat energetically
homogeneous surface of a solid adsorbent. It is worth noting that, as early as
in 1938, in \cite{ref.Zeld}, Zeldovich based on the idea of a change in
adsorption properties of the adsorbent surface in adsorption, predicted a
hysteresis of adsorption isotherms if the typical time of adsorption and
desorption is much less than the relaxation time of the surface.

In recent years, it has been established an essential influence of memory
effects on the surface diffusion of adparticles over the adsorbent surface in
the case where the relaxation time of the adsorbent is comparable (or greater
than) with typical times for moving adparticles (see, e.g., the review
\cite{ref.AFY} and references therein). Dynamical changes in properties of the
surface by moving particles are taken into account in some models (e.g., in
\cite{ref.TZHZJ,ref.HZJ}), which, to some extent, is similar to the Zeldovich
idea of an absorbent varying its adsorption properties in adsorption.
\looseness=-1

Since an actual adsorbate includes several species of particles, in adsorption,
particles of different species compete for adsorption sites. This leads not
only to a decrease in the number of adparticles of a species relative to that
for one-component adsorption \cite{ref.Rut,ref.Do,ref.KSt,ref.Yang,ref.Gun} but
also to the qualitative change in the shape of adsorption isotherms with regard
for lateral interactions between adparticles \cite{ref.Tov}. In view of
hysteresis-shaped isotherms of localized adsorption of a one-component gas on
the flat surface of a solid adsorbent due to the adsorption-induced deformation
of an adsorbent \cite{ref.Usenko}, it is of interest to investigate the
influence of this factor on changes in the classical extended Langmuir
adsorption isotherms of a multicomponent gaseous system. \looseness=-1

In the present paper, we study specific features of adsorption isotherms of a
two-component gas on the surface of a solid adsorbent whose adsorption
properties vary in adsorption.

A model of adsorption of a two-component gas taking into account variations in
adsorption properties of an adsorbent caused by its deformation in adsorption
is proposed in Sec.~2. We obtain a system of equations that describes the
kinetics of the surface coverage and the displacement of adsorption sites. For
each species of adparticles, it is introduced the dimensionless coupling
parameter equal to the normalized maximum increment of the activation energy
for desorption due to the adsorbent deformation in one-component adsorption.
The influence of the adsorbent deformation on the adsorption isotherms of
adparticles of both species is investigated in Sec.~3. It is established a
considerable redistribution of the amount of adsorbed substances as compared
with that in the classical case even for a negligible quantity of particles of
one species in a gas mixture. The obtained adsorption isotherms essentially
depend on the coupling parameters and differ both from the Langmuir adsorption
isotherms of a two-component gas and from the adsorption isotherms of a
one-component gas for an adsorbent whose adsorption properties vary in
adsorption. We establish bistability and tristability of the system caused by
variations in adsorption properties of the adsorbent in competitive adsorption.
Conditions under which adsorption isotherms of a binary gas mixture have two
stable asymptotes are derived. In Sec.~4, within the framework of the
overdamped approximation and essential difference in the linear relaxation
times of the dynamical variables, the behavior of the system under study is
described in terms of a potential whose explicit form is obtained. The specific
features of isotherms of competitive adsorption are explained with the use of
the (single-, two-, or three-well) potential. \looseness=-1



\section{General Relations}  \label{Model}

We consider localized monolayer competitive adsorption of particles of a
two-component gas on the flat surface of a solid adsorbent using the classical
Langmuir model generalized to the case of variations in adsorption properties
of the adsorbent in adsorption--desorption of gas particles \cite{ref.Usenko}.
Gas particles are adsorbed on adsorption sites located at the adsorbent surface
and total number of sites $N$ does not change in time. All adsorption sites
have equal adsorption activity (energetically homogeneous surface) and each
adsorption site can be bound only with one gas particle. We introduce the
Cartesian coordinate system with the origin on the adsorbent surface and the
$0X$-axis directed into the adsorbent so that the adsorbent and the gas occupy
the regions $x \ge 0$ and $x < 0$, respectively.

Following \cite{ref.Usenko}, we simulate each vacant adsorption site by a
one-dimensional linear oscillator of mass $m_0$ that oscillates perpendicularly
to the surface about its equilibrium position  $x = 0$. \looseness=-1

The binding of a gas particle with an adsorption site is accompanied by a
change in the spatial distribution of the charge density of the bound
adsorption site as compared with that of a vacant one. This change depends on
the nature of adsorption bonds and specific features of both the adsorbent and
gas particles (see, e.g.,
\cite{ref.Nau_78,ref.Zhd,ref.LNP,ref.IEr,ref.Imb_1,ref.Imb_2,ref.Ert,ref.Nau_94}).
\looseness=-1

This leads to a change in the interaction of the bound adsorption site with
neighboring atoms of the adsorbent located both on the surface and in the
nearest subsurface region.  As a result, the resulting force acting on the
bound adsorption site changes as against that acting on the vacant adsorption
site. This can be regarded as the appearance of a certain adsorption-induced
force $\vec{F}_n(\vec{r},t)$ acting on the adsorption site occupied by an
adparticle of species  $n = 1, 2$ (here and below, the subscript $n = 1, 2$
denotes the species of particles), where $\vec{r}$ is the running coordinate of
the adsorption site.  Under the action of this force, the bound adsorption site
tends to a new equilibrium position. However, as soon as the adparticle leaves
the adsorption site, the last becomes vacant and relaxes to its nonperturbed
equilibrium position $x = 0$. For the subsequent adsorption of other gas
particle on this vacant adsorption site, two essentially different situations
are possible: a gas particle occupies the site after or before it reaches the
nonperturbed equilibrium position. In the first case, a new adparticle on the
adsorption site does not ``fill'' its earlier occupation by previous
adparticles. In the second, a particle is adsorbed on the surface locally
deformed by the previous adparticle (not necessarily of the same species),
i.e., the retardation of relaxation of the surface occurs or, in other words,
adsorption with memory takes place.  \looseness=-1

We consider the case where the force $\vec{F}_n(\vec{r},t)$ is normal to the
boundary and depends only on the coordinate  $x$: \  $\vec{F}_n(\vec{r},t)=
\vec{e}_x\, F_n(x,t)$, where $\vec{e}_x$ is the unit vector along the
$0X$-axis.

The force  $\vec{F}_n(\vec{r},t)$  acts on the adsorption site only during
discrete time intervals where the site is bound. Thus, at any instant, the
adsorption site is in one of the three states: vacant or bound with adparticle
of species 1 or 2. Instead, we consider the approximation of a time-continuous
adsorption-induced force  $\vec{F}(\vec{r},t)$, which corresponds to an
adsorption site permanently bound with an adparticle with the time-dependent
probability (the mean occupancy of adparticle on an adsorption site) equal to
the surface coverage by adparticles of species  $n$, $\theta_n = N_n(t)/N$,
where $N_n(t)$ is the number of adsorption sites occupied by adparticles of
species  $n$ at the time $t$. Since an adsorption site can be bound only with
one adparticle, $\vec{F}(\vec{r},t)= \vec{F}_1(x)\, \theta_1 + \vec{F}_2(x)\,
\theta_2$, where $\vec{F}_n(x) = \vec{e}_x\, F_n(x)$, and, hence,
$\vec{F}(\vec{r},t)= \vec{e}_x\, F(x,t), \ F(x,t) = F_1(x)\, \theta_1 +
F_2(x)\, \theta_2$.  This approximation is similar to the mean-field
approximation used in the adsorption theory taking into account lateral
interactions between adparticles (see, e.g., \cite{ref.Zhd,ref.Do}). Expanding
$F_n(x)$ in the Taylor series in the neighborhood of $x = 0$ and keeping only
the first term of the expansion, and expressing the adsorption-induced force
$\vec{F}_n(x)$  in terms of the potential, $F_n(x) = - \frac{dV_n(x)}{dx}$, we
get \looseness=-1
\begin{equation}
 V_n(x) \approx -\chi_n \, x \, ,  \qquad  n = 1, 2,
\end{equation}
\noindent  where

$$
\left. \chi_n = - \frac{dV_n(x)}{dx} \right|_{x = 0}
$$
is the constant adsorption-induced force acting on the adsorption site occupied
by an adparticle of species $n$.

We introduce the dimensionless quantity  $G = \chi_2/\chi_1$, which is positive
or negative for parallel ($\sign\,\chi_1 = \sign\,\chi_2 $) or antiparallel
($\sign\,\chi_1 = -\sign\,\chi_2$) adsorption-induced forces, respectively.

Disregarding the internal motions in the adparticle--adsorption site system,
i.e., considering the motion of the bound adsorption site as a whole, and
taking into account a change in the mass of the oscillator in adsorption within
the framework of this approximation, we obtain the following equation of motion
of an oscillator of variable mass in the adsorption-induced force field:
\looseness=-1
\begin{equation}
 \frac{d}{dt} \biggl( m_{eff}(\Theta) \frac{dx}{dt} \biggr) + \alpha
  \frac{dx}{dt} + \varkappa\, x = \chi_1 \, \theta_1 +  \chi_2 \, \theta_2 \, ,
\end{equation}
\noindent  where $\varkappa$ is the restoring force constant, $\alpha$ is the
friction coefficient, $m_{eff}(\Theta) = m_0 + m_1\, \theta_1 + m_2\, \theta_2$
is the effective mass of the oscillator that varies in adsorption,  $m_n$ is
the mass of an adparticle of species $n$, and the symbol $\Theta \equiv
\{\theta_1,\, \theta_2\}$ denotes a collection of the surface coverages. Since
$\theta_n \leq 1$, the effective mass of the oscillator is lesser than $M = m_0
+ m_1 + m_2$. \looseness=-1

It follows from Eq.~(2) that, due to adsorption, the equilibrium position of
the oscillator $x = 0$ shifts to the new one $x^{eq}(\Theta)$ defined by the
relation
\begin{equation}
 x^{eq}(\Theta) = x_1^{eq}(\theta_1) +  x_2^{eq}(\theta_2) \, ,
\end{equation}
\noindent where  $x_n^{eq}(\theta_n) = x_n^{max}\, \theta_n$ is the equation
for determination of the equilibrium position of the oscillator in adsorption
of a one-component gas of species $n$ and $x_n^{max} \equiv x_n^{eq}(1)=
\chi_n/ \varkappa$ is the maximum stationary displacement of the oscillator
from its nonperturbed equilibrium position $x = 0$ in the case of the total
surface coverage ($\theta_n = 1$).

Within the framework of the used approximation, the forces of lateral
interactions between adparticles are parallel to the adsorbent surface and the
adsorption-induced forces $\vec{F}_n(x)$ are perpendicular to the surface,
which means that the forces $\vec{F}_n(x)$ are caused by the interaction of
bound adsorption sites with the nearest subsurface atoms of the adsorbent.
Nevertheless, the lateral interactions between adparticles affect the
adsorption-induced force $\vec{F}(\vec{r},t)$ (and, hence, a normal
displacement of the plane of adsorption sites) via the surface coverages
$\theta_1$ and $\theta_2$.  \looseness=-1

In the Langmuir theory of kinetics on a nondeformable adsorbent ($\chi_n = 0, \
n = 1,2$) neglecting interactions between adparticles, the rate constants for
adsorption and desorption $k_n^a$ and $k_n^d$ of particles of species  $n$,
respectively, do not depend on the concentration of particles in the gas phase
and are defined by the Arrhenius relations
\begin{equation}
 k_n^a = k_n^+ \exp{\biggl(-\frac{E_n^a}{k_B T} \biggr)}\, , \qquad
 k_n^d = k_n^- \exp{\biggl(-\frac{E_n^d}{k_B T} \biggr)}\, ,  \qquad   n = 1,2,
\end{equation}
\noindent  where $E_n^a$ and $E_n^d$  are the activation energies for
adsorption and desorption, respectively, $k_n^+$ and $k_n^-$  are the
pre-exponential factors, $T$ is the absolute temperature, and $k_B$ is the
Boltzmann constant.

The Hamiltonian of the adparticles--adsorbent system contains the term $-\chi_1
\, x \, N_1 -\chi_2 \, x \, N_2 $ caused by the adsorbent deformation in
adsorption due to the adsorption-induced force field $F(x, t)$. This implies
that an adparticle of species  $n$ is not only in a potential well of constant
depth $E_n^d$ but also in the adsorption-induced potential $V_n(x)$. For
parallel adsorption-induced forces, an adparticle of any species is in a deeper
potential well than on a nondeformable adsorbent. As a result, in the case at
hand, for desorption of an adparticle of species $n$, it must get an energy
greater than $E_n^d$ by the value $|V_n(x)| = \chi_n \, x $, which can be
regarded as the increment of the activation energy for desorption $E_n^d$ of an
adparticle of species $n$ caused by the adsorbent deformation. For antiparallel
adsorption-induced forces, the increments $\chi_n \, x $  of the activation
energies for desorption $E_n^d$ of adparticles of different species $n$ have
opposite signs. Thus, the adsorbent deformation increases the activation energy
for desorption of adparticles of one species and decreases the activation
energy for desorption of adparticles of another species. Note that the
quantities $E_n^d$ and $\chi_n \, x$ can be interpreted as the first and second
terms, respectively, of the Taylor series of the coordinate-dependent
activation energy for desorption $E_n^d(x)$. \looseness=-1

It is well known that lateral interactions between adparticles essentially
change adsorption isotherms of a binary gas mixture (see, e.g.,
\cite{ref.Tov}). In the present paper, to illustrate that there is another
factor (the adsorption-induced deformation of the adsorbent) leading to
qualitative changes in isotherms of competitive adsorption of a two-component
gas, we do not take into account lateral interactions between adparticles.

The adsorbent deformation in adsorption affects the desorption rates of
adparticles and, hence, the surface coverage. Assuming that the pre-exponential
factors $k_n^-$ are not changed, we obtain the following expression for the
rate coefficients for desorption:
\begin{equation}
 k_n^d(x) = k_n^d \exp{\biggl(-\frac{\chi_n \, x}{k_B T} \biggr)}.
\end{equation}
\noindent Thus, the rate coefficients for desorption (5) are
coordinate-dependent functions, and gas particles are adsorbed on the surface
whose adsorption characteristics vary with time.

According to (5), for $G > 0$, the desorption rates of adparticles of both
species decrease due to the adsorbent deformation in adsorption. For $G < 0$,
the joint action of adparticles of both species on the adsorbent leads to the
opposite results: the desorption rate of adparticles decreases for one species
and increases for another.

With regard for variations in adsorption properties of the adsorbent in
adsorption, the kinetics of the surface coverages is described by the equations
\begin{equation}
 \frac{d \theta_n}{dt} = k_n^a C_n \, \theta_0
  - k_n^d \, \theta_n \, \exp{\biggl(-\frac{\chi_n\, x}{k_B T} \biggr)},
  \qquad   n = 1,2,
\end{equation}
\noindent where  $C_n$ is the concentration of particles of species  $n$ in the
gas phase that is kept constant, $\theta_0 = 1 - \theta_+ = N_0(t)/N$ is the
vacant part of the surface, $\theta_+ = \theta_1 + \theta_2 = N_b(t)/N$ is the
surface coverage by adparticles of both species, $N_b(t) = N_1(t) + N_2(t)$ and
$N_0(t)$ are, respectively, the numbers of occupied and vacant adsorption sites
at the time $t$, $N_b(t) + N_0(t) = N$. \looseness=-1

Setting in (6) $\chi_n = 0$, we obtain the known system of two linear equations
that describes the Langmuir kinetics of adsorption of a two-component gas
\cite{ref.FrK}.

Introducing the dimensionless coordinate of oscillator $\xi = x/x_1^{max}$, we
obtain the following autonomous system of three nonlinear differential
equations that describes the kinetics of the surface coverages and the normal
displacement of adsorption sites in localized adsorption with regard for
variations in adsorption properties of the adsorbent in adsorption:
\begin{eqnarray}
 && \left\{
\begin{array}{l}
 \dfrac{d \theta_n}{dt} = k_n^a C_n \, \theta_0
  - k_n^d \, \theta_n \exp{\left(-\dfrac{g_n}{G_n}\,\xi\, \right)},
  \qquad\quad   n = 1,2,  \vspace{-2mm}  \\
  \\ \dfrac{d}{dt} \biggl( m_{eff}(\Theta) \dfrac{d\xi}{dt} \biggr)
  + \alpha \dfrac{d\xi}{dt}  =
  \varkappa \, \bigl( \theta_1 + G\,\theta_2 - \xi \bigr).
  \qquad \quad \
\end{array}\right.
\end{eqnarray}
\noindent Here, the dimensionless quantity
\begin{equation}
 g_n = |V_n|/k_B T,   \qquad   n = 1,2,
\end{equation}
\noindent called a coupling parameter, is the maximum increment of the
activation energy for desorption (normalized by $k_B\,T$) due to the adsorbent
deformation in adsorption of a one-component gas of species $n$, $\ V_n \equiv
V_n(x_n^{max})= - \chi_n^2/\varkappa$,  $\,G_1 = 1$, $\,G_2 \equiv G$, $\,g_2 =
g_1\, G^2$.

Setting in (7) $C_2 = 0$ and $\theta_2 = 0$, we obtain the system of two
differential equations that describes the kinetics of the amount of a
one-component gas of species 1 adsorbed on a deformable adsorbent
\cite{ref.Usenko}.  \looseness=-1

The average coordinate-dependent residence times of adparticles on the surface
of a deformable adsorbent $\tau_n^d(\xi) = 1/k_n^d(\xi)\, ,  \  n = 1,2$,
\begin{equation}
 \tau_1^d(\xi) = \tau_1^d \exp{\left(g_1\,\xi\, \right)},   \qquad\qquad
 \tau_2^d(\xi) = \tau_2^d \exp{\left(g_1\,G\,\xi\, \right)},
\end{equation}

\noindent  increase for  $G > 0$  as against the classical residence times
\begin{equation}
 \tau_n^d = \frac{1}{k_n^d}\, ,   \qquad\qquad    n = 1,2,
\end{equation}
\noindent and, furthermore, the greater the displacement of adsorption sites
from their nonperturbed equilibrium position, the more this increase. Since the
residence time of adparticles with a greater value of  $|\chi_n|$ increases
greater, the surface is more intensively occupied by particles of this species
and this process rapidly grows with $\xi$. Denoting the ratio of the residence
times of adparticles of different species on the adsorbent surface by
\begin{equation}
 R(\xi) = \frac{\tau_2^d(\xi)}{\tau_1^d(\xi)}\, ,
\end{equation}
\noindent we obtain
\begin{equation}
 R(\xi) = R_0\, w(\xi),
\end{equation}
\noindent where
\begin{equation}
 R_0 \equiv R(0) = \frac{\tau_2^d}{\tau_1^d}
\end{equation}
\noindent  is the ratio of the residence times of adparticles of different
species on the nondeformable adsorbent and the quantity
\begin{equation}
 w(\xi) = \exp{\Bigl(g_1 \left(G - 1 \right) \xi \Bigr)}
\end{equation}
\noindent  characterizes a variation in ratio (13) due to the different action
of adparticles of both species on the adsorbent. In the special case of the
identical action of all adparticles on the adsorbent ($\chi_1 = \chi_2$), we
have $w(\xi) = 1$. According to (14), for $G > 1$, the quantity  $w(\xi)$ can
reach large values, which essentially affects the surface coverages $\theta_1$
and $\theta_2$.

Expressions (9), (11), (12), and (14) are also true for  $G < 0$. However, in
this case, the adsorbent deformation caused by the joint action of adparticles
of both species leads to an increase in the residence time of adparticles of
one species and a decrease in the residence time of adparticles of other
species as against the classical residence times (10).




\section{Stationary Case}  \label{Stationary}

\subsection{General Relations}  \label{Stationary general}

In the stationary case, system (7) is reduced to the system
\begin{eqnarray}
 && \left\{
\begin{array}{llll}
 \ell_1 &=& \dfrac{\theta_1}{\theta_0}\,
             \exp{\left(-g_1\, \xi \right)}\, ,  \vspace{2mm}  \\
 \ell_2 &=& \dfrac{\theta_2}{\theta_0}\,
             \exp{\left(-g_1\, G\, \xi \right)} \, , \vspace{2mm}  \\
 \xi &=& \theta_1 + G\, \theta_2 \, ,
\end{array}\right.
\end{eqnarray}
\noindent where  $\ell_n = C_n\, K_n$ is the dimensionless concentration of gas
particles of species $n = 1, 2$ and  $K_n = k_n^a/k_n^d$ is the adsorption
equilibrium constant for a one-component gas of species $n$ in the linear case
($\chi_n = 0$).

After simple transformations, we obtain the following expressions for the
surface coverages:
\begin{eqnarray}
 \theta_1 &=& \frac{\xi}{1 + G\, S(\xi)}\, ,  \\ [\smallskipamount]
 \theta_2 &=& S(\xi)\, \theta_1
\end{eqnarray}
\noindent  as functions of the coordinate $\xi$, which is determined from the
transcendental equation
\begin{equation}
 \ell_1 = \frac{\xi\, \exp{\left(-g_1\, \xi \right)}}{D(\xi)}\, ,
\end{equation}
\noindent  where
\begin{eqnarray}
 D(\xi)  &=&  \left(1 - \xi \right) + \left(G - \xi \right)\, S(\xi),  \\ [\smallskipamount]
 S(\xi)  &=&  \frac{\theta_2}{\theta_1} = S_0\, w(\xi),    \\
 S_0 \quad\  &\equiv& S(0) = \frac{\ell_2}{\ell_1}\, .
\end{eqnarray}

Thus, the problem under study is reduced to the investigation of the
equilibrium position of oscillator $\xi$ in an adsorption-induced force field,
i.e., dependence of a solution of Eq.~(18) on the control parameters $\ell_1,
\chi_1$ and $\ell_2, \chi_2$. In what follows, as control parameters, we use
$\ell_1, g_1$ (for particles of species 1) and $S_0,\, G$ (for particles of
species 2) equal to, respectively, $\ell_2$ and $\chi_2$ normalized by $\ell_1$
and $\chi_1$. For the classical adsorption of a binary gas mixture, the
quantity $S_0$  for $C_2 = C_1$ called the separation factor
\cite{ref.Rut,ref.KSt} (or the adsorbent selectivity of particles of species 2
in relation to particles of species 1 \cite{ref.Do,ref.Yang}) is independent of
the gas concentration. Thus, $w(\xi)$ characterizes the deviation of the
quantity  $S(\xi)$ from its classical analog $S_0$ due to the adsorbent
deformation in adsorption. \looseness=-1

To pass to the case of adsorption of a one-component gas of species 1, we set
$C_2 = 0$ in relations (16)--(21), which yields $\xi = \theta_1$ and the
following equation for the surface coverage $\theta_1$ on a deformable
adsorbent \cite{ref.Usenko}:
\begin{equation}
 \ell_1 = \frac{\theta_1}{1 - \theta_1} \,
  \exp{\left( - g_1\, \theta_1 \right)}\, .
\end{equation}

According to (20), the quantity $S(\xi)$  depends on both the dimensional
concentrations of gas particles of both species and the adsorption-induced
forces.

Passing in relations (14), (16)--(20) to the limit $\chi_1,\, \chi_2
\rightarrow 0$, we obtain the classical extended Langmuir (Markham--Benton)
isotherms of a binary gas mixture \cite{ref.AdC,ref.FrK,ref.Do}
\begin{equation}
 \theta_n^L = \frac{\ell_n}{1 + \ell_1 + \ell_2}\, ,   \qquad n = 1,2,
\end{equation}
\noindent and $\lim\limits_{\chi_1,\, \chi_2 \rightarrow 0}\,S(\xi) = S_0$.
Since the adsorbent surface is more intensively occupied by gas particles with
a greater dimensionless concentration, for  $S_0 \ll 1$, we can neglect the
presence of particles of species 2 in the binary gas mixture, and the problem
under study can be regarded as the problem of adsorption of a one-component
gas.

It follows from (20) that  $S(\xi)$ is equal to $S_0$ only for $\chi_1 =
\chi_2$. In this special case, the adsorbent deformation in adsorption leads to
an increase in the numbers of adparticles of each species not changing their
ratio $S_0$.

For  $\chi_1 \neq \chi_2$, the quantity $S(\xi)$ nonlinearly depends on the
concentrations $\ell_1$ and $\ell_2$ and the parameters  $g_1$ and $G$, and the
problem of neglect of gas particles of the second species in a binary gas
mixture in adsorption for $S_0 \ll 1$ remains open. In the general case, to
substantiate the passage from the two-component adsorption to the one-component
adsorption, it is necessary to investigate in detail the behavior of $S(\xi)$
as a function of the control parameters in the entire range of their variation.
Nevertheless, several qualitative conclusions can be drawn without awkward
calculations. To this end, for $\chi_1 \neq \chi_2$, we consider the case of
the total coverage ($\theta_+ = 1$), which is realized for large (infinite, in
the limit) concentrations of gas particles provided that $S_0 \neq 0$. Using
relations (16)--(20), we obtain the following asymptotic values of the surface
coverages $\theta^a_n = \lim\limits_{\ell_1 \rightarrow \infty}\, \theta_n, \ n
= 1, 2$,
\begin{equation}
 \theta^a_1  = \frac{G - \xi^a}{G - 1}\, , \qquad
 \theta^a_2  = \frac{\xi^a - 1}{G - 1}\, , \qquad
\end{equation}
\noindent  where  $\xi^a$ is a root of the equation
\begin{equation}
 D(\xi) = 0\, ,
\end{equation}
\noindent  which belongs to the interval  $(1,\, G)$ if $G > 1$ or $(G, 1)$ if
$G < 1$. Since the concentration  $\ell_1$ is positive, the coordinate $\xi$
tends to its asymptotic value $\xi^a$ in a half neighborhood of the point
$\xi^a$ in which $\sign\,D(\xi) =\sign\,\xi$, which yields
\begin{equation}
 \lim\limits_{\xi \rightarrow \, \xi^a}\ell_1(\xi)\, = + \infty.
\end{equation}

Using (24), we obtain the simple expression for the asymptotic ratio of surface
coverages
\begin{equation}
 S(\xi^a) \equiv \frac{\theta^a_2}{\theta^a_1} = \frac{\xi^a - 1}{G - \xi^a}\, .
\end{equation}
\noindent  Thus, for the total coverage, under the condition
\begin{equation}
 \xi^a > \frac{G + 1}{2} \quad \mbox{if}  \quad  G > 1 \qquad  \mbox{or} \qquad
 \xi^a < \frac{G + 1}{2} \quad \mbox{if}  \quad  G < 1 \, ,
\end{equation}
\noindent the number of adparticles of species 2 is greater than the number of
adparticles of species 1 even if $S_0 \ll 1$, which indicates the necessity of
taking account of particles of both species in problems of adsorption of binary
gas mixtures. In what follows, the realization of condition (28) for $S_0 \ll
1$ will be shown for specific systems.

For given values of the control parameters $g_1$, $G$, and $S_0$, Eq.~(25) can
have several roots that belong to the above-mentioned interval and satisfy
condition (26). In this case, the quantities $\xi^a$ and $\theta^a_n$ have an
additional subscript indicating the number of the root, and the functions
$\xi(\ell_1)$ and $\theta_n(\ell_1)$ have several horizontal asymptotes in the
limit $\ell_1 \rightarrow + \infty$.

Analysis shows that, for  $G \neq 1$, the function  $\xi(\ell_1)$ has three
horizontal asymptotes  $\xi = \xi^a_1$,\, $\xi = \xi^a_2$,\, and $\xi =
\xi^a_3$ if  $g_1 > g^a_c$, where $g^a_c = 4/(G-1)^2$, and $S_0 \in (I^a_-,
I^a_+)$, where  $I^a_\pm = S^a_\pm$  if  $G > 1$ or $I^a_\pm = S^a_\mp$  if $G
< 1$,
\begin{equation*}
 S^a_\pm = \frac{1 \mp w_a\,\sign(G-1)}{1 \pm w_a\,\sign(G-1)}\, \exp{\left(2\,q_a\,\beta_\pm\right)},
\end{equation*}
\begin{equation}
 w_a = \sqrt{1 - \frac{1}{q_a}}\,,  \qquad q_a = \frac{g_1}{g^a_c}\,, \qquad
 \beta_\pm = \beta \pm w_a\,\sign(G-1)\,, \qquad   \beta = \frac{1+G}{1-G}\,.
 \end{equation}

For the interval $[I^a_-, I^a_+]$, its width  $I_a(g_1, G) = I^a_+ - I^a_-$ and
the coordinate of its center $S^a(g_1, G) = (S^a_+ + S^a_-)/2$ are equal to
\begin{eqnarray}
 I_a(g_1, G) &=& 2\, \bigl\{\left( 2q_a - 1 \right)\,\sinh{h} - h\,\cosh{h} \bigr\}\,
  (S^a_c)^{q_a}, \\
 S^a(g_1, G) &=& \bigl\{\left( 2q_a - 1 \right)\,\cosh{h} - h\, \sinh{h}\bigr\}\,
  (S^a_c)^{q_a},
\end{eqnarray}
\noindent  where $h = 2\, w_a\,q_a = 2\, \sqrt{q_a(q_a-1)}$\, and $S^a_c =
\exp{\left(2\beta\right)}$ is the critical value of $S_0$ for which the
interval $[I^a_-, I^a_+]$ degenerates into a point ($S^a_+ = S^a_- = S^a_c$)
for $g_1 = g^a_c$.  The  interval $[I^a_-, I^a_+]$ exists for $g_1
> g^a_c$ and lies from the left (if $ G > 1$) or from the right (if $G \in [0,
1)$) of  $S^a_c$; for $G < 0$, depending on  $q_a > 1$, the interval can both
contain and not contain  $S^a_c$.

If the coupling parameter  $g_1$ is close to the critical  $g^a_c$, i.e., $q_a
= 1 + \varepsilon, \ 0 < \varepsilon \ll 1$, then
\begin{equation}
 I_a(g_1, G) \approx \frac{8}{3}\, \varepsilon^{3/2}\, S^a_c,  \qquad
 S^a_c(g_1, G) \approx (1 + 2\beta\, \varepsilon) \, S^a_c.
\end{equation}

For a very strong coupling, $g_1 \gg g^a_c$ ($q_a \gg 1$),
\begin{equation}
 I_a(g_1, G) \approx 2\, S^a_c(g_1, G),  \qquad
 S^a_c(g_1, G) \approx \frac{1}{8q_a} \exp{\left(\frac{4q_a}{1-G}\right)}.
\end{equation}

For   $S_0 \notin [I^a_-, I^a_+]$, the function $\xi(\ell_1)$ has only one
horizontal asymptote $\xi = \xi^a_1$, whereas, for $S_0 \in (I^a_-, I^a_+)$, it
has three horizontal asymptotes. Furthermore, the appearance of two additional
asymptotes and their specific features essentially depend on the value of $G$.

For $G > 1$, as $S_0$  increases from a value lesser than $S^a_-$, for $S_0 =
S^a_-+0$, there appear two infinitely close asymptotes  $\xi = \xi^a_2$ and
$\xi = \xi^a_3$ above the asymptote  $\xi = \xi^a_1$\ ($\xi^a_1 < \xi^a_2 <
\xi^a_3$); furthermore, the asymptote  $\xi = \xi^a_3$, along with the
asymptote $\xi = \xi^a_1$, is stable and the asymptote $\xi = \xi^a_2$ is
unstable, which means that they are, respectively, asymptotes of the
corresponding stable and unstable branches of the function $\xi(\ell_1)$. In
the limiting case  $S_0 = S^a_-$, the asymptotes $\xi = \xi^a_2$ and $\xi =
\xi^a_3$ coalesce into one line  $\xi = \xi^a_-, $ where $\xi^a_- =
\beta_-\,(1-G)$/2, which is already not an asymptote of $\xi(\ell_1)$ because
$\ell_1(\xi)$ does not satisfy condition (26) for $\xi = \xi^a_-$. The distance
between the asymptotes $\xi = \xi^a_2$ and $\xi = \xi^a_3$ increases with  $S_0
\in (S^a_-, S^a_+)$. Moreover, the unstable $\xi = \xi^a_2$ and stable $\xi =
\xi^a_1$ asymptotes approach each other and, for $S_0 = S^a_+$, coalesce into
one doubly degenerate asymptote $\xi = \xi^a_+$, where $\xi^a_+ =
\beta_+\,(1-G)/2$, which, for $S_0 > S^a_+$, disappears, and the function
$\xi(\ell_1)$ again has one asymptote but $\xi = \xi^a_3$. \looseness=-1

For $G < 1$, the function $\xi(\ell_1)$ have three horizontal asymptotes if
$S_0 \in (S^a_+, S^a_-)$. However, its behavior with variation in $S_0$ differs
from that considered above for $G > 1$. For $G \in [0, 1)$, as $S_0$ increases,
for $S_0 = S^a_+$, the doubly degenerate asymptote $\xi = \xi^a_+$ appears
below the asymptote $\xi = \xi^a_1$. As  $S_0$ negligibly increases, this
asymptote splits into two infinitely close asymptotes: stable  $\xi = \xi^a_3$
and unstable $\xi = \xi^a_2$ ($\xi^a_1 > \xi^a_2 > \xi^a_3$). As  $S_0 \in
(S^a_+, S^a_-)$ increases, the distance between the asymptotes $\xi = \xi^a_2$
and $\xi = \xi^a_3$ grows and the unstable  $\xi = \xi^a_2$ and stable $\xi =
\xi^a_1$ asymptotes approach each other and, for  $S_0 = S^a_-$, coalesce into
one line $\xi = \xi^a_-$, which is already not an asymptote of $\xi(\ell_1)$
because $\ell_1(\xi)$  does not satisfy condition (26) for $\xi = \xi^a_-$. As
a result, for $S_0 > S^a_-$, the function $\xi(\ell_1)$ again has one asymptote
but $\xi = \xi^a_3$.

Thus, for $G \ge 0$, the function $\xi(\ell_1)$ has one horizontal doubly
degenerate asymptote  $\xi = \xi^a_+$  if the value of $S_0$ coincides with the
right end point (for $C > 1$) or the left end point (for $G \in [0, 1))$ of the
interval $[I^a_-, I^a_+]$.

For $G < 0$, the behavior of $\xi(\ell_1)$ depends on signs of $\xi^a_+$ and
$\xi^a_-$. Note that $\xi^a_+ < \xi^a_-$ for any $G$ and $g_1 > g^a_c$. In the
special case $g_1 = g^a_c$, $\xi^a_+ = \xi^a_- \equiv \xi^a_c = (1+G)/2$. If
$\xi^a_+ > 0$, then the behavior of the function $\xi(\ell_1)$ is similar to
its behavior for $G \in [0, 1)$. If $\xi^a_+ < 0$ and $\xi^a_- > 0$, then this
behavior of $\xi(\ell_1)$ remains valid except for the case $S_0 = S^a_+$ for
which the line $\xi = \xi^a_+$ is already not a doubly degenerate asymptote of
$\xi(\ell_1)$. Thus, for these values of $G$, the function $\xi(\ell_1)$ does
not have a horizontal doubly degenerate asymptote. If $\xi^a_- < 0$, then the
function $\xi(\ell_1)$ has the doubly degenerate asymptote  $\xi = \xi^a_-$ for
$S_0 = S^a_-$  if $G < -1$ and $g_1 \in (g^a_c, 1/|G|)$; otherwise, the
function $\xi(\ell_1)$ does not have a doubly degenerate asymptote.

According to (16) and (17), for $S_0 \in (I^a_-, I^a_+)$, the functions
$\theta_1(\ell_1)$ and $\theta_2(\ell_1)$ also have three horizontal asymptotes
$\theta_1 = \theta^a_{1,k}$ and $\theta_2 = \theta^a_{2,k},\ k = 1, 2, 3$, two
of which are stable and one is unstable. At the end points of this interval,
the doubly degenerate asymptotic values of the surface coverages
$\theta^a_{1,-}$ and $\theta^a_{2,-}$ (for $S_0 = S^a_-$) or $\theta^a_{1,+}$
and $\theta^a_{2,+}$ (for $S_0 = S^a_+$) are equal to
\begin{equation}
 \theta^a_{1,\pm} = \frac{1 \pm w_a\,\sign(G-1)}{2}\,  \qquad
 \theta^a_{2,\pm} = \frac{1 \mp w_a\,\sign(G-1)}{2}\,,
\end{equation}
\noindent  and $|\theta^a_{1,\pm} - \theta^a_{2,\pm}| = w_a$ increases with the
coupling parameter $g_1$. It follows from (34) that, for $g_1 = g^a_c$, the
quantities $\theta^a_{1,+} = \theta^a_{1,-} \equiv \theta^a_{1,c}$ and
$\theta^a_{2,+} = \theta^a_{2,-} \equiv \theta^a_{2,c}$, where $\theta^a_{n,c}
= 1/2, \ n = 1, 2$, are equal each other and, unlike $\xi^a_c $, independent of
$G$. Since only the surface coverages (34) consistent with condition (26) have
a physical meaning, we obtain that, e.g., for $G \geq 0$, these quantities are
$\theta^a_{1,+}$ and $\theta^a_{2,+}$, which yields $\theta^a_{1,+} >
\theta^a_{2,+}$ for $G > 1$ and $\theta^a_{1,+} < \theta^a_{2,+}$ for $G \in
[0, 1)$.  \looseness=-1


\subsection{Identical Action of Adparticles on the Adsorbent: $\chi_1 = \chi_2$}  \label{Identical action}

In this simplest case, $g_1 = g_2 \equiv g,\  G = 1$, and the required
quantities $\xi$  and  $\theta_n$ are defined only by one quantity $\theta_+$
\begin{equation}
 \xi = \theta_+\, ,  \qquad \quad
 \theta_1 = \frac{\theta_+}{1 + S_0}\, , \qquad \quad
 \theta_2 = S_0\, \theta_1 \, .
\end{equation}

The surface coverage  $\theta_+$ is a solution of the equation
\begin{equation}
 \ell_+ = \frac{\theta_+}{1 - \theta_+} \,
  \exp{\left( - g\, \theta_+ \right)}\, ,
\end{equation}
\noindent  where
\begin{equation}
 \ell_+ = \ell_1 + \ell_2 = \left(1 + S_0 \right)\ell_1
\end{equation}
is the summary dimensionless concentration.

Since Eq.~(36) coincides with the equation for one-component adsorption (22)
with replacements of $\theta_1$ by $\theta_+$ and $\ell_1$ by $\ell_+$, the
problem of adsorption of a two-component gas is reduced to the problem of
adsorption of a one-component gas with the dimensionless concentration  $l_+$
and the coupling parameter  $g$. This enables us to directly use the results
obtained in \cite{ref.Usenko} for the one-component adsorption.

First, consider the case of a small coupling parameter, $g \ll 1$. Using (35)
and (36), we get
\begin{equation}
 \theta_n \approx \theta_n^L \, \biggl\{ 1 + g \,\frac{\ell_+}{(1 + l_+)^2} \biggr\},
  \qquad n = 1,2.
\end{equation}

Since the second term on the right-hand side of (38) is positive, the adsorbent
deformation in adsorption increases the number of adparticles of both species.
This result agrees with the general conclusion presented below of an increase
in the number of adparticles due to the adsorbent deformation, which is true
for any value of $g$. Indeed, rewriting (36) in the form
\begin{equation}
 \frac{\theta_+}{1 - \theta_+} = \ell_+ \, \exp{\left(g\, \theta_+ \right)}
\end{equation}
\noindent  and taking into account that the quantities $\theta_+/(1 -
\theta_+)$ and $\ell_+$ are equal to the ratios of the number of bound
adsorption sites to the number of vacant adsorption sites, respectively, with
and without regard for the adsorbent deformation in adsorption, we immediately
establish that the surface coverage $\theta_+$  is greater than that in the
Langmuir case for any gas concentration. The difference between the number of
adparticles in the nonlinear ($g \neq 0$) and linear ($g = 0$) cases increases
with the coupling parameter $g$.

Using analysis of adsorption isotherms in \cite{ref.Usenko}, we obtain that the
surface coverage  $\theta_+$ essentially depends on values of $g$. For $g < g_c
=4$, as in the Langmuir case, the system is monostable: there is a
single-valued correspondence between the concentration $\ell_+$ and the surface
coverage  $\theta_+$. For $g > g_c$, the situation cardinally changes: if
$\ell_+ \notin [\ell^b_{+,1}, \, \ell^b_{+,2}]$, where $\ell^b_{+,1}$  and
$\ell^b_{+,2}$ are the bifurcation concentrations whose explicit expressions
are given below, then, as before, for every $\ell_+$, there is a unique
$\theta_+$, whereas, for any $\ell_+ \in (\ell^b_{+,1}, \, \ell^b_{+,2})$,
there are three values of $\theta_+$: \ $\theta_{+,1} < \theta_{+,2} <
\theta_{+,3}$. Furthermore, the stationary solutions $\theta_{+,1}$ and
$\theta_{+,3}$ of system (7) are asymptotically stable and the stationary
solution $\theta_{+,2}$ is unstable.

If the concentration  $\ell_+ \in [\ell^b_{+,1}, \, \ell^b_{+,2}]$ tends to the
end point $\ell^b_{+,1}$ (or $\ell^b_{+,2}$) of the interval, then the stable
$\theta_{+,3}$ (or $\theta_{+,1}$) and unstable $\theta_{+,2}$ solutions
approach each other and, in the limit $\ell_+ = \ell^b_{+,1}$ (or $\ell_+ =
\ell^b_{+,2}$), coalesce into the two-fold solution $\theta^b_{+,1}$ (or
$\theta^b_{+,2}$)
\begin{equation}
 \theta^b_{+,1} = \frac{1 + w_+}{2}\,  \qquad  \mbox{or}  \qquad
 \theta^b_{+,2} = \frac{1 - w_+}{2}\, ,
\end{equation}
\noindent  where the quantity
\begin{equation}
 w_+ = \sqrt{1 - \frac{4}{g}}
\end{equation}
\noindent is the width of the interval of instability for $\theta_+$ symmetric
about $1/2$.

The bifurcation concentrations  $\ell^b_{+,1}$ and $\ell^b_{+,2}$ for which the
dynamical system (7) has two stationary solutions one of which
($\theta^b_{+,1}$ or $\theta^b_{+,2}$) is two-fold are equal to
\cite{ref.Usenko}
\begin{equation}
 \ell^b_{+,n} = \bigl( g \, \theta^b_{+,n} - 1 \bigr)\, \exp{(-g \, \theta^b_{+,n})},
  \quad  n = 1,2.
\end{equation}

Thus, for $g > g_c$, there is an interval of values of $\ell_+$ whose end
points  $\ell^b_{+,1}$, $\ell^b_{+,2}$ and width
\begin{equation}
 I_+(g) = \ell^b_{+,2} - \ell^b_{+,1}
        = \Bigl\{\left( g - 2 \right)\, \sinh{\frac{gw_+}{2}} - gw_+\, \cosh{\frac{gw_+}{2}}
          \Bigr\} \exp{\left(-\frac{g}{2}\right)}
\end{equation}
\noindent  depend on the coupling parameter $g$ so that the system is bistable
if the concentration $\ell_+$ belongs to this interval. We call this interval
of concentrations  $\ell_+$ the bistability interval of the system. Note that
relation (43) coincides with the width of the interval of pump intensity
obtained in \cite{ref.Chr} for bistability of a macromolecular in repeating
cycles of reactions.

If the coupling parameter $g$  is close to the critical value $g_c$, i.e., $g =
4\,(1 + \varepsilon), \ 0 < \varepsilon \ll 1$, then the bistability interval
is very narrow
\begin{equation}
 I_+(g) \approx \frac{8}{3}\, \exp{(-2)}\, \varepsilon^{3/2}\, ,
\end{equation}
\noindent the stationary solutions $\theta_{+,1}, \theta_{+,2}$, and
$\theta_{+,3}$ are close to each other, and  $w_+ \approx \varepsilon^{1/2}$.
In the limit $\varepsilon \rightarrow +0\,$, the bistability interval
disappears and three stationary solutions coalesce into the three-fold solution
$\theta_+^c = 1/2$. Thus, for the critical values of the control parameters ($g
= 4$ and  $\ell_+ = \ell_+^c = \exp{(-2)} \approx 0.135$), the dynamical system
(7) has one three-fold stationary solution.  \looseness=-1

\begin{figure}[!ht]
 \centering{
 \includegraphics[width=140mm, height=115mm]{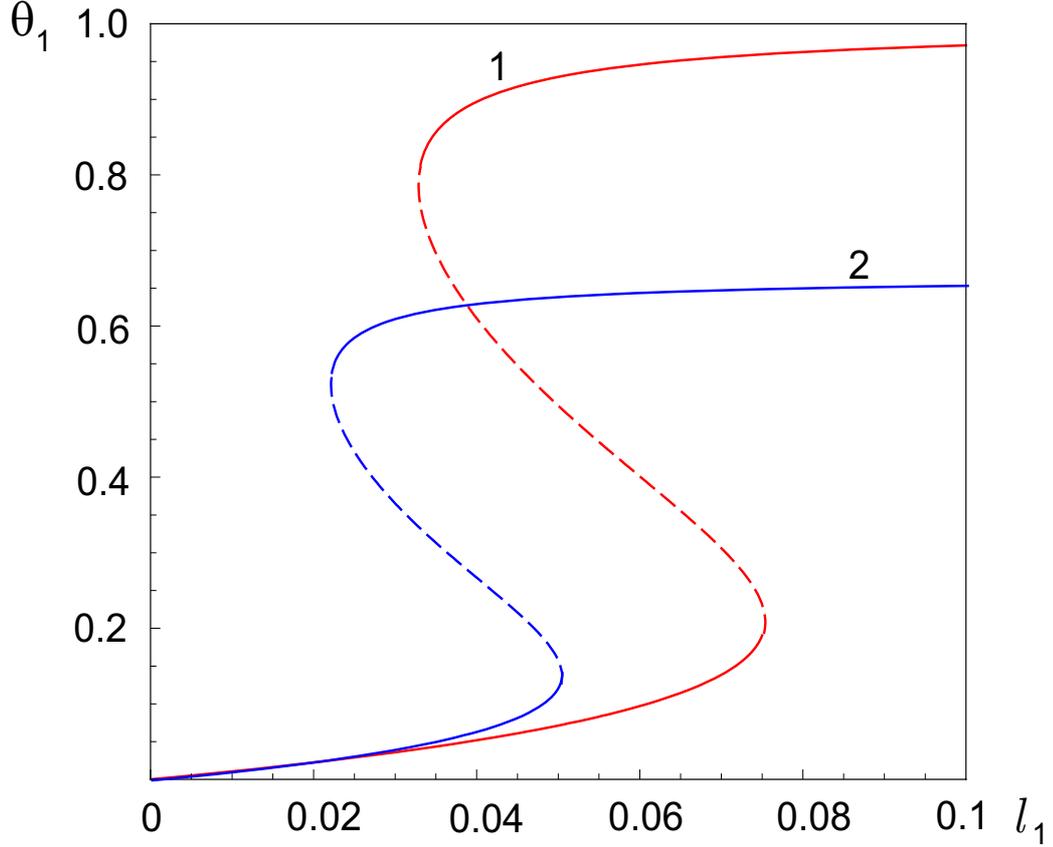}
 \caption {Adsorption isotherms of adparticles of species 1 for
  one-component (curve 1) and two-component (curve 2) gas
  for  $g$ = 6,  $G$ = 1, $S_0$ = 0.5.}
 }
 \label{pRGBfigc._1}
\end{figure}

The comparison of the $S-$shaped adsorption isotherms of adparticles of species
1 in Fig.~1 for one-component (curve 1) and two-component (curve 2) gas for $g
= 6 > g_c$ clearly illustrate the influence of particles of species 2 in a gas
mixture on the behavior of the surface coverage $\theta_1$. In this and
subsequent figures for the surface coverages $\theta_n(\ell_1)$ and the
equilibrium position of oscillator $\xi(\ell_1)$, parts of curves corresponding
to stable and unstable stationary solutions are shown, respectively, by solid
and broken lines.

The obtained adsorption isotherms essentially differ from the classical
Langmuir isotherms. At the same, for the model taking into account variations
in adsorption properties of the adsorbent in adsorption, the presence of
particles of species 2 in the gas phase leads only to quantitative changes in
adsorption isotherms of a one-component gas \cite{ref.Usenko}: a decrease in
the amount of adsorbed substance and displacement and decrease in the
bistability interval of the system, which completely agrees with relations
(35), (37), and (42).

As the concentration $\ell_1$ increases from zero, the surface coverage
$\theta_1$ increases along the lower stable branch of the isotherm and the
increment of the surface coverage is determined by both an increase in the gas
concentration and a change in adsorption properties of the adsorbent due to its
deformation. Since the lower stable branch of the isotherm ends at $\ell_1 =
\ell^b_{1,2}$, a negligible excess of the bifurcation concentration
$\ell^b_{1,2}$ is accompanied by the jump to the upper stable branch of the
isotherm, i.e., a stepwise increase in the surface coverage solely due to a
change in adsorption properties of the adsorbent. This transition can include
many gas particles (furthermore, of both species) successively taking part in
the process of adsorption--desorption on the same adsorption site. Thus, at
this stage, a certain interaction between the particle leaving the adsorption
site and the particle binding with it occurs.

A subsequent increase in the concentration $\ell_1$ slightly affects the
surface coverage $\theta_1$ varying along the upper stable branch because the
majority of adsorption sites are already bound either with particles of species
1 or with particles of species 2.

In passing through the bifurcation concentration  $\ell^b_{1,2}$, desorption of
adparticles essentially decreases due to a considerable increase in their
activation energy for desorption. As a result, for returning from the upper
stable branch of the isotherm to its lower stable branch, the concentration
must be lower than $\ell^b_{1,2}$. As the concentration  $\ell_1$ decreases
from a value greater than  $\ell^b_{1,2}$, the surface coverage $\theta_1$
decreases along the upper stable branch of the isotherm up to its end at
$\ell_1 = \ell^b_{1,1}$.  In passing through the bifurcation concentration
$\ell^b_{1,1}$, the surface coverage jumps down to the lower stable branch of
the isotherm and then decreases along this branch.

The behavior of the surface coverage $\theta_1$ vs $\ell_1$ agrees with the
principle of perfect delay \cite{ref.PoS, ref.Gil}:  a system, which is in a
stable state at the initial time, remains in this state with variation in a
parameter (the concentration  $\ell_1$ in the case at hand) until the state
exists.

According to (35), the specific features of adsorption isotherms are also true
for the coordinate  $\xi$  characterizing the displacement of the plane of
adsorption sites from its nonperturbed position. For example, the curves in
Fig.~1 also describe the equilibrium position of oscillator vs $\ell_1$ in
adsorption of one-component and two-component gas if, instead of $\theta_1$,
$\xi$ (for one-component adsorption) and $(2/3)\, \xi$ (for two-component
adsorption) are laid off along the ordinate axis. \looseness=-1



\subsection{Equilibrium Position of Oscillator}  \label{Equilibrium position}

The equilibrium position of oscillator $\xi(\ell_1)$ is a solution of Eq.~(18).
To analyze solutions of this transcendental equation, we plot the function
$\ell_1(\xi)$  inverse to the required $\xi(\ell_1)$, i.e., the right-hand side
of Eq.~(18). The abscissas of the points of intersection of the graph of the
function $\ell_1(\xi)$ with a horizontal line corresponding to the given
concentration $\ell_1 > 0$ are solutions of Eq.~(18). Thus, the problem under
study is reduced to the investigation of the function  $\ell_1(\xi) \ge 0$
depending on the control parameters $\, g_1, G$, and $S_0$.

Points of possible finite local extrema of the function $\ell_1(\xi)$ are
solutions of the equation
\begin{equation}
 L(\xi; g_1, G, S_0) = 0,
\end{equation}
\noindent  where
\begin{equation}
L(\xi; g_1, G, S_0) = L_1(\xi; g_1) + G\, S(\xi)\,L_2(\xi; g_1, G),
\end{equation}
the quantity
\begin{equation}
 L_1(\xi; g_1) = 1 + g_1 \xi \left( \xi - 1 \right)
\end{equation}
\noindent  is associated with adsorption of a one-component gas of species 1
and the quantity
\begin{equation}
 L_2(\xi; g_1, G) = 1 + g_1 \xi \left( \xi - G \right)
\end{equation}
\noindent  is caused by the presence of particles of species 2 in the binary
gas mixture.

In the special case where adparticles of species 2 do not affect the adsorbent
deformation, $\chi_2 = 0 \ (G = 0)$, Eq.~(45) coincides with the equation
\begin{equation}
 L_1(\xi; g_1) = 0
\end{equation}
\noindent  for points of possible extrema of $\ell_1(\xi)$  in adsorption of a
one-component gas of species 1.

Note that Eq.~(45) is also reduced to Eq.~(49) in other special case
investigated in Sec.~3.2 of the identical action of all adparticles on the
adsorbent, $\chi_1 = \chi_2 \ (G = 1)$.

We denote roots of Eq.~(45) for $S_0 > 0$ by $\xi^b_k$, where $k = 1, 2, \ldots
$, and call roots  $\xi^b_k$ for which the bifurcation concentrations
$\ell^b_{1,k} = \ell_1(\xi^b_k) > 0 $ bifurcation coordinates. Using (16),
(17), and (45)--(48), for $G \neq 0, 1$, we obtain the following relations for
the bifurcation surface coverages $\theta^b_{1,k} = \theta_1(\xi^b_k)$ and
$\theta^b_{2,k} = \theta_2(\xi^b_k)$:
\begin{equation}
 \theta^b_{1,k} = \frac{L_2(\xi^b_k; g_1, G)}{g_1 \left(1 - G \right)}, \qquad\quad
  \theta^b_{2,k} = \frac{L_1(\xi^b_k; g_1)}{g_1 G \left(G -1 \right)}.
\end{equation}

The function  $\ell_1(\xi)$ has a finite local extremum at the point   $\xi =
\xi^b_k$ if the function
\begin{equation}
 L_c(\xi; g_1, G) = 2\xi - 1 + \left(\xi - G \right)\, L_1(\xi; g_1)
\end{equation}
\noindent  is not equal to zero at this point. Otherwise, for
\begin{equation}
 L_c(\xi^b_k; g_1, G) = 0\, ,
\end{equation}
\noindent  the investigation of  $\ell_1(\xi)$ at this point must be continued.
By $\xi^c_k, \ k = 1, 2, 3$, we denote real roots of the cubic equation (52).
Both the number of these roots and their values depend on the parameters $g_1$
and $G$. We call roots $\xi^c_k$ for which the critical concentrations
$\ell^c_{1, k} = \ell(\xi^c_k) > 0 $  critical coordinates. The critical values
of $S_0 > 0$ denoted by $S^c_k$ are determined from Eq.~(45) for $\xi =
\xi^c_k$. The critical surface coverages $\theta^c_{1,k}$ and $\theta^c_{2,k}$
are defined by relations (50) with $\xi^b_k$ replaced by $\xi^c_k$.
\looseness=-1

The more detailed analysis shows that Eq.~(18) has a maximum (five-fold)
multiple root for three values of the parameter $G$ equal to 2\,, 1/2\,, and -1
and the corresponding values of  the other parameters $g_1,\, \ell_1,\,S_0$. In
the four-dimensional space of control parameters $\{\ell_1,\, S_0,\, g_1,\,
G\}$, a point with coordinates $\ell_1^{\,but},\, S_0^{\,but},\,g_1^{\,but},\,
G^{\,but}$ gives a five-fold stationary solution of system (7). In the
three-dimensional space of solutions $\{\xi,\, \theta_1,\, \theta_2\}$, this
five-fold solution is a point with coordinates $\xi^{\,but},\,
\theta_1^{\,but},\, \theta_2^{\,but}$. The values of three five-fold stationary
solutions of system (7) and the corresponding control parameters are given in
Table~1.

\begin{center}
 Table~1. Control parameters for five-fold solutions.  \\[7pt]
 \setlength{\extrarowheight}{3pt}
\begin{tabular}{|c|c|c|c|c|c|c|c|}
 \hline
 No. & \multicolumn{4}{c|}{Control parameters}
     & \multicolumn{3}{c|}{Solutions}\\[2pt] \cline{2-8}
     & $G^{\,but}$ & $g_1^{\,but}$ & $\ell_1^{\,but}$ & $S_0^{\,but}$ &
 $\xi^{\,but}$ & $\theta_1^{\,but}$ & $\theta_2^{\,but}$  \\[2pt] \hline
 1 & 2   &  3 & $4 \exp{(-3)}$ & $  \exp{(-3)}/4$ &  1   &  2/3 & 1/6  \\
 2 & 1/2 & 12 & $  \exp{(-6)}$ & $4 \exp{(3)}$    &  1/2 &  1/6 & 2/3  \\
 3 & -1  &  3 & 1/4            & 1                &  0   &  1/6 & 1/6  \\
 \hline
\end{tabular}
\end{center}

In the general case, analysis of stationary solutions of system (7) depending
on control parameters is a complicated problem. First, we decrease the
dimension of the space of control parameters by fixing a value of the parameter
$G$, i.e., select a three-dimensional subspace of control parameters
$\{\ell_1,\, S_0,\, g_1\}$ from the original four-dimensional space. Among all
three-dimensional subspaces thus obtained, there are only three subspaces for
$G = 2\,, 1/2\,, -1$ each of which contains a unique point with coordinates
$\ell_1^{\,but},\, S_0^{\,but},\,g_1^{\,but}$ giving a five-fold stationary
solution of system (7). Moreover, in these three cases, analytic expressions
are relatively simple. Then, using the results of analysis of stationary
solutions of system (7) in these special cases, we can draw the corresponding
conclusions for values of $G$ for which system (7) does not have five-fold
stationary solutions. Since the case of negative values of $G$ is of interest
in its own right, the case  $G = -1$ is not investigated here.  In view of the
fact that the cases $G = 2$ and $G = 1/2$ are similar (see Table~1), below, we
consider the case $G = 2$.



\subsection{Case $ G = 2$}  \label{G 2}

In this case, Eq.~(52) has three roots
\begin{equation}
 \xi^c_1 = 1, \qquad
 \xi^c_2 \equiv \xi^c_+ = 1 + q, \qquad
 \xi^c_3 \equiv \xi^c_- = 1 - q, \qquad
 q = \sqrt{1 - \dfrac{3}{g_1}}\,,
\end{equation}
\noindent  which are horizontal points of inflection of the function
$\ell_1(\xi)$ and, furthermore, at the points  $\xi = \xi^c_+$  and  $\xi =
\xi^c_-, \  d^3\, \ell_1(\xi)/d\xi^3 \neq 0$, whereas, at the point  $\xi =
\xi^c_1, \ d^3\, \ell_1(\xi)/d\xi^3 \neq 0$ if $g_1 \neq 3$ and $d^3\,
\ell_1(\xi)/d\xi^3 = d^4\, \ell_1(\xi)/d\xi^4 = 0, \ d^5\, \ell_1(\xi)/d\xi^5
\neq 0$ if $g_1 = 3$. According to (53), the function $\ell_1(\xi)$ has one
horizontal point of inflection  $\xi^c_1$ if $g_1 < 3$ and three horizontal
points of inflection if $g_1 > 3$; furthermore, only two of them ($\xi^c_+$ and
$\xi^c_-$) depend on the coupling parameter $g_1$. This result essentially
differs from results of adsorption of a one-component gas or a two-component
gas for  $G = 1$ for which the function $\ell_1(\xi)$ has only one horizontal
point of inflection, $\xi^c = 1/2$, for $g_1 = 4$. If $\lim\limits_{g_1
\rightarrow 3\, +0}\,$, then three roots (53) coalesce into one triple root.

The critical parameters $\ell^c_{1,k}$ and $S^c_k$ (redenoted as follows:
$\ell^c_{1,2} \equiv \ell^c_{1,+}, \ \ell^c_{1,3} \equiv \ell^c_{1,-}, \ S^c_2
\equiv S^c_+, \ S^c_3 \equiv S^c_-$) are equal to
\begin{eqnarray}
 \ell^c_{1,1}   &=& 2 \left(g_1 - 1 \right)\, \exp{(-g_1)}, \qquad
 \ell^c_{1,\pm}  = \frac{4 }{g_1 \xi^c_{\mp} - 2}\, \exp{(-g_1 \xi^c_{\pm})}\,,  \\
 S^c_1    &=& \frac{\exp{(-g_1)}}{2 \left(g_1 - 1 \right)}\,,
 \qquad\qquad\qquad\
 S^c_{\pm} = \frac{g_1 \xi^c_{\pm} - 2}{4}\, \exp{(-g_1 \xi^c_{\pm})}\,.
\end{eqnarray}

Nonnegativity of the quantities $\ell_1$ and $S_0$ imposes the following
restrictions on  $g_1$:  $\ g_1> 1$ for $\xi^c_1$ and $g_1 \in [3, 4)$ for
$\xi^c_{\pm}$. Thus, the function $\ell_1(\xi) \ge 0$ has three horizontal
points of inflection only for $ g_1 \in (3, 4)$ and, hence, an essential
difference between adsorption isotherms of two-component and one-component
gases are expected precisely in this range of values of $g_1$. The quantities
$\ell^c_{1,1}$, $\ell^c_{1,\pm}$ and $S^c_1$, $S^c_{\pm}$ as functions of the
coupling parameter $g_1$ are arranged as follows: $\ell^c_{1,+} > \ell^c_{1,-}
> \ell^c_{1,1}$ and $S^c_1 > S^c_+ > S^c_-$ for any  $g_1 \in (3, 4)$ and
coincide ($\ell^c_{1,1} = \ell^c_{1,\pm} = \ell_1^{\,but} \approx 0.199$ and
$S^c_1 = S^c_{\pm} = S_0^{\,but} \approx 0.0124$) for $g_1 = 3$.

Substituting (54) and (55) into (50), we obtain the following critical surface
coverages $\theta^c_{n,k}$ (redenoted as follows: $\theta^c_{n,2} \equiv
\theta^c_{n,+},\ \theta^c_{n,3} \equiv \theta^c_{n,-}$,\ where $n = 1, 2$):
\begin{equation}
 \theta^c_{1,1}   = 1 - \frac{1}{g_1}, \quad
 \theta^c_{1,\pm} = \frac{2}{g_1}, \quad
 \theta^c_{2,1}   = \frac{1}{2 g_1}, \quad
 \theta^c_{2,\pm} = \frac{g_1 \xi^c_{\pm} - 2}{2 g_1}\,.
\end{equation}
\noindent  In the degenerate case $g_1 = 3$, we have  $\theta^c_{1,1} =
\theta^c_{1,\pm} = \theta_1^{\,but} = 2/3$ \ and \ $\theta^c_{2,1} =
\theta^c_{2,\pm} = \theta_2^{\,but} = 1/6$.

In the case $g_1 = 3.5$ considered below, $\xi^c_+ \approx 1.378, \ \xi^c_-
\approx 0.622$ and $\ell^c_{1,1} \approx 0.151,\ \ell^c_{1,+} \approx 0.182,\
\ell^c_{1,-} \approx 0.161,\ S^c_1 \approx 0.00604,\ S^c_+ \approx 0.00568,\
S^c_- \approx 0.00502$.

The graphs of the function $\ell_1(\xi)$ for different values of $S_0$ are
shown in Fig.~2. The required solutions $\xi$ of Eq.~(18) are the abscissas of
the points of intersection of a dashed horizontal line corresponding to the
given concentration $\ell_1$ with the graph of the function $\ell_1(\xi)$.

\begin{figure}[!ht]
 \centering{
 \includegraphics[width=75mm, height=100mm]{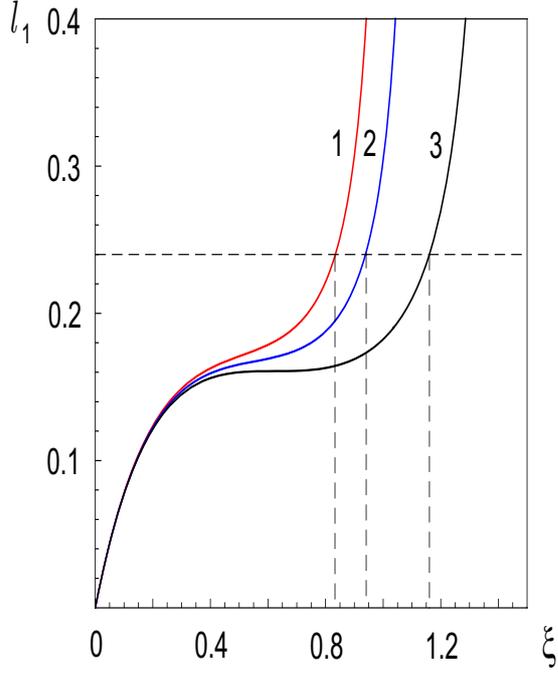} \hfill
 \includegraphics[width=75mm, height=100mm]{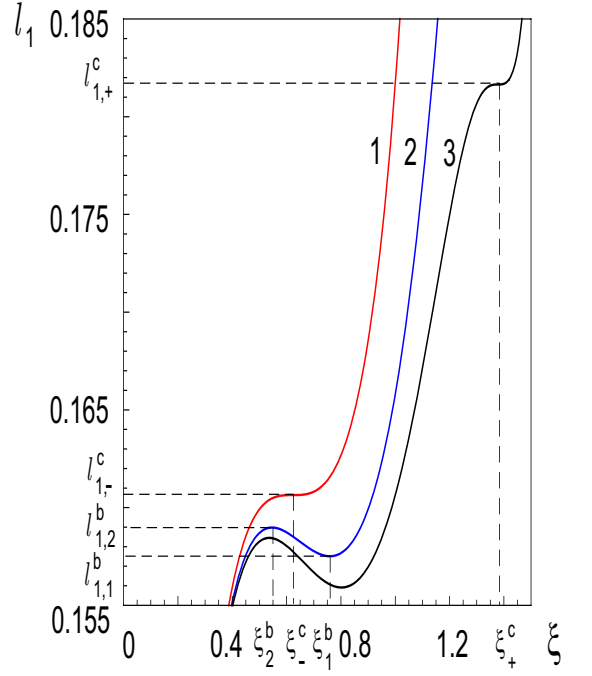}  \\
  \vfill
 \includegraphics[width=75mm, height=100mm]{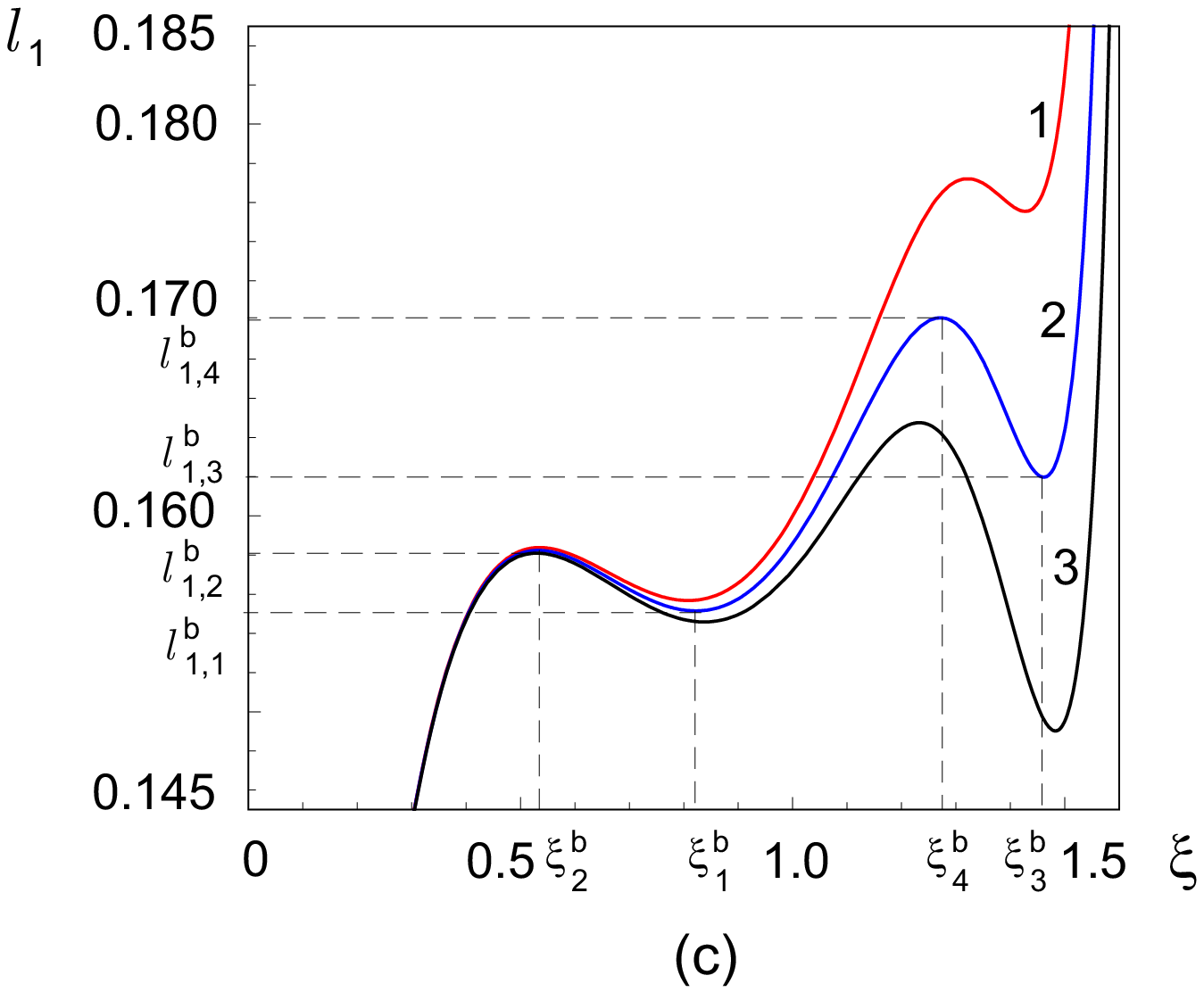} \hfill
 \includegraphics[width=75mm, height=100mm]{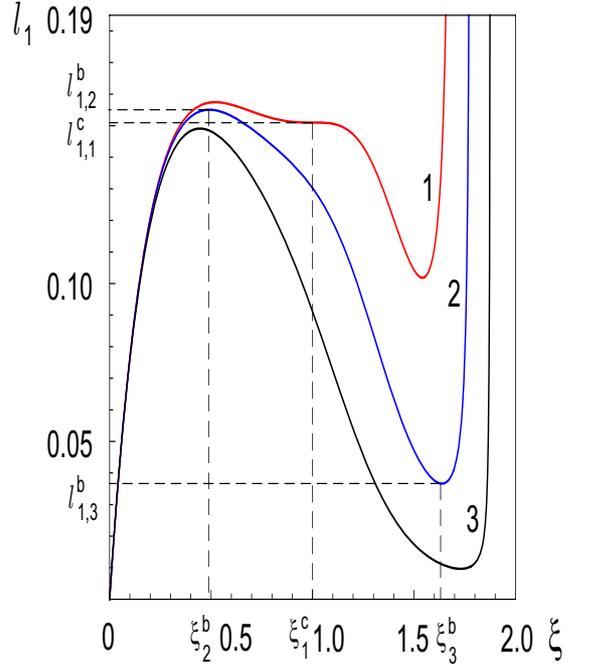}  \\
  \vfill
 \caption {Graphical solution of Eq.~(18) for different values of  $S_0$:
  (a) $S_0$ = 0.001~(1),    0.003~(2),    0.005~(3);
  (b) $S_0$ = $S^c_-$~(1),  0.0055~(2),   $S^c_+$~(3);
  (c) $S_0$ = 0.0057~(1),   0.00575~(2),  0.0058~(3);
  (d) $S_0$ = $S^c_1$~(1),  0.007~(2),    0.01~(3);
  $G = 2,  g_1 = 3.5$.
  Horizontal dashed lines correspond to constant values of the
  concentration $\ell_1$.}
 }
 \label{pRGBfigc._2}
\end{figure}

For low concentrations  $\ell_2$ such that  $S_0 < S^c_-$, the curve of the
function $\ell_1(\xi)$ intersects any horizontal line of the given
concentration  $\ell_1$ at one point, which gives a unique value of $\xi$ for
any $\ell_1$ (Fig.~2a). An increase in $S_0$ is accompanied by an increase in
the number of adparticles of species 2 and, hence, the adsorption-induced force
acting on adsorption sites, which increases their displacement from the
nonperturbed equilibrium position $\xi = 0$. For the least critical $S_0 =
S^c_-$ (curve 1 in Fig.~2b), the function $\ell_1(\xi)$ has a horizontal point
of inflection at $\xi = \xi^c_-$ and its value at this point is equal to the
critical concentration  $\ell^c_{1,-}$ ($\ell^c_{1,-}$ and $\xi^c_-$ are
depicted in Fig.~2b). A negligible increase in $S_0$ leads to the deformation
of the curve in the neighborhood of the point $\xi^c_-$ so that there appear a
minimum and a maximum of the function $\ell_1(\xi)$ equal to the bifurcation
concentrations $\ell^b_{1,1}$ and $\ell^b_{1,2}$ ($\ell^b_{1,1} < \ell^b_{1,2}
< \ell^c_{1,-}$), respectively, at the points $\xi = \xi^b_1$ and $\xi =
\xi^b_2$. As $S_0$ increases, the bifurcation concentrations $\ell^b_{1,1}$ and
$\ell^b_{1,2}$ decrease and the width $I_{2,1} = \ell^b_{1,2} - \ell^b_{1,1}$
of the interval $[\ell^b_{1,1}, \,\ell^b_{1,2}]$ called the first bistability
interval of the system increases (cf. $I_{2,1}$ for curves 2 and 3 in Fig.~2b).
In Fig.~2b, the bifurcation concentrations $\ell^b_{1,1}$ and $\ell^b_{1,2}$
and the bifurcation coordinates  $\xi^b_1$ and $\xi^b_2$ are shown for $S_0 =
0.0055$, $S_0\in (S^c_-,\, S^c_+)$. The situation is similar to that in
adsorption of a one-component gas \cite{ref.Usenko} or a two-component gas for
$G = 1$ if values of the coupling parameter are greater than critical: For
$\ell_1 \notin[\ell^b_{1,1}, \, \ell^b_{1,2}]$, there is a single-valued
correspondence between the concentration $\ell_1$ and the coordinate $\xi$; for
any $\ell_1\in (\ell^b_{1,1}, \, \ell^b_{1,2})$, there are three values of the
coordinate $\xi$: \ $\xi_1 < \xi_2 < \xi_3$; furthermore, the stationary
solutions $\xi_1$ and $\xi_3$ of system (7) are asymptotically stable and the
stationary solution $\xi_2$ is unstable. If the concentration $\ell_1\in
[\ell^b_{1,1}, \, \ell^b_{1,2}]$ tends to the end point $\ell^b_{1,1}$ (or
$\ell^b_{1,2}$) of the interval, then the stable $\xi_3$ (or $\xi_1$) and
unstable $\xi_2$ solutions approach each other and, in the limit $\ell_1 =
\ell^b_{1,1}$ (or $\ell_1 = \ell^b_{1,2}$), coalesce into the two-fold solution
$\xi^b_1$ (or $\xi^b_2$). Thus, for  $S_0 \in (S^c_-,\, S^c_+)$, the system is
monostable if $\ell_1 \notin [\ell^b_{1,1}, \, \ell^b_{1, 2}]$ and bistable if
$\ell_1 \in [\ell^b_{1,1}, \, \ell^b_{1, 2}]$. \looseness=-1

For the second critical value  $S_0 = S^c_+$ (curve 3 in Fig.~2b), the function
$\ell_1(\xi)$ has a horizontal point of inflection at $\xi = \xi^c_+$ and its
value at this point is equal to the maximum critical concentration
$\ell^c_{1,+}$ ($\ell^c_{1,+}$ and $\xi^c_+$ are shown in Fig.~2b). As $S_0 \in
(S^c_+,\, S^c_1)$ increases, the behavior of the function $\ell_1(\xi)$
(Fig.~2c) is similar to that in Fig.~2b for  $S_0\in (S^c_-,\, S^c_+)$. First,
the function changes the shape in the neighborhood of the point $\xi^c_+$ so
that $\ell_1(\xi)$ has a minimum and a maximum equal to the bifurcation
concentrations $\ell^b_{1,3}$ and $\ell^b_{1,4}$ ($\ell^b_{1,3} < \ell^b_{1,4}
< \ell^c_{1,+}$), respectively, at the points  $\xi = \xi^b_3$ and $\xi =
\xi^b_4$. This yields the second bistability interval $[\ell^b_{1,3}, \,
\ell^b_{1, 4}]$ of the system of width $I_{4,3} = \ell^b_{1,4} - \ell^b_{1,3}$
nonintersecting with the first. As $S_0$ increases, the bifurcation
concentrations $\ell^b_{1,3}$ and $\ell^b_{1,4}$ decrease and the width
$I_{4,3}$ increases (cf. $I_{4,3}$ for the curves in Fig.~2c). The bifurcation
concentrations $\ell^b_{1,1}$, $\ell^b_{1,2}$ for the first bistability
interval and $\ell^b_{1,3}, \ell^b_{1,4}$ for the second and the bifurcation
coordinates $\xi^b_1, \xi^b_2$ and $\xi^b_3, \xi^b_4$ for them are shown in
Fig.~2c for $S_0 = 0.00575$, $S_0\in (S^c_+,\, S^c_1)$. An increase in $S_0 \in
(S^c_+,\, S^c_1)$ leads, first, to the partial overlapping of the first and
second bistability intervals and then to their complete overlapping where the
second bistability interval includes the first (curve 3 in Fig.~2c). In the
case of the overlapping bistability intervals, for any concentration $\ell_1
\in (\ell^t_{1,1}, \, \ell^t_{1, 2})$, where $[\ell^t_{1,1}, \, \ell^t_{1, 2}]
= [\ell^b_{1,1}, \, \ell^b_{1, 2}] \cap [\ell^b_{1,3}, \, \ell^b_{1, 4}]$,
$\ell^t_{1,1} = \max\,(\ell^b_{1,1}, \, \ell^b_{1,3})$, $\ell^t_{1,2} =
\min\,(\ell^b_{1,2}, \, \ell^b_{1,4})$, there are five values of the coordinate
$\xi$: \ $\xi_1 < \xi_2 < \xi_3 < \xi_4 < \xi_5$; furthermore, the stationary
solutions $\xi_1$, $\xi_3$, and $\xi_5$ of system (7) are asymptotically stable
and the stationary solutions  $\xi_2$ and $\xi_4$ are unstable. Thus, for
$\ell_1 \in [\ell^t_{1,1}, \, \ell^t_{1, 2}]$, the system is tristable. We call
the concentration interval $[\ell^t_{1,1}, \, \ell^t_{1, 2}]$ a tristability
interval of the system.  \looseness=-1

If the concentration $\ell_1 \in [\ell^t_{1,1}, \, \ell^t_{1, 2}]$ tends to the
end point of the interval, then a stable solution and an unstable solution
approach each other and, in the limit $\ell_1 = \ell^t_{1,1}$ (or $\ell_1 =
\ell^t_{1,2}$), coalesce into a two-fold solution. In this case, system (7) has
four stationary solutions (two asymptotically stable, one unstable, and one
two-fold). For $\ell_1 = \ell^t_{1,1}$, the two-fold solution is $\xi^b_1$ if
$\ell^b_{1,1} > \ell^b_{1,3}$ or $\xi^b_3$ if $\ell^b_{1,1} < \ell^b_{1,3}$.
For  $\ell_1 = \ell^t_{1,2}$, the two-fold solution is $\xi^b_2$ if
$\ell^b_{1,2} < \ell^b_{1,4}$ or $\xi^b_4$ if $\ell^b_{1,2} > \ell^b_{1,4}$.
\looseness=-1

The arguments for a two-fold solution for the end points of the tristability
interval must be corrected for two values of $S_0 \in (S^c_+,\, S^c_1)$ denoted
by  $S_d$ and $S_t$ and corresponding, respectively, to the equality of the
lower (for $S_0 = S_d \approx 0.00578$) or upper (for $S_0 = S_t \approx
0.00589$) end points of the bistability intervals, i.e., the cases of equal
minima $\ell^b_{1,1} = \ell^b_{1,3} \equiv \ell^b_d \approx 0.155$
($\ell^t_{1,1} = \ell^b_d$) or maxima  $\ell^b_{1,2} = \ell^b_{1,4} \equiv
\ell^b_t \approx 0.158$ ($\ell^t_{1,2} = \ell^b_t$) of the function
$\ell_1(\xi)$. For $\ell^t_{1,1} = \ell^b_d$ (or $\ell^t_{1,2} = \ell^b_t$), if
the concentration $\ell_1 \in [\ell^t_{1,1}, \ell^t_{1, 2}]$ tends to the end
point $\ell^b_d$ (or $\ell^b_t$) of the interval, then simultaneously two pairs
of stable and unstable solutions approach each other and, in the limit $\ell_1
= \ell^b_d$ (or $\ell_1 = \ell^b_t$), coalesce into two two-fold solutions. In
these two cases, system (7) has three stationary solutions (one asymptotically
stable and two two-fold). As soon as the concentration $\ell_1 $ leaves the
tristability interval, both two-fold solutions disappear and the system becomes
monostable.  \looseness=-1

It is worth noting one more  case where system (7) has one asymptotically
stable and two two-fold stationary solutions. This case occurs for the value of
$S_0 \in (S^c_+,\, S^c_1)$ denoted by  $S_u$ ($S_u \approx 0.005764$) for which
two bistability intervals have only one common point $\ell^b_u$ such that
$\ell^b_{1,2} = \ell^b_{1,3} \equiv \ell^b_u \approx 0.1582$, i.e., the maximum
of $\ell_1(\xi)$ at $\xi =\xi^b_2$ is equal to the minimum of this function at
$\xi =\xi^b_3$: $ \ell_1(\xi^b_2) = \ell_1(\xi^b_3) \equiv \ell^b_u$. Unlike
the cases considered above for two-fold solutions, in this case, a negligible
variation (furthermore, in any side) in $\ell_1$ from $\ell^b_u$ is accompanied
by the disappearance of one two-fold solution and split of the second into two
(stable and unstable) solutions. \looseness=-1

Three values  $S_d$, $S_t$, and $S_u$ of the parameter $S_0 \in (S^c_+,\,
S^c_1)$ for which the system has two two-fold solutions are arranged as
follows: $S_t > S_d > S_u$.

As $S_0 \in (S^c_+,\, S^c_1)$ increases, the first bistability interval
decreases and the points  $\xi^b_1$ and $\xi^b_4$ at which the function
$\ell_1(\xi)$ has the minimum and the maximum, respectively, approach each
other. For $S_0 = S^c_1$, the points coalesce into one  $\xi^c_1$ at which the
function $\ell_1(\xi)$ has a horizontal point of inflection and is equal to the
least critical concentration $\ell^c_{1,1}$ (curve 1 in Fig.~2d). For $S_0 >
S^c_1$, the inflection of the function $\ell_1(\xi)$ disappears and the
function has one minimum and one maximum at the points  $\xi = \xi^b_3$ and
$\xi = \xi^b_2$, respectively (curves 2 and 3 in Fig.~2d). Thus, for $S_0 >
S^c_1$, the system has only one bistability interval $[\ell^b_{1,3}, \,
\ell^b_{1, 2}]$ of width $I_{2,3} = \ell^b_{1,2} - \ell^b_{1,3}$. The
bifurcation concentrations $\ell^b_{1,3}$ and $\ell^b_{1,2}$ and the
corresponding bifurcation coordinates $\xi^b_3$ and $\xi^b_2$ are shown  in
Fig.~2d for $S_0 = 0.007 > S^c_1$.

\begin{figure}[!htb]
 \centering{
 \includegraphics[width=0.9\textwidth]{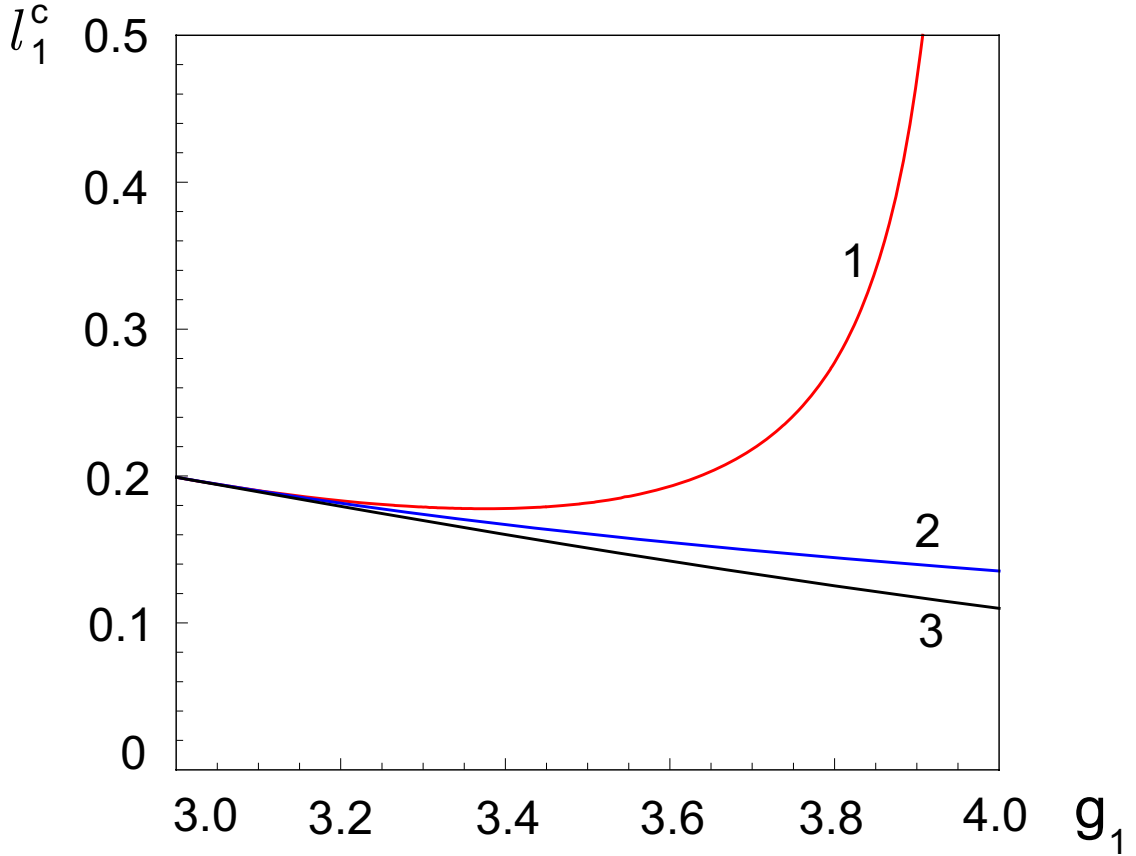}
 \caption {Dependence of the critical concentrations  $\ell^c_{1,+}$ (curve 1),
  $\ell^c_{1,-}$ (curve 2), and  $\ell^c_{1,1}$ (curve 3) on the coupling
  parameter  $g_1$ for  $G$ = 2.}
 }
 \label{pRGBfigc._3}
\end{figure}

Thus, to investigate specific features of stationary solutions of system (7),
first, it is necessary to determine the critical values of $S_0$ for which the
function $\ell_1(\xi)$ has horizontal points of inflection and the critical
concentrations at these points. The critical concentration $\ell_1^c$ and
parameter $S_0$ as functions of the coupling parameter $g_1$ defined by
relations (54) and (55) are shown in Figs.~3 and 4, respectively, in the range
of values of $g_1$ where the function $\ell_1(\xi) \ge 0$ has three horizontal
points of inflection.

Note that the behavior of the function  $\ell_1(\xi)$ shown in Fig.~2 for $g_1
= 3.5$ remains true for other values of  $g_1 \in (3,\, 4)$. Thus, qualitative
analysis of specific features of stationary solutions of system (7) can be made
using the graphs in Fig.~2 and the curves for the critical parameters $\ell_1$
and $S_0$ in Figs.~3 and 4.

\begin{figure}[!htb]
 \centering{
 \includegraphics[width=0.9\textwidth]{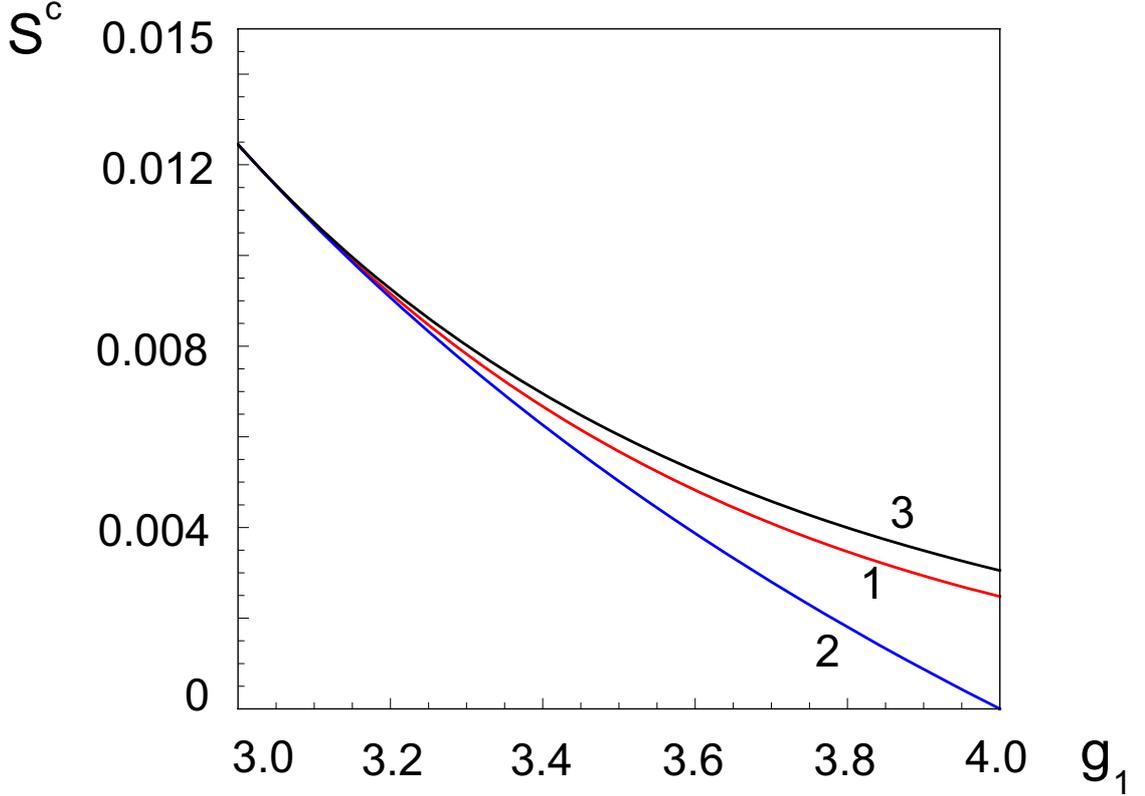}
 \caption {Dependence of $S^c_+$ (curve 1), $S^c_-$ (curve 2) and $S^c_1$
 (curve 3) on the coupling parameter $g_1$ for  $G$ = 2.}
 }
 \label{pRGBfigc._4}
\end{figure}

Specific features of the behavior of system (7) for $G = 2$ can be clearly
illustrated by plotting a bifurcation surface, which is a set of multiple roots
of Eq.~(18), in the three-dimensional space of control parameters $\{\ell_1,\,
S_0,\, g_1\}$. Instead of this surface, we plot the bifurcation curve
$\ell_1(S_0)$, which is the projection of the section of this bifurcation
surface by a plane of a fixed value of the coupling parameter $g_1$ onto the
plane ($S_0, \, \ell_1$). To this end, using relations (18) and (45), we obtain
the following representation of this bifurcation curve in the parametric form,
which is true for any $G \neq 0, 1$:  \looseness=-1

\begin{equation}
 S_0    =  -\frac{1}{G} \, \frac{L_1(\xi; g_1)}{L_2(\xi; g_1,G)}\, \frac{1}{w(\xi)} \, , \qquad
 \ell_1 = \frac{G}{1-G} \, \frac{L_2(\xi; g_1,G)\, \exp{(-g_1 \xi)}}{1 + g_1 (1-\xi)(G-\xi)}\, .
\end{equation}

The bifurcation curve in Fig.~5 plotted for the same values of $g_1$ and $G$ as
in Fig.~2 has several singular points, which are shown in Fig.~5b where a part
of the curve is scaled up for the most interesting range $S_0 \in (S^c_-,
S^c_1$), $\ell_1 \in (\ell^c_{1,1}, \ell^c_{1,+})$. The points $P_1 \equiv
(S^c_1, \, \ell^c_{1,1}), \ P_+ \equiv (S^c_+, \, \ell^c_{1,+})$, and $P_-
\equiv (S^c_-, \, \ell^c_{1,-})$ are the cusps of the bifurcation curve
corresponding to the horizontal points of inflection of the function
$\ell_1(\xi)$ at $\xi = \xi^c_1, \ \xi^c_+$, and  $\xi^c_-$, respectively. The
self-intersection points of the bifurcation curve  $P_d \equiv (S_d, \,
\ell^b_d), \ P_t \equiv (S_t, \, \ell^b_t)$, and  $P_u \equiv (S_u, \,
\ell^b_u)$ correspond to the system with two two-fold stationary solutions. For
any point of the first quadrant of this plane lying outside the bifurcation
curve, system (7) has one asymptotically stable stationary solution, i.e., it
is a domain of monostability of the system. For a point lying in the
curvilinear quadrangle $P_t\,P_u\,P_d\,P_1$, which is the domain of
intersection of two curvilinear triangles $P_1\,P_+\,P_d$ and $P_1\,P_-\,P_t$,
system (7) has five stationary solutions (three asymptotically stable and two
unstable), i.e., it is the domain of tristability of the system. If a point
lies in one of the domains: the curvilinear triangles $P_t\,P_+\,P_u$ and
$P_u\,P_-\,P_d$ and the domain $B_t\,P_t\,P_1\,P_d\,B_d$ ($B_t$ and $B_d$ are
symbolic notations for points of the upper and lower branches of the
bifurcation curve, respectively, in the limit  $S_0 \rightarrow \infty$), then
system (7) has three stationary solutions (two asymptotically stable and one
unstable), i.e., they are domains of bistability of the system. At any point of
the bifurcation curve, except for the points of the boundary of the curvilinear
quadrangle $P_t\,P_u\,P_d\,P_1$ and the singular points $P_+$ and $P_-$, system
(7) has two stationary solutions (one asymptotically stable and one two-fold).
At a non vertex point of the boundary of the quadrangle $P_t\,P_u\,P_d\,P_1$,
system (7) has four stationary solutions (three structurally stable
(furthermore, two asymptotically stable and one unstable) and one two-fold). At
the singular points $P_d,\ P_t$, and $P_u$, system (7) has three stationary
solutions (one asymptotically stable and two two-fold). At the singular points
$P_+$ and $P_-$, system (7) has one three-fold stationary solution. At the
singular point $P_1$, system (7) has three stationary solutions (two
asymptotically stable and one three-fold). \looseness=-1

\begin{figure}[!ht]
 \centering{
 \includegraphics[width=75mm, height=90mm]{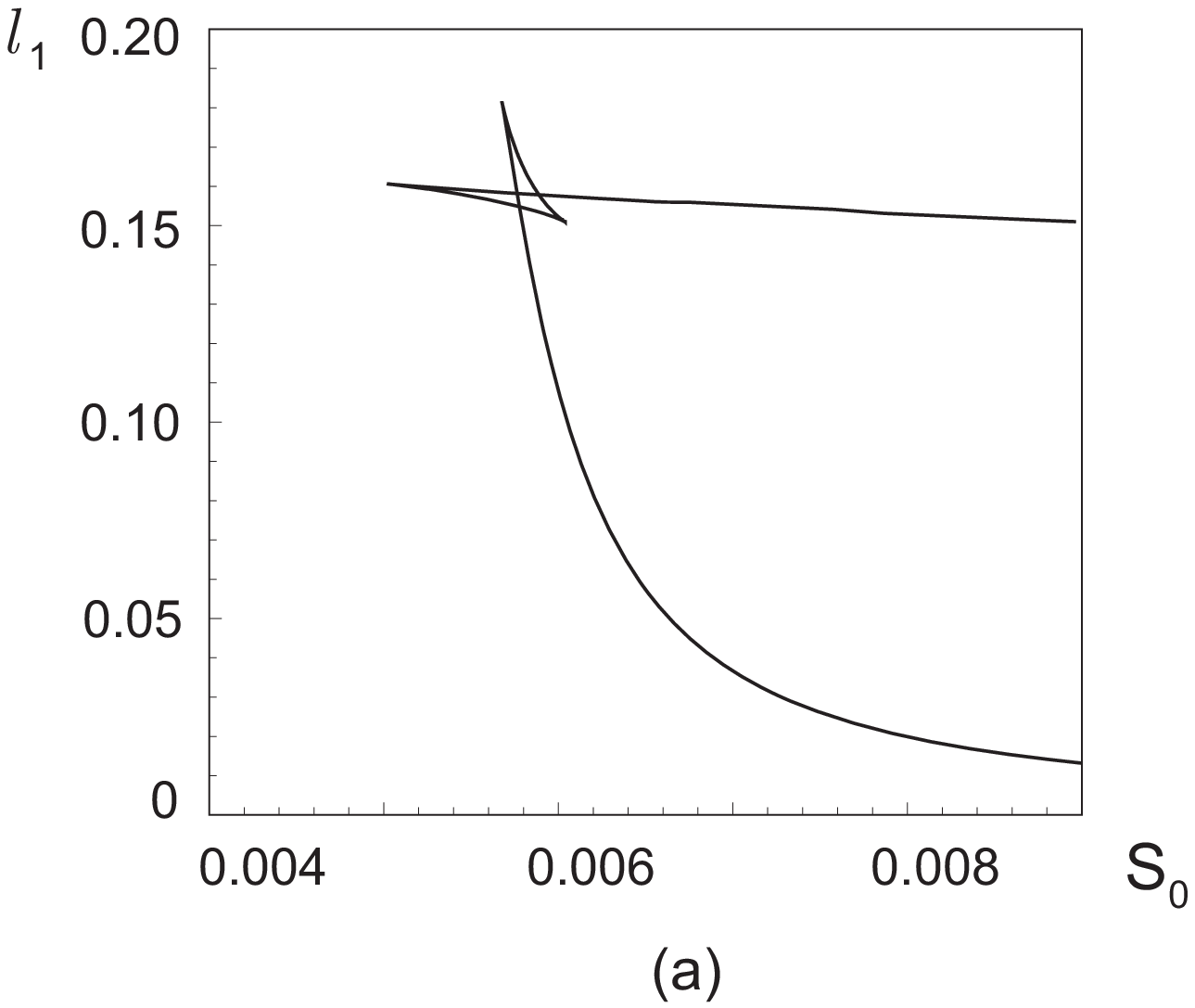}  \hfill
 \includegraphics[width=75mm, height=90mm]{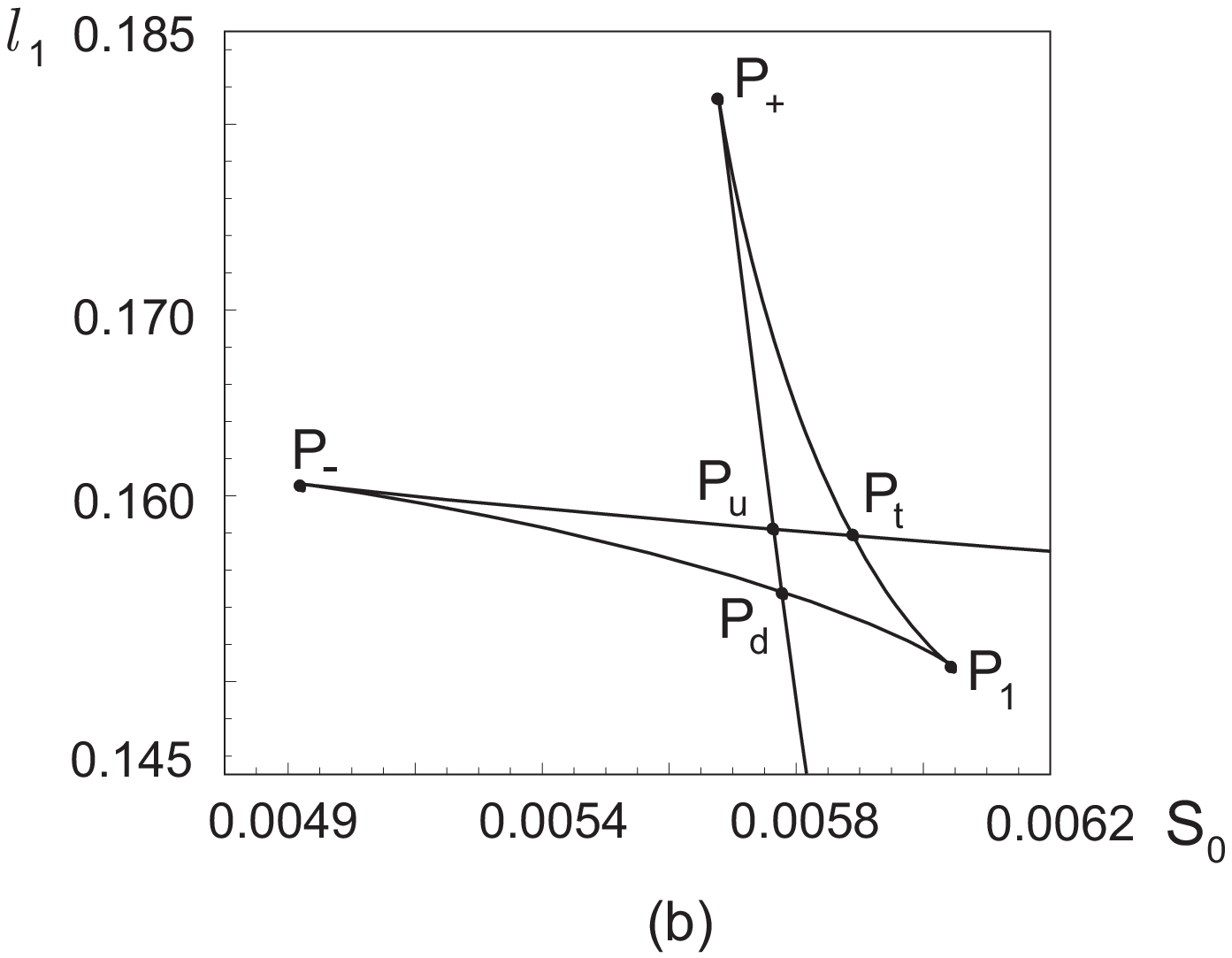}  \\
 \caption {Bifurcation curve for $G = 2,  g_1 = 3.5$.}
 }
 \label{pfigc._5p}
\end{figure}

Motion in the plane of control parameters ($S_0, \, \ell_1$) along a line can
be accompanied by the appearance of new solutions, disappearance of existing
solutions, and a change in solution stability in intersecting the bifurcation
curve. This depends on both the point of intersection and the line itself if it
intersects the bifurcation curve at a singular point and the direction of
motion. Independently of the line, its intersection with the bifurcation curve
at a nonsingular point is accompanied by the appearance/disappearance
(depending on the direction of motion) of a pair of stationary (stable and
unstable) solutions of system (7). In entering the domain $P_t\,P_+\,P_u$ or
$P_u\,P_-\,P_d$ through the cusp  $P_+$ or $P_-$\,, respectively, a stable
solution splits into three solutions (two stable and one unstable) and changes
its stability. In entering the domain $P_t\,P_+\,P_u$ $P_1$ through the cusp
$P_1$, an unstable solution splits into three solutions (two unstable and one
stable). In leaving these domains along a line passing through the cusp, three
solutions (two stable and one unstable in the domains $P_t\,P_+\,P_u$ and
$P_u\,P_-\,P_d$ or two unstable and one stable in the domain $P_t\,P_+\,P_u$
$P_1$) coalesce into a three-fold solution at the cusp with its subsequent
transformation outside the point into a simple stable (for the points $P_+$ and
$P_-$) or unstable (for the point $P_1$) solution. In entering/leaving the
domain of tristability $P_t\,P_u\,P_d\,P_1$ from/for the domain of
monostability through the points $P_d$, $P_t$, or $P_u$, which are unique
common points of these domains, two two-fold solutions  simultaneously
appear/disappear. If a line enters the domain $P_t\,P_+\,P_u$ from the domain
$P_u\,P_-\,P_d$ (or, conversely, enters the domain $P_u\,P_-\,P_d$ from the
domain $P_t\,P_+\,P_u$) through their common point $P_u$, then a stable
solution and an unstable solution coalesce into a two-fold solution at the
point $P_u$ that disappears with moving away from the point, whereas another
two-fold solution appears at this point and then splits into a pair of stable
and unstable solutions. A similar behavior of the system occurs in the motion
from one domain of bistability to another through their common point ($P_d$ for
the domains of bistability $P_u\,P_-\,P_d$ and $B_t\,P_t\,P_1\,P_d\,B_d$ or
$P_t$ for the domains of bistability $P_t\,P_+\,P_u$  and
$B_t\,P_t\,P_1\,P_d\,B_d$). \looseness=-1

The bifurcation curve for other values of the coupling parameter $g_1 \in (3,\,
4)$ is similar to the curve in Fig.~5. As $g_1$ decreases, the triangles
$P_u\,P_-\,P_d$ and $P_t\,P_+\,P_u$ and the quadrangle, $P_t\,P_u\,P_d\,P_1$
decrease and, in the limit  $g_1 \rightarrow 3$, shrink to the point $P_{\,but}
\equiv (S_0^{\,but}, \, \ell_1^{\,but}$), and tristability of the system is
impossible for $g_1 \leq 3$. As $g_1$ increases, the triangles
$P_u\,P_-\,P_d$ and $P_t\,P_+\,P_u$ elongate toward the ordinate axis and along
it, respectively, and, furthermore, in the limit $g_1 \rightarrow 4$, the
vertex  $P_-$ of the triangle $P_u\,P_-\,P_d$ lies on the ordinate axis ($P_-
\equiv (0, \ell_1^c$), where $\ell_1^c = \exp{(-2)}$ is the critical
concentration in adsorption of a one-component gas of species 1, and the vertex
$P_+$ of the triangle $P_t\,P_+\,P_u$ is at infinity ($P_+ \equiv (\exp{(-6)},
\infty$). \looseness=-1

The graphs of the equilibrium position of oscillator $\xi(\ell_1)$ and the
surface coverage $\theta_n(\ell_1)$ in Figs.~6 and 7, respectively, plotted on
the basis of relations (16)--(20) clearly illustrate their essential dependence
on the value of $S_0$. In the most interesting range $S_0 \in (S^c_-,\, S^c_1)$
(Figs.~6b--h and 7b--h), these characteristics are shown only in a small
interval of the concentration $\ell_1$ in which the adsorption isotherms
essentially differ from the classical Langmuir ones.

For $S_0 < S^c_-$, the coordinate $\xi$ increases with the concentration
$\ell_1$ and tends to its asymptotic value  $\xi^a$ determined from Eq.~(25)
(Fig.~6a). According to analysis in Sec.~3.1, $\xi^a\in(1, 2)$; the numerical
analysis shows that $\xi^a$ increases with $S_0$. The adsorption isotherms in
Fig.~7a are similar to the classical Langmuir isotherms. However, unlike the
Langmuir case for which the ratio $\theta_2/\theta_1$ is the constant equal to
$S_0$, variations in adsorption properties of the adsorbent in the competitive
adsorption leads to the dependence of this ratio on the concentration $\ell_1$.
As $\ell_1$ increases, the ratio $\theta_2/\theta_1$ increases and considerably
exceeds the Langmuir one (for large values of $\ell_1$, approximately by a
factor of 50).

For  $S_0 \in (S^c_-,\, S^c_+)$, the coordinate  $\xi(\ell_1)$ (Fig.~6b) and
the surface coverage $\theta_n(\ell_1)$ (Fig.~7b) have a hysteresis in the
first bistability interval of the system.

As the concentration $\ell_1$ increases from zero, both the coordinate $\xi$
and the surface coverages $\theta_n$ increase along their lower stable branches
ending at the bifurcation concentration $\ell_1 = \ell^b_{1,2}$. At this
concentration, $\xi$ and $\theta_n$ jump up to their upper stable branches
solely due to a change in adsorption properties of the adsorbent. For
convenience, transitions between stable branches of $\xi(\ell_1)$ are shown in
Fig.~6 by light vertical lines with arrows indicating the direction of
transition. Arrows under and above stable branches of $\xi(\ell_1)$ indicate
the direction of variation in $\ell_1$.  As the concentration $\ell_1$
increases from $\ell^b_{1,2}$, the coordinate $\xi$ and the surface coverages
$\theta_n$ increase along the upper stable branches and tend to their
asymptotic values $\xi^a$ and $\theta^a_n$ defined by relations (25) and (24),
respectively.

The transition of the coordinate $\xi$  from the lower stable branch to the
upper one at the bifurcation concentration  $\ell^b_{1,2}$ is accompanied by an
increase in the activation energy for desorption of adparticles, which hampers
their desorption. As a result, as the concentration $\ell_1$ decreases from a
value greater than $\ell^b_{1,2}$, the reverse transition of $\xi$ and
$\theta_n$ from their upper stable branches to the lower ones occurs at the
lower bifurcation concentration  $\ell^b_{1,1} < \ell^b_{1,2}$.

The curves in Figs.~6c and 7c correspond to the special case  $S_0 = S_u$ where
the system has two bistability intervals with common point $\ell_1 = \ell^b_u$.
Each of the functions  $\xi(\ell_1)$ and $\theta_n(\ell_1)$ has three stable
and two unstable branches (the $j$th unstable branch connects the $j$th and $(j
+ 1)$th stable branches, where $j = 1,2$). However, the behaviors of these
functions are different. The coordinate $\xi(\ell_1)$ has two successive
hystereses in the touching bistability intervals (Fig.~6c). As the
concentration $\ell_1$ increases from zero, the coordinate $\xi$ increases
along the first stable branch up to its end at $\ell_1 = \ell^b_u$; then jumps
up to the second stable branch and increases with $\ell_1$ along this branch up
to its end at $\ell_1 = \ell^b_{1,4}$; then jumps up to the third stable branch
and increases along it with $\ell_1$ tending to its asymptotic value $\xi^a$.
As the concentration $\ell_1$ decreases from a value greater than
$\ell^b_{1,4}$, the coordinate $\xi$ successively jumps down from the third
stable branch to the second and from the second stable branch to the first,
respectively, at the bifurcation concentrations $\ell^b_u$ and $ \ell^b_{1,1}$
at which these branches end, furthermore, the transitions from the first and
third stable branches to the second go along the same vertical straight line
$\ell_1 = \ell^b_u$. \looseness=-1

The behavior of the surface coverage $\theta_2(\ell_1)$ in Fig.~7c is similar
to the behavior of the coordinate $\xi(\ell_1)$ in Fig.~6c. Note that a similar
behavior of $\theta_2(\ell_1)$ and $\xi(\ell_1)$ also occurs for other values
of $S_0$ (cf. curve 2 in Fig.~7c--i with curve in Fig.~6c--i).

The surface coverage $\theta_1(\ell_1)$ has another behavior (curve 1 in
Fig.~7c). The different location of the second and third stable branches of
$\theta_1(\ell_1)$ and $\theta_2(\ell_1)$ illustrates the essentially different
behavior of the surface coverages $\theta_1$ and $\theta_2$ in transition
between stable branches at bifurcation concentrations of $\ell_1$. The
transition of the surface coverages $\theta_n$ from the first stable branch to
the second leads to their stepwise increase. However, in transition of the
surface coverages $\theta_n$ from the second stable branch to the third, the
value of $\theta_2$ stepwise increases, whereas the value of $\theta_1$
stepwise decreases. Thus, as the concentration $\ell_1$ increases, the surface
coverage $\theta_2$ continuously increases with $\ell_1$ along its stable
branches and stepwise increases in transition between stable branches at a
bifurcation concentration of $\ell_1$, whereas the surface coverage $\theta_1$
can continuously both increase and decrease with $\ell_1$ along its stable
branches and stepwise both increase and decrease in transition between stable
branches at a bifurcation concentration of $\ell_1$. This different behavior of
the surface coverages $\theta_1(\ell_1)$ and $\theta_2(\ell_1)$ is caused by
the different growth of the residence times of adparticles of different species
on the deformable adsorbent in adsorption and, furthermore, quantity (14)
characterizing this difference exponentially increases with displacement of
adsorption sites from their nonperturbed equilibrium position. This leads to a
greater amount of adparticles of species 2 relative to that of species 1 (cf.
the third stable branches of the surfaces coverages $\theta_1(\ell_1)$ and
$\theta_2(\ell_1)$), whereas, in the classical case, $\theta_2 \ll \theta_1$.
This result agrees with condition (28) according to which, for $\xi^a > 1.5$,
the asymptotic ratio $S(\xi^a)$ defined by relation (27) is greater than 1.
Indeed, in the considered case, $\xi^a \approx 1.65$, which yields $S(\xi^a)
\approx 1.85$. \looseness=-1

One more specific feature is a self-tangency point of $\theta_1(\ell_1)$ (point
of contact of four branches of the function: two stable (first and third) and
two unstable branches) at $\ell_1 = \ell^b_u$. As was discussed above, in this
special case, there are three stationary coordinates: two two-fold $\xi^b_2$
and $\xi^b_3$, $\xi^b_2 < \xi^b_3$, and one stable lying between them and equal
to the ordinate of the point of intersection of the second stable branch of
$\xi(\ell_1)$ with the vertical straight line $\ell_1 = \ell^b_u$ (Fig.~6c).
Taking into account the principle of perfect delay \cite{ref.PoS, ref.Gil} and
the condition for transition between stationary solutions of the system
(according to which all components ($\xi$, $\theta_1$, and $\theta_2$) of a
stationary three-component solution simultaneously go from their stable
branches at a bifurcation concentration at which these branches end to the
corresponding other stable branches), as $\ell_1$ increases, a discontinuous
transition of $\theta_1$ to the second stable branch occurs at $\ell_1 =
\ell^b_u$ rather than a continuous transition to the third stable branch
touching with the first stable branch at $\ell_1 = \ell^b_u$. Further, as
$\ell_1$ increases, the surface coverage $\theta_1$ decreases along the second
stable branch up to its end at $\ell_1 = \ell^b_{1,4}$; then jumps down to the
third stable branch and decreases along this branch tending to its asymptotic
value $\theta^a_1$. As the concentration $\ell_1$ decreases from a value
greater than $\ell^b_{1,4}$, the surface coverage $\theta_1$ varies along the
third stable branch up to its end at $\ell_1 = \ell^b_u$; then jumps up to the
second stable branch, furthermore, along the same vertical straight line as for
increasing $\ell_1$, rather than continuously goes to the first stable branch
touching with the second stable branch at $\ell_1 = \ell^b_u$, and then varies
along the second stable branch up to its end at $\ell_1 = \ell^b_{1,1}$; then
jumps down to the first stable branch and decreases along this branch.
\looseness=-1

It turns out that the equality of two bifurcation values of the surface
coverage $\theta_1$ for  $\ell_1 = \ell^b_u$ and  $S_0 = S_u\ $
($\theta^b_{1,2} =\theta^b_{1,3} \equiv \theta^u_1$) shown in Fig.~7c also
occurs for other values of the parameters $g_1$ and $G$; the values of
$\ell^b_u$ and $S_u$ depend on $g_1$ and $G$. Using relations (50) and (48), we
obtain that, in this case, the bifurcation coordinates $\xi^b_2$ and $\xi^b_3$
are symmetrically located about $G/2$: $\ \xi^b_2 = G/2 - \eta$ and $\xi^b_3 =
G/2 + \eta$.  For $G = 2$, the quantity  $\eta$  is a solution the equation
\looseness=-1
\begin{equation}
 \frac{g_1 \eta}{1 + g_1 \eta^2} = \tanh{g_1 \eta}\,,
\end{equation}
\noindent  the quantities  $S_u$  and $\ell^b_u$ are expressed in terms of
$\eta$ as follows:
\begin{equation}
 S_u = \frac{\cosh{g_1 \eta}}{\cosh{2 g_1 \eta}}\,
       \frac{\eta\,(1 - \eta\,\tanh{g_1 \eta})}{(1+\eta^2)\,\tanh{2 g_1 \eta}- 2 \eta}\,
       \exp{(-g_1)}\,, \quad
 \ell^b_u = 2\, \frac{g_1 (1-\eta^2)-1}{1+g_1\eta (1 + \eta)}\,
            \exp{(g_1 (\eta - 1))}\,,
\end{equation}
and the bifurcation surface coverages have the form
\begin{equation}
 \theta^u_1 = 1 - \eta^2 - \frac{1}{g_1}\,, \qquad
 \theta^b_{2,2} = \frac{1}{2}\, \biggl\{\frac{1}{g_1}
                + \eta \left( \eta - 1 \right)\biggr\}\,, \quad
 \theta^b_{2,3} = \frac{1}{2}\, \biggl\{\frac{1}{g_1}
                + \eta \left( \eta + 1 \right)\biggr\}\,.
\end{equation}

Equation (58) has a nonzero solution $\eta$ for  $g_1 > 3 $ and its value
increases with $g_1$: $\eta \in (0, 1)$ for $g_1 \in (3, \infty)$. According to
(60), as  $g_1$ increases, the difference $\delta^u = \theta^u_1 -
\theta^b_{2,3}$ decreases from the maximum value $\lim\limits_{g_1 \rightarrow
3}\delta^u = 0.5$ to the minimum value $\lim\limits_{g_1 \rightarrow
\infty}\delta^u = -1$. For example, $\delta^u \approx 0.453$ for $g_1 = 3.01$
and $\delta^u \approx -0.774$ for $g_1 = 10$. For $g_1 = 3.5$ (Figs.~6c and
7c), $\eta \approx 0.468$ and the values of $\theta^u_1$ and $\theta^b_{2,3}$
are close to each other ($\delta^u \approx 0.01$).

The curves in Figs.~6d--h and 7d--h correspond to different cases of
tristability of the system: partial (Figs.~6d,h and 7d,h) and complete
(Figs.~6e--g and 7e--g) overlapping of the bistability intervals
$[\ell^b_{1,1}, \, \ell^b_{1,2}]$ and $[\ell^b_{1,3}, \,\ell^b_{1,4}]$. In the
domain $\ell_1 \in [\ell^t_{1,1}, \, \ell^t_{1, 2}]$, the coordinate
$\xi(\ell_1)$ and the surface coverage $\theta_2(\ell_1)$ have two ``parallel''
hystereses.

\begin{figure}
 \centering{
 \includegraphics[width=50mm, height=70mm]{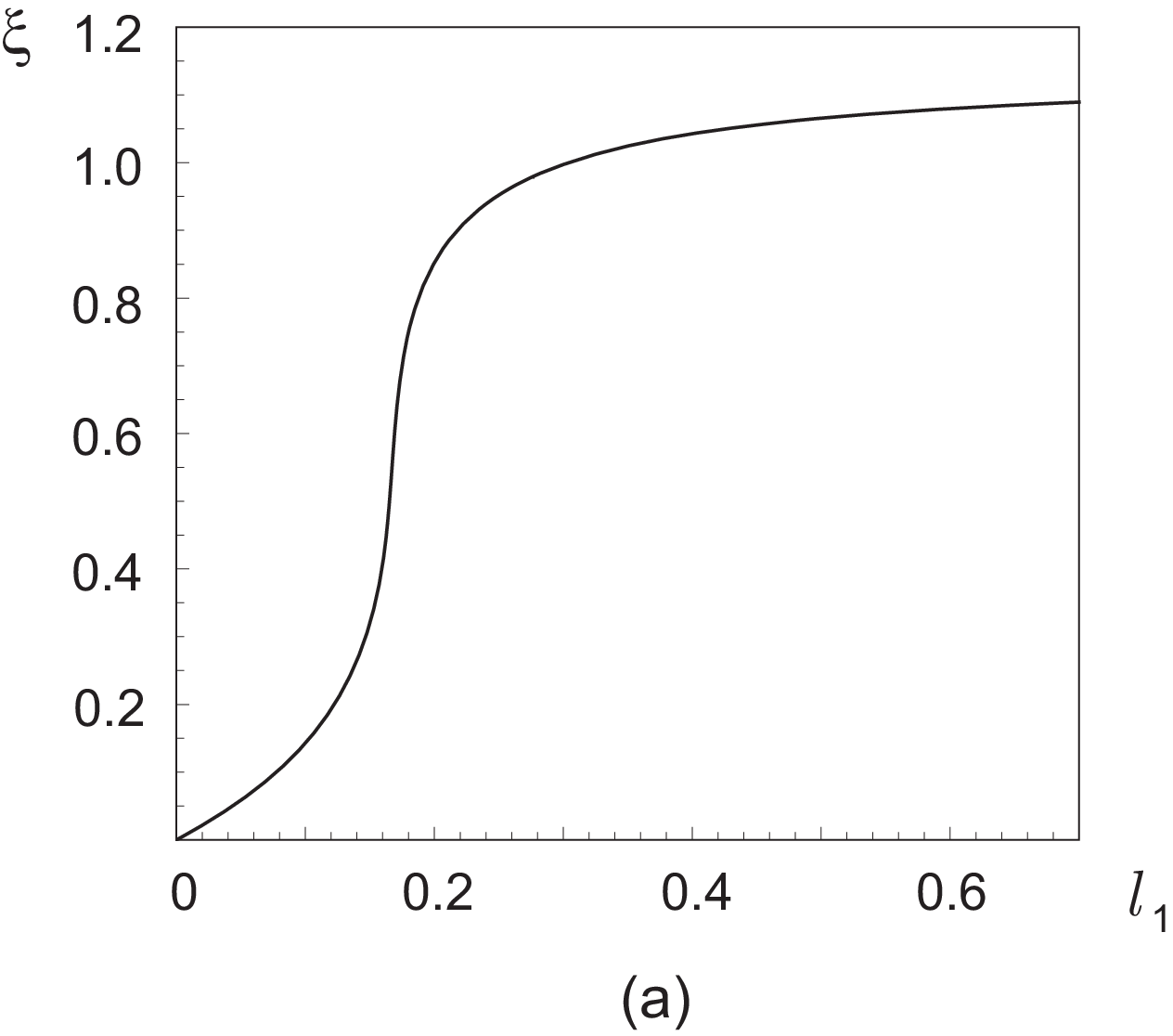}    \hfill
 \includegraphics[width=50mm, height=70mm]{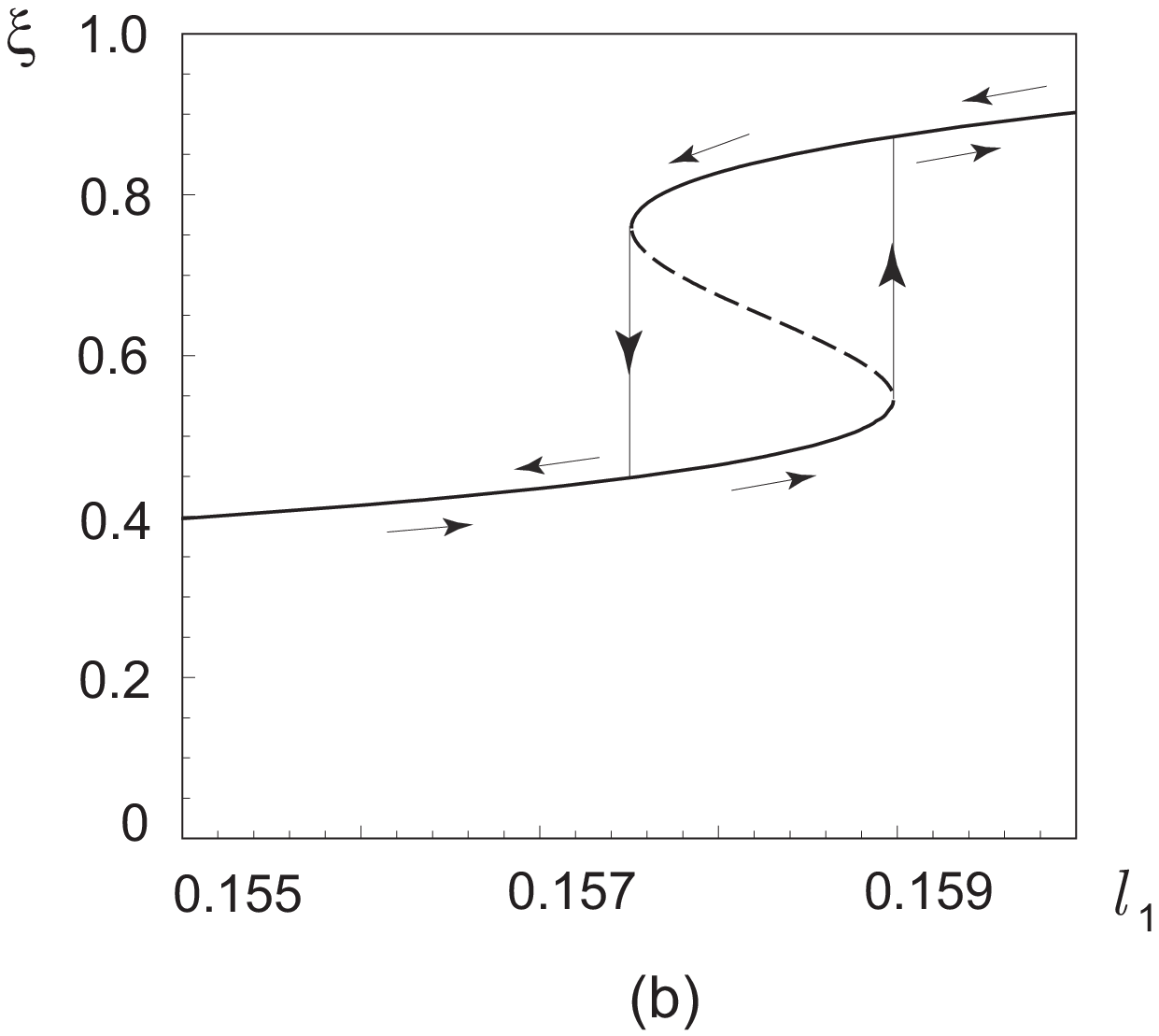}  \hfill
 \includegraphics[width=50mm, height=70mm]{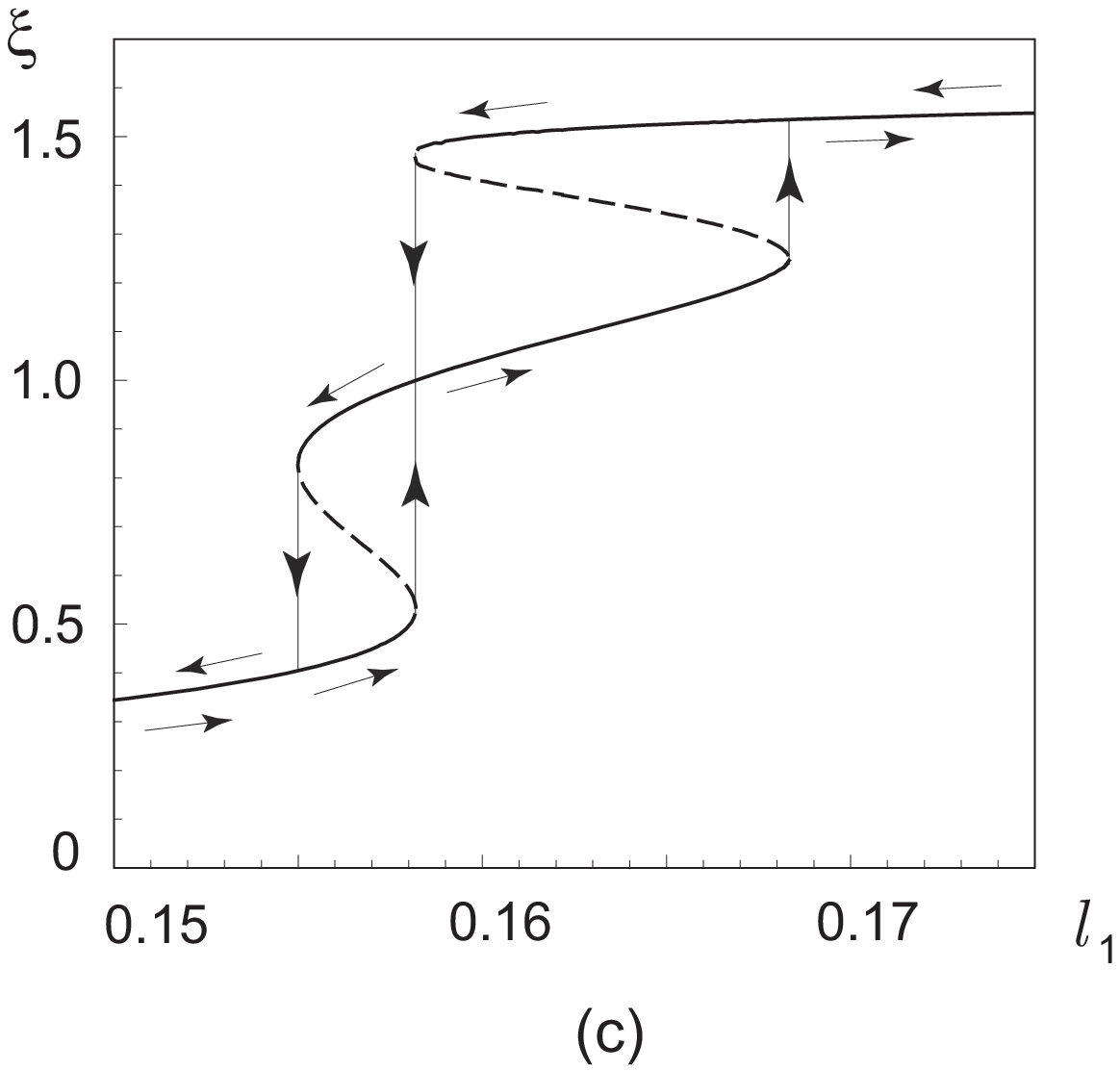}       \\
  \vfill
 \includegraphics[width=50mm, height=70mm]{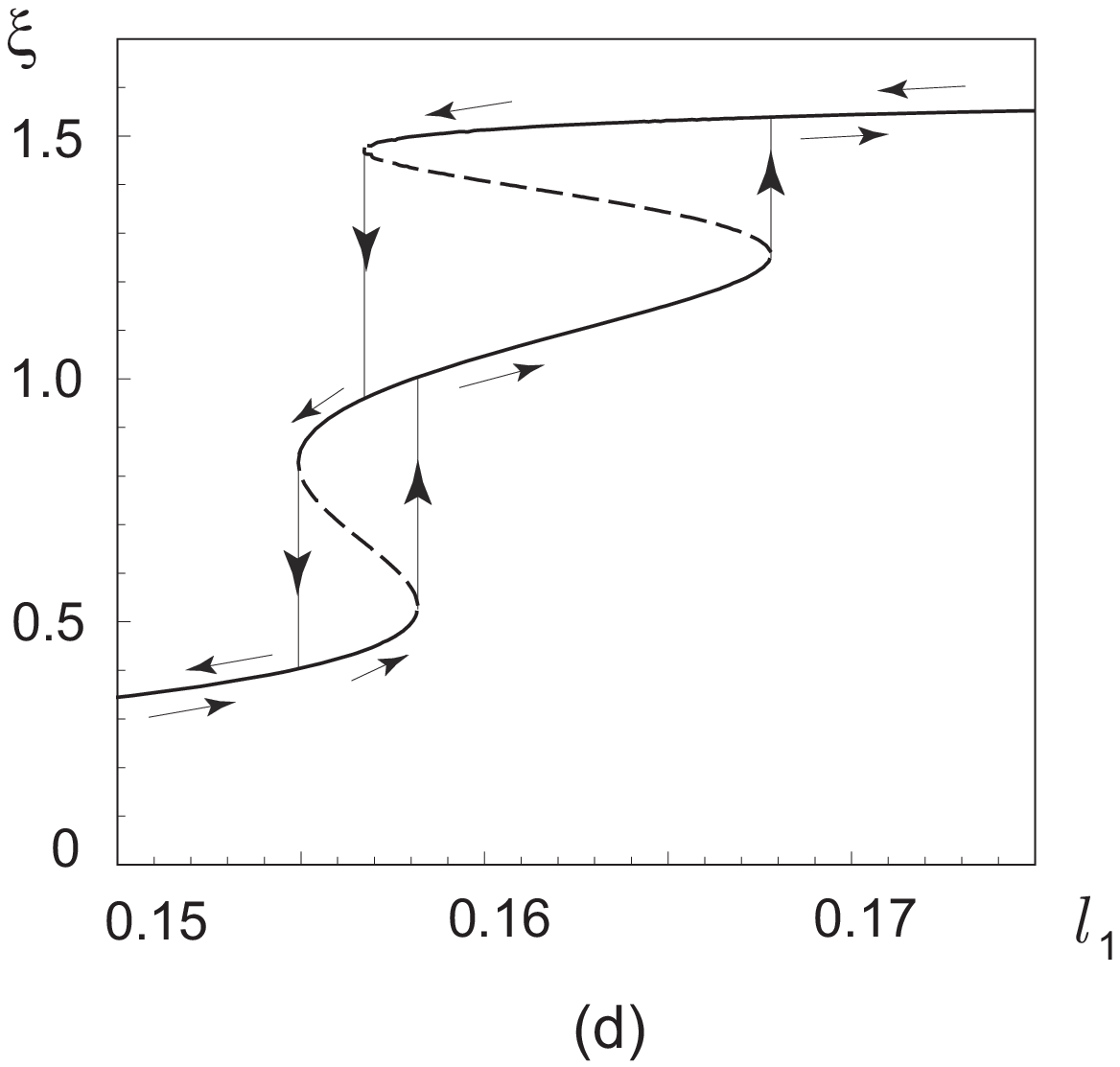}  \hfill
 \includegraphics[width=50mm, height=70mm]{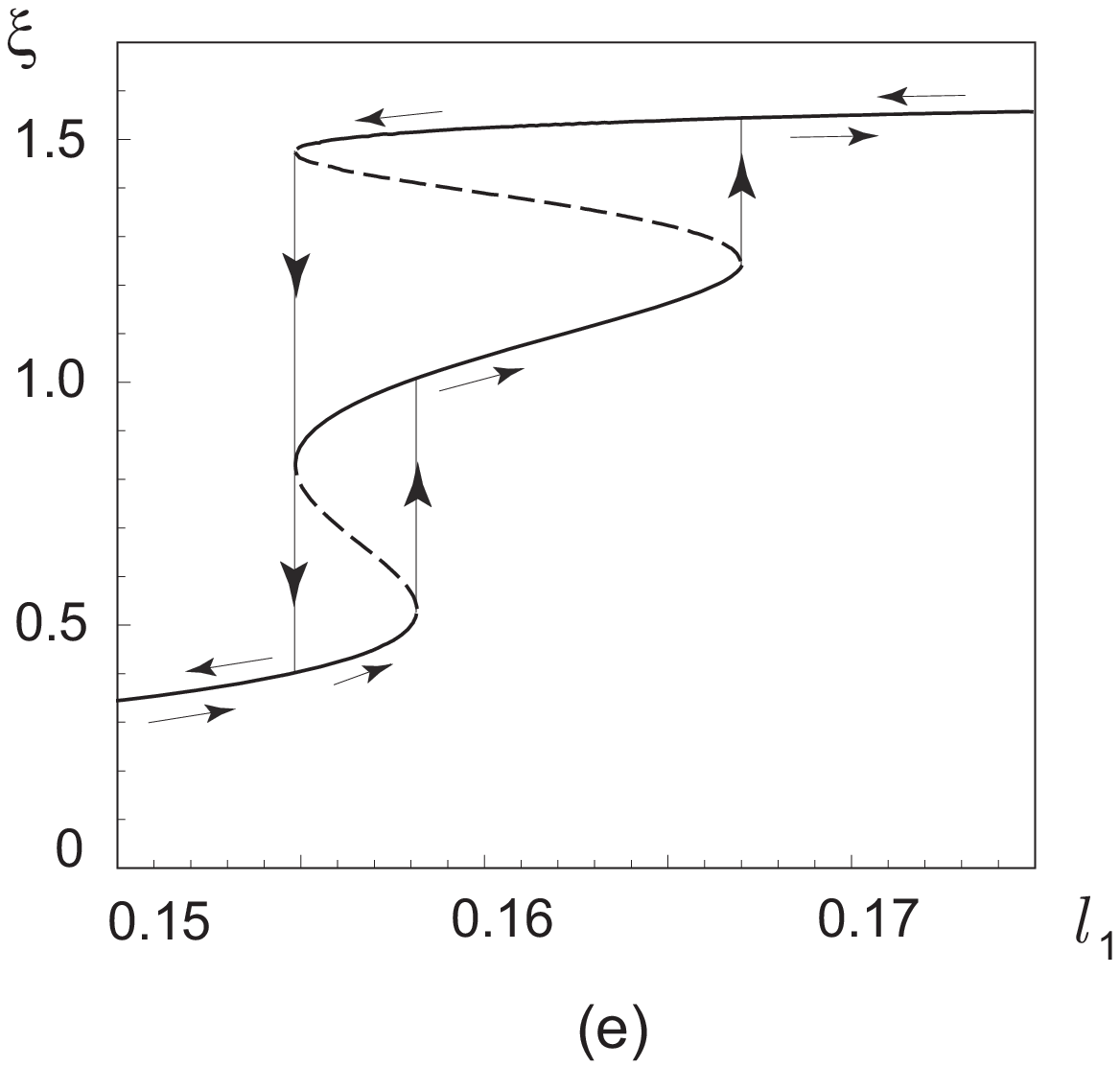}  \hfill
 \includegraphics[width=50mm, height=70mm]{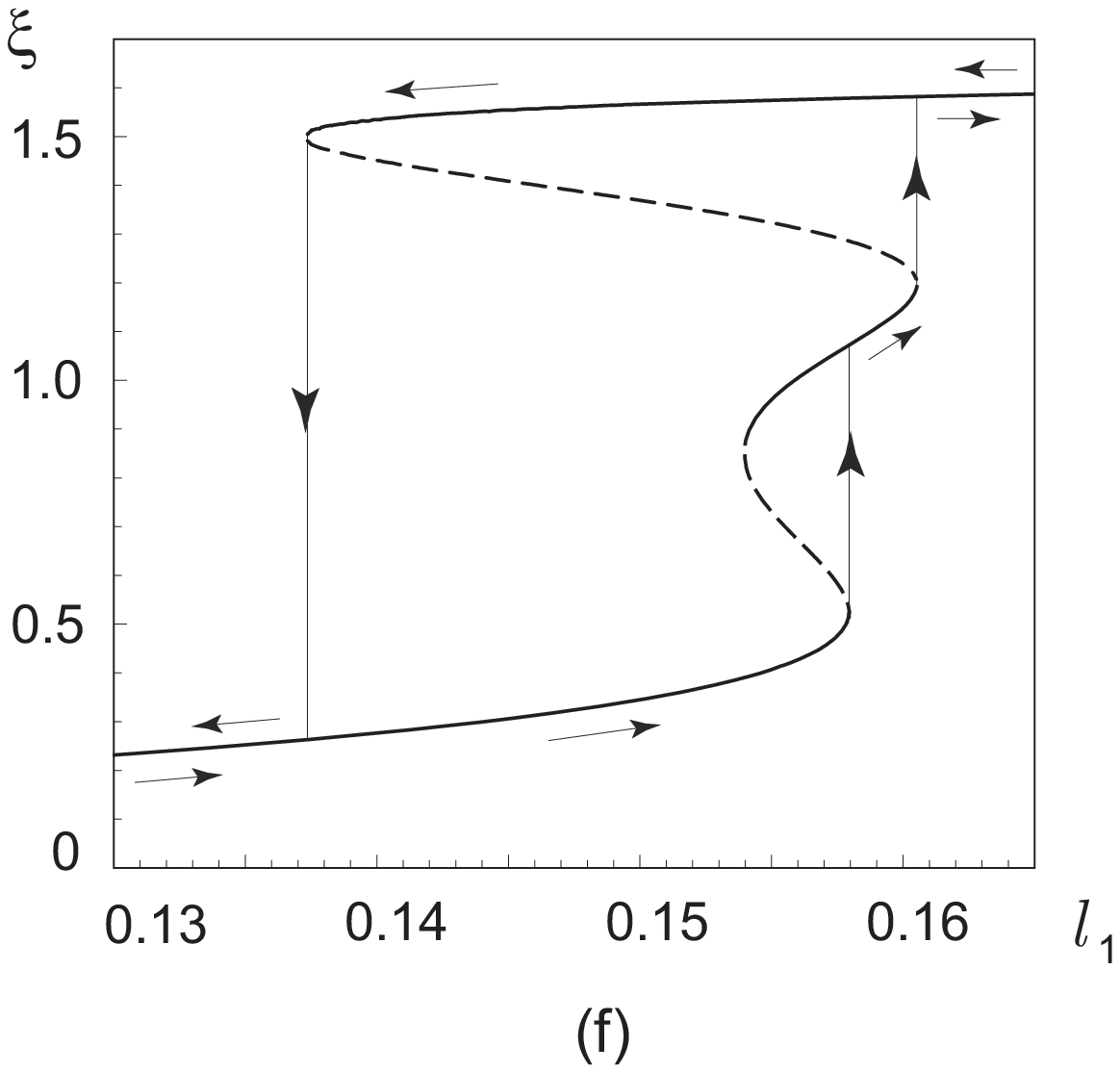}  \\
  \vfill
 \includegraphics[width=50mm, height=70mm]{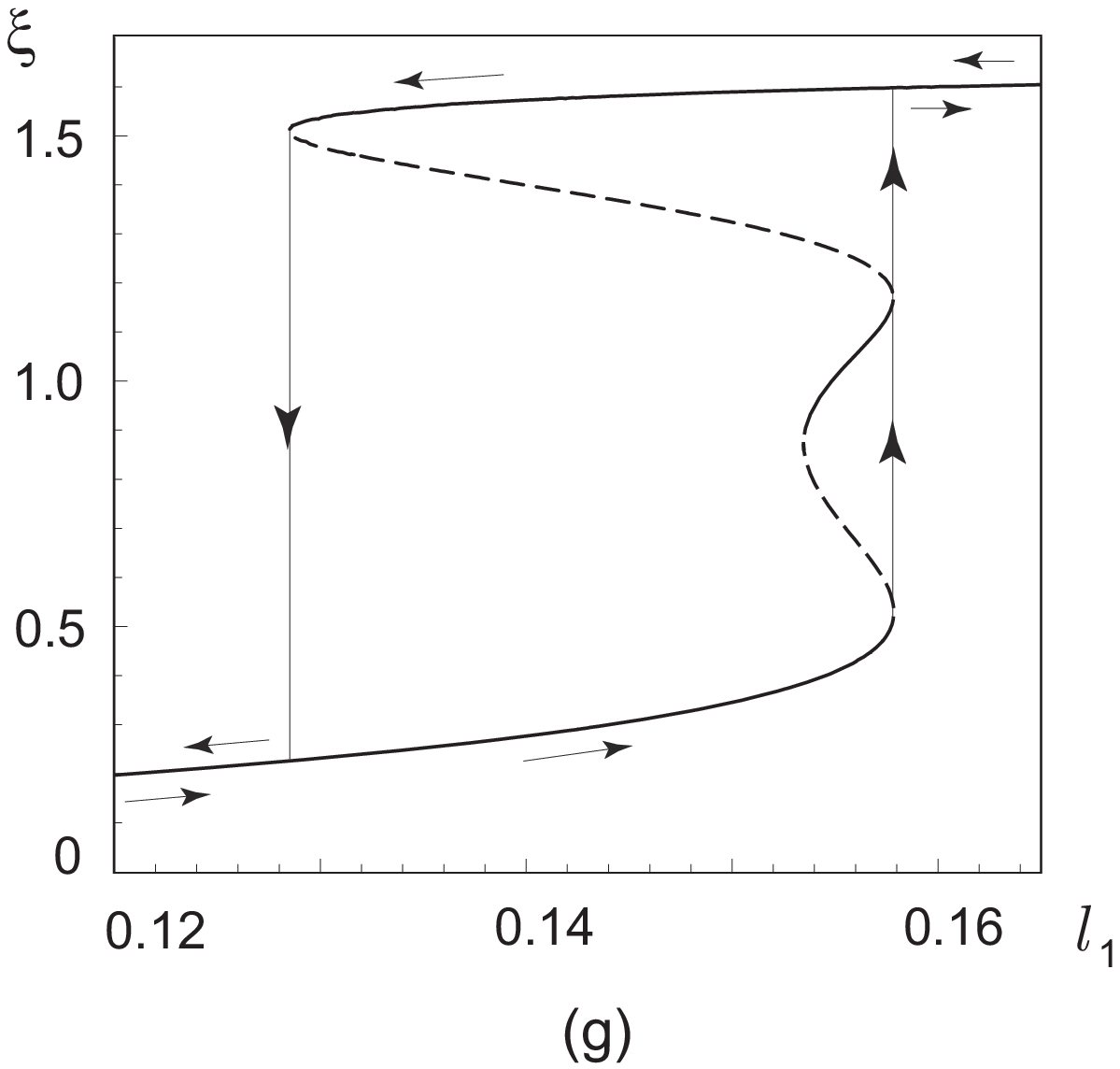}  \hfill
 \includegraphics[width=50mm, height=70mm]{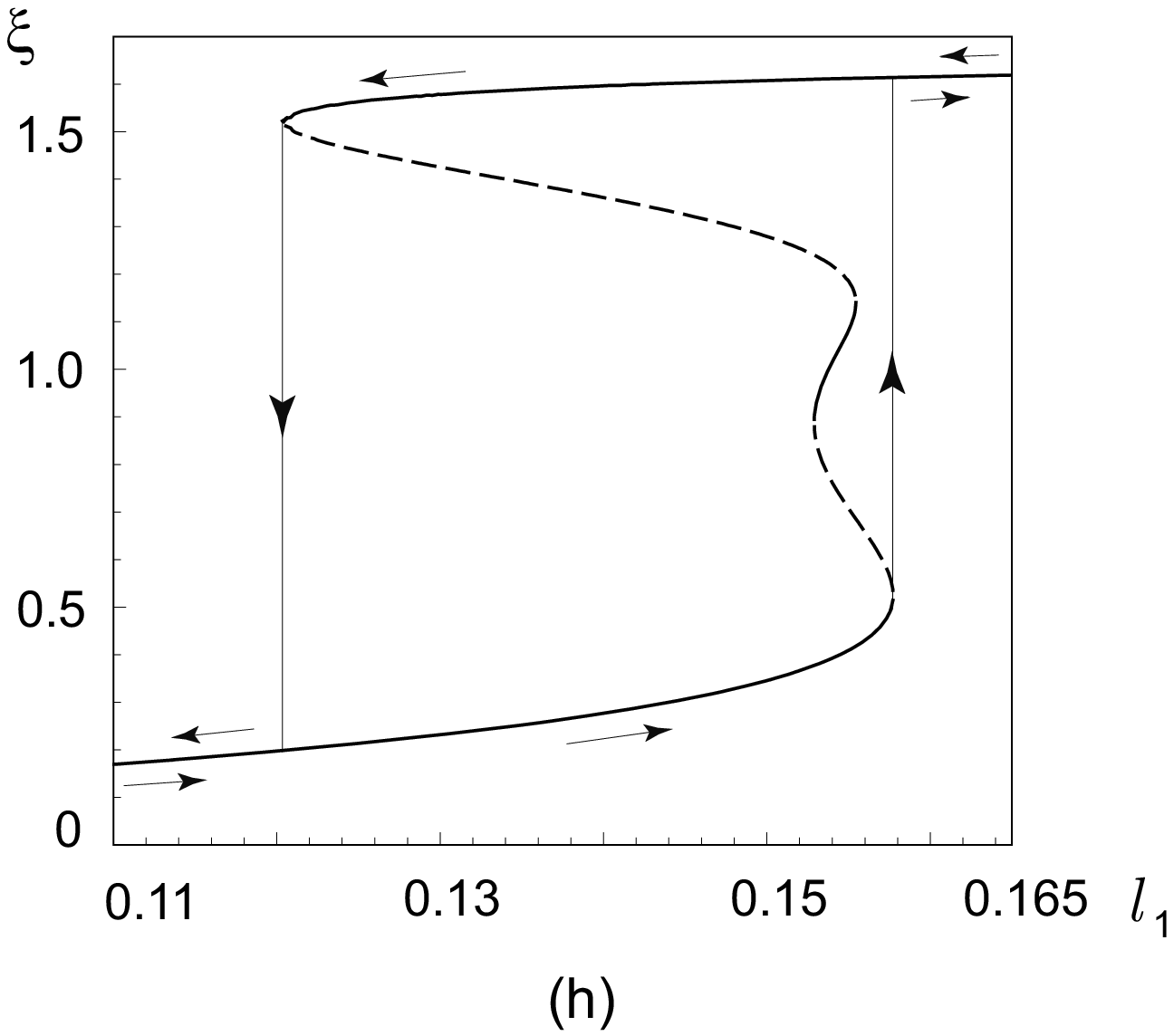}  \hfill
 \includegraphics[width=50mm, height=70mm]{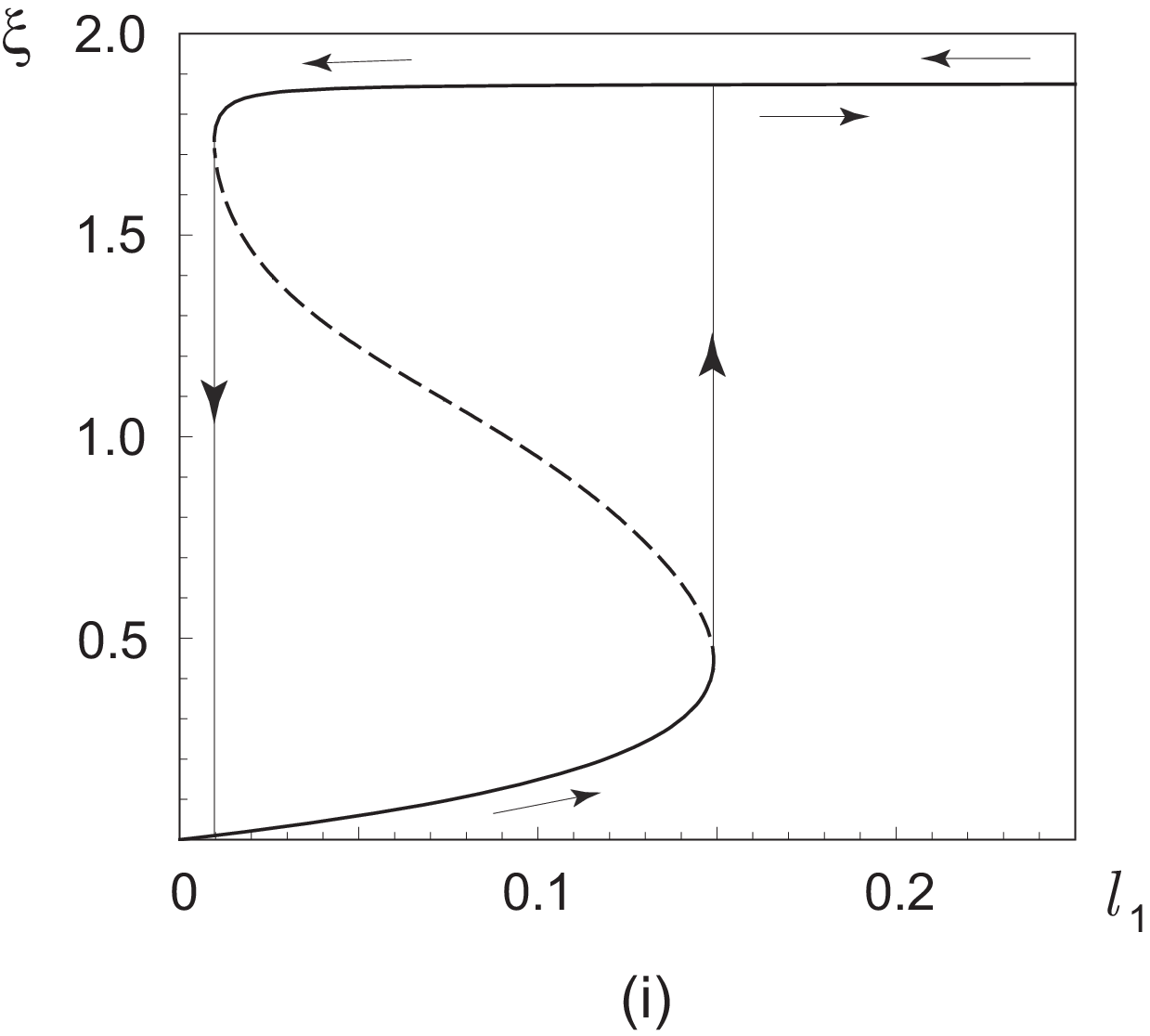}  \\
  \vfill
 \caption {Equilibrium position of oscillator  $\xi$  vs  the concentration
  $\ell_1$ for different values of  $S_0$:
  $S_0$ = 0.003~(a),    0.0055~(b),   $S_u$~(c),
          0.00577~(d),  $S_d$~(e),    0.00585~(f),
          $S_t$~(g),    0.00593~(h),  0.01~(i);
  $G = 2,  g_1 = 3.5$.}
 }
\label{pafigc._6_003}
\end{figure}

\begin{figure}
 \centering{
 \includegraphics[width=50mm, height=70mm]{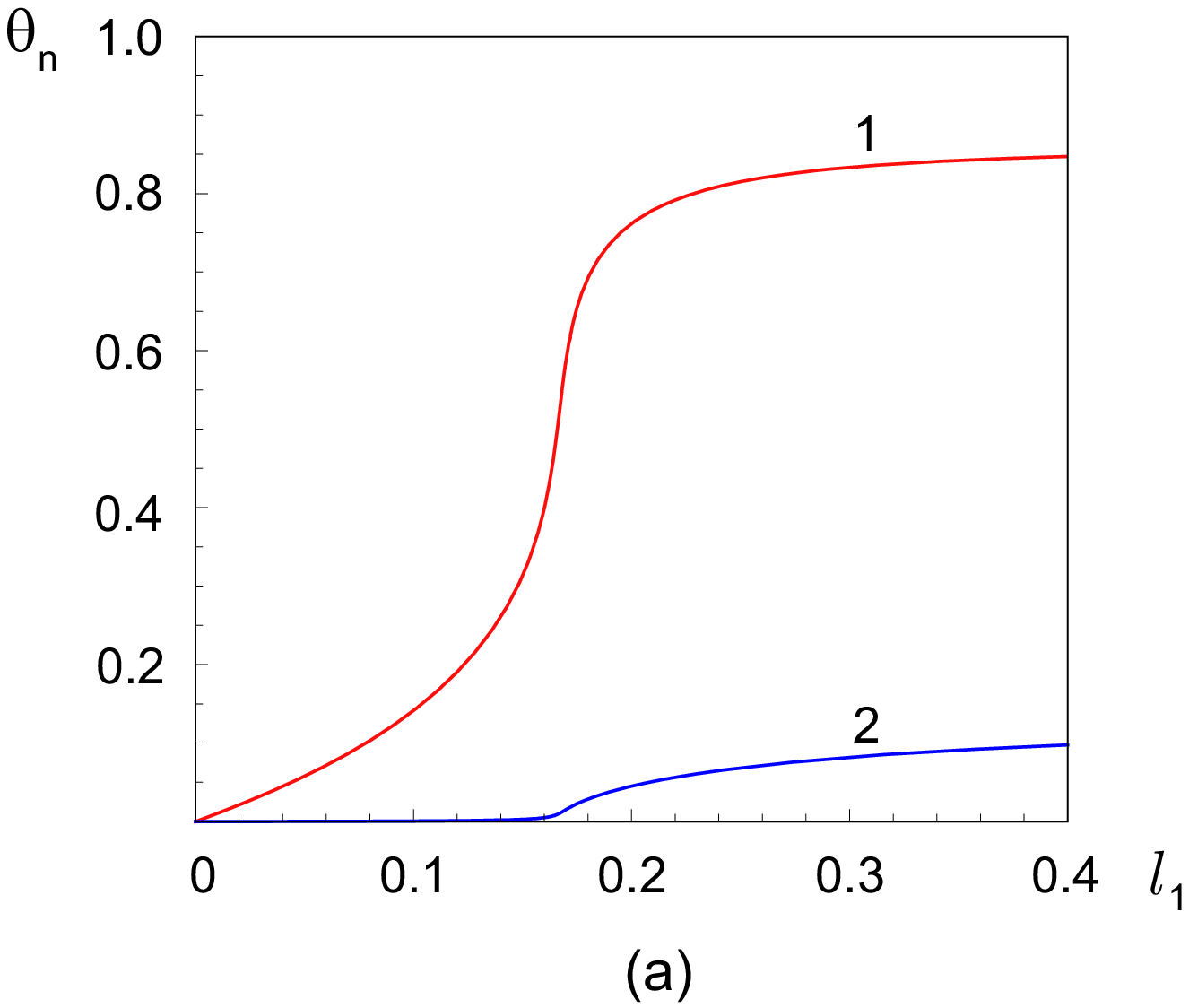}   \hfill
 \includegraphics[width=50mm, height=70mm]{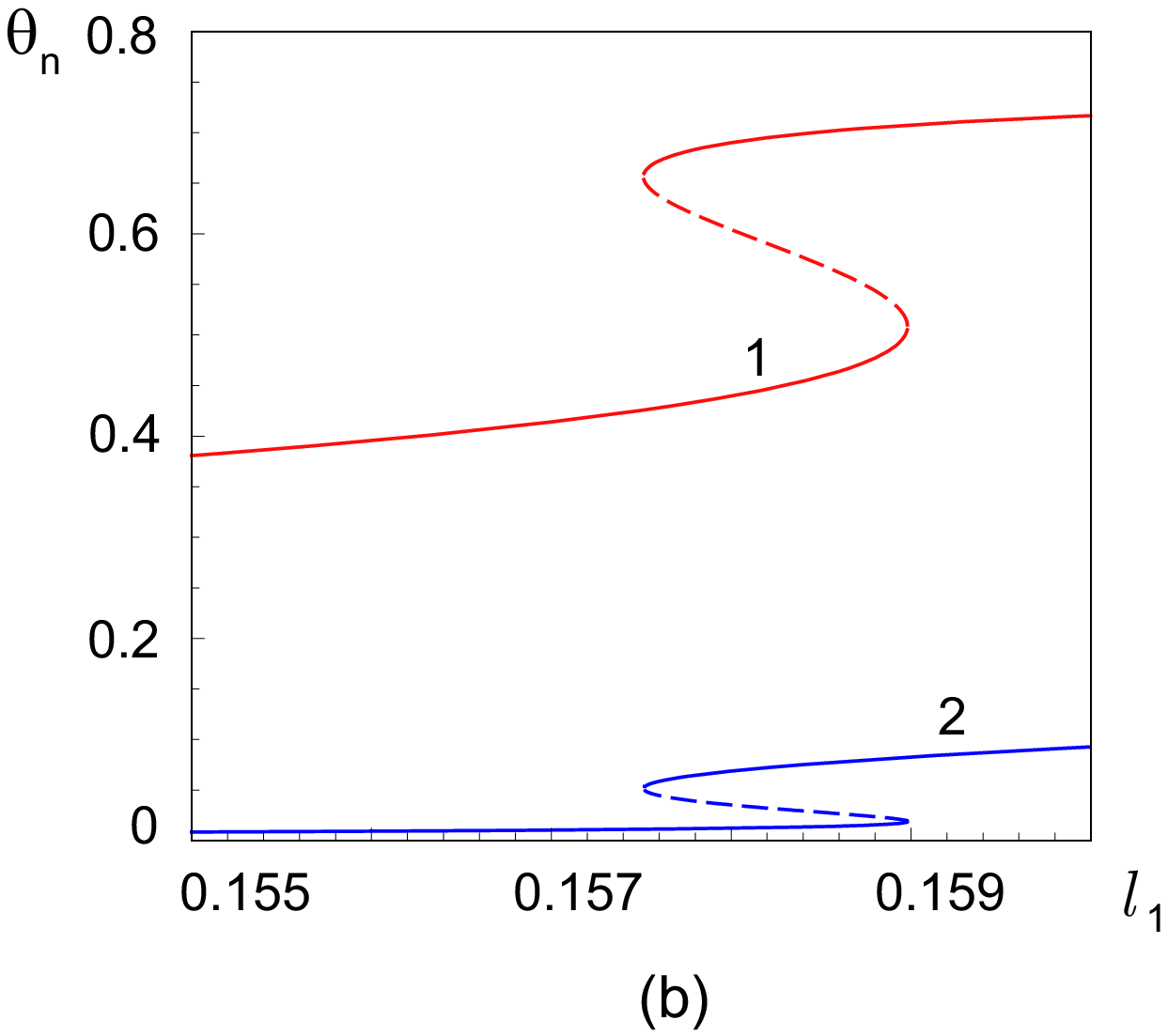}  \hfill
 \includegraphics[width=50mm, height=70mm]{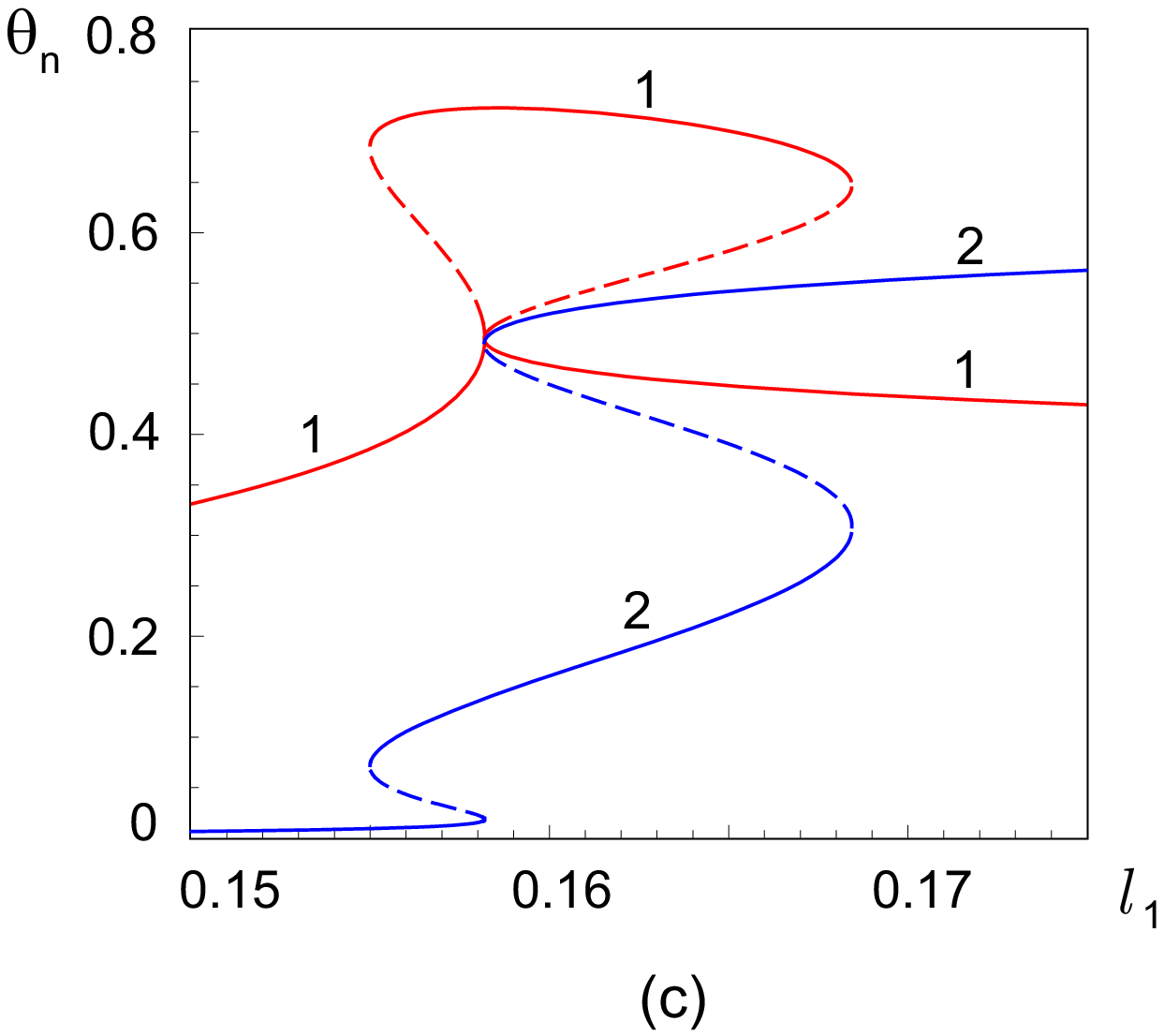}       \\
  \vfill
 \includegraphics[width=50mm, height=70mm]{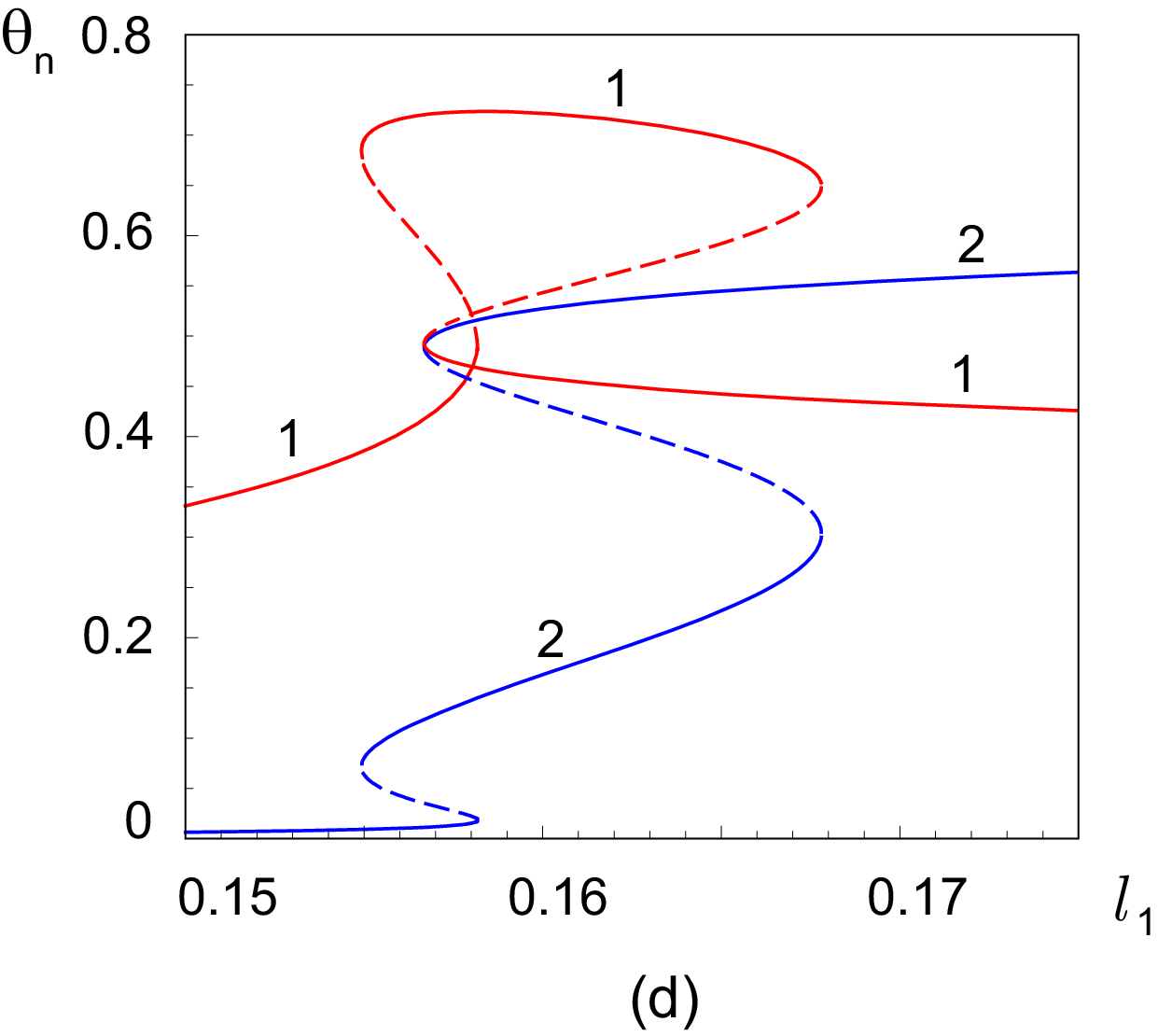}  \hfill
 \includegraphics[width=50mm, height=70mm]{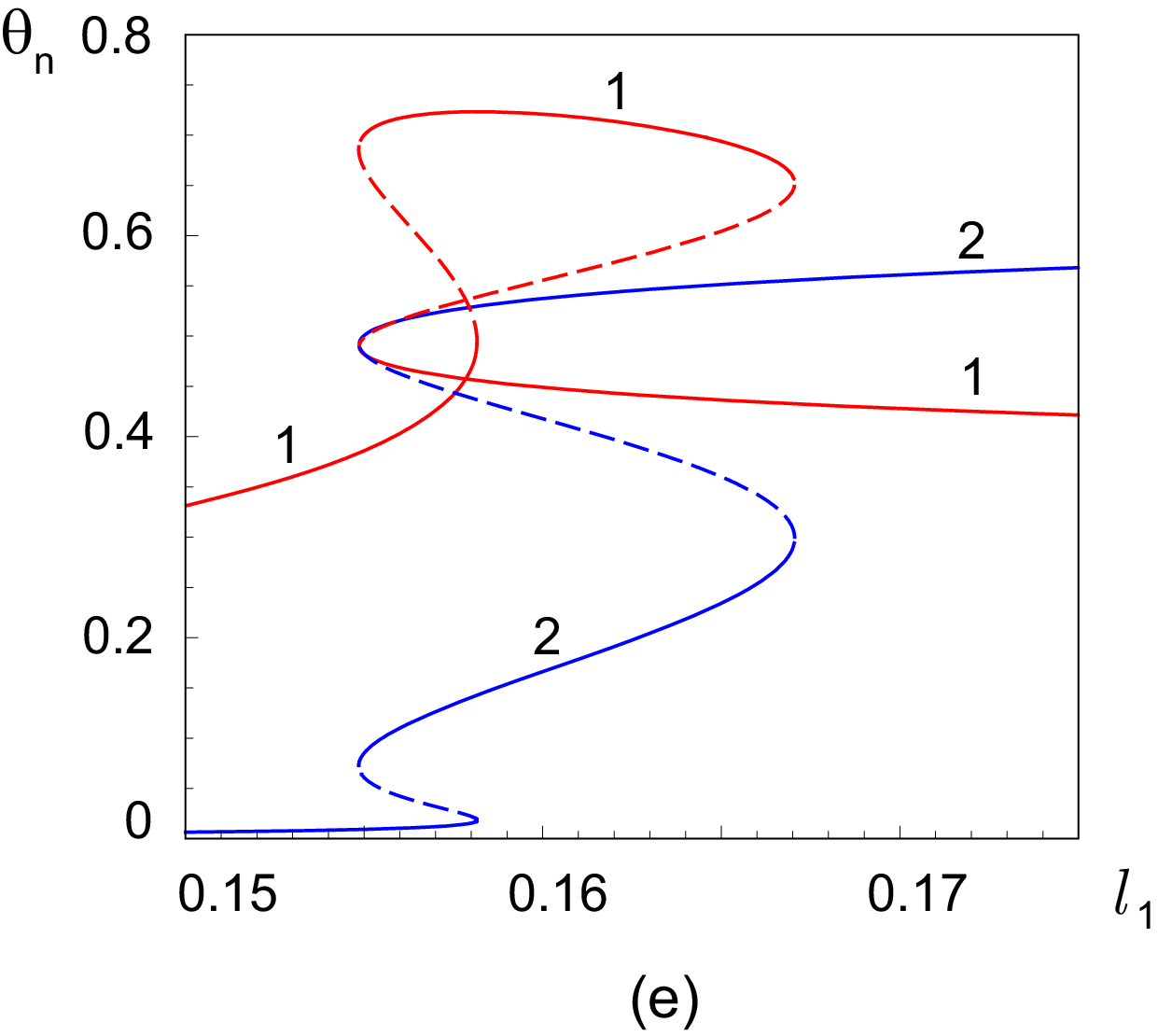}  \hfill
 \includegraphics[width=50mm, height=70mm]{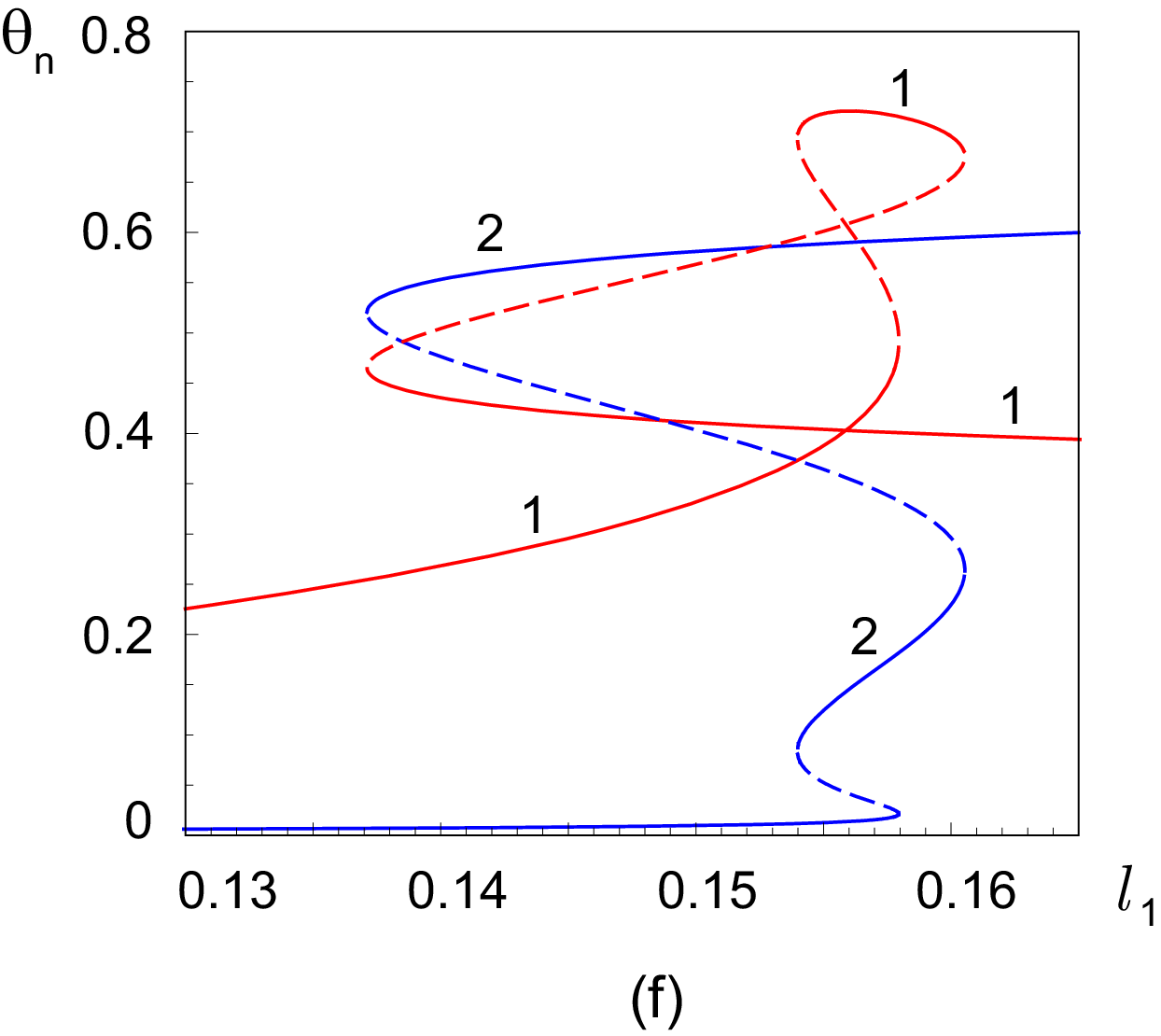}  \\
  \vfill
 \includegraphics[width=50mm, height=70mm]{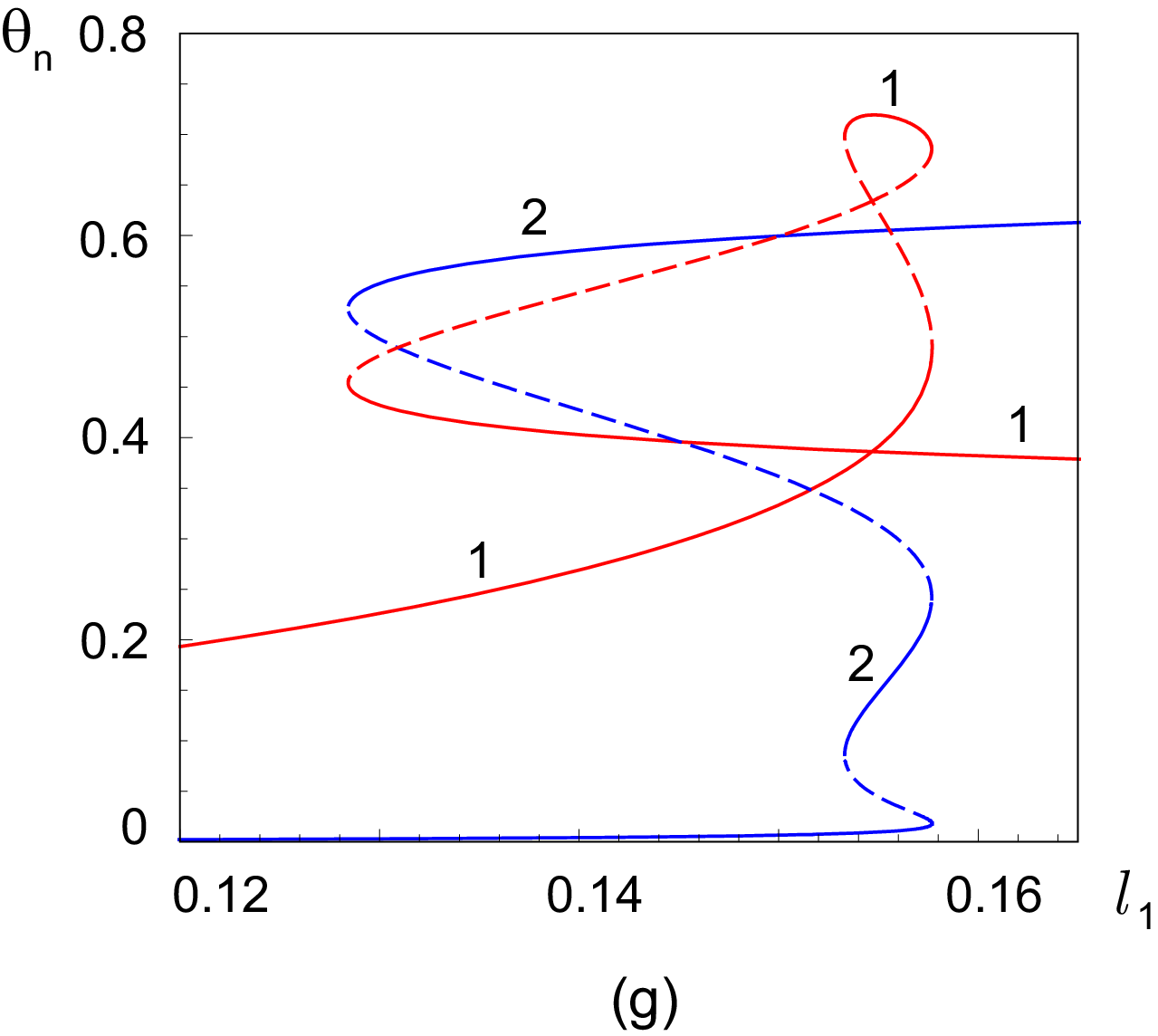}  \hfill
 \includegraphics[width=50mm, height=70mm]{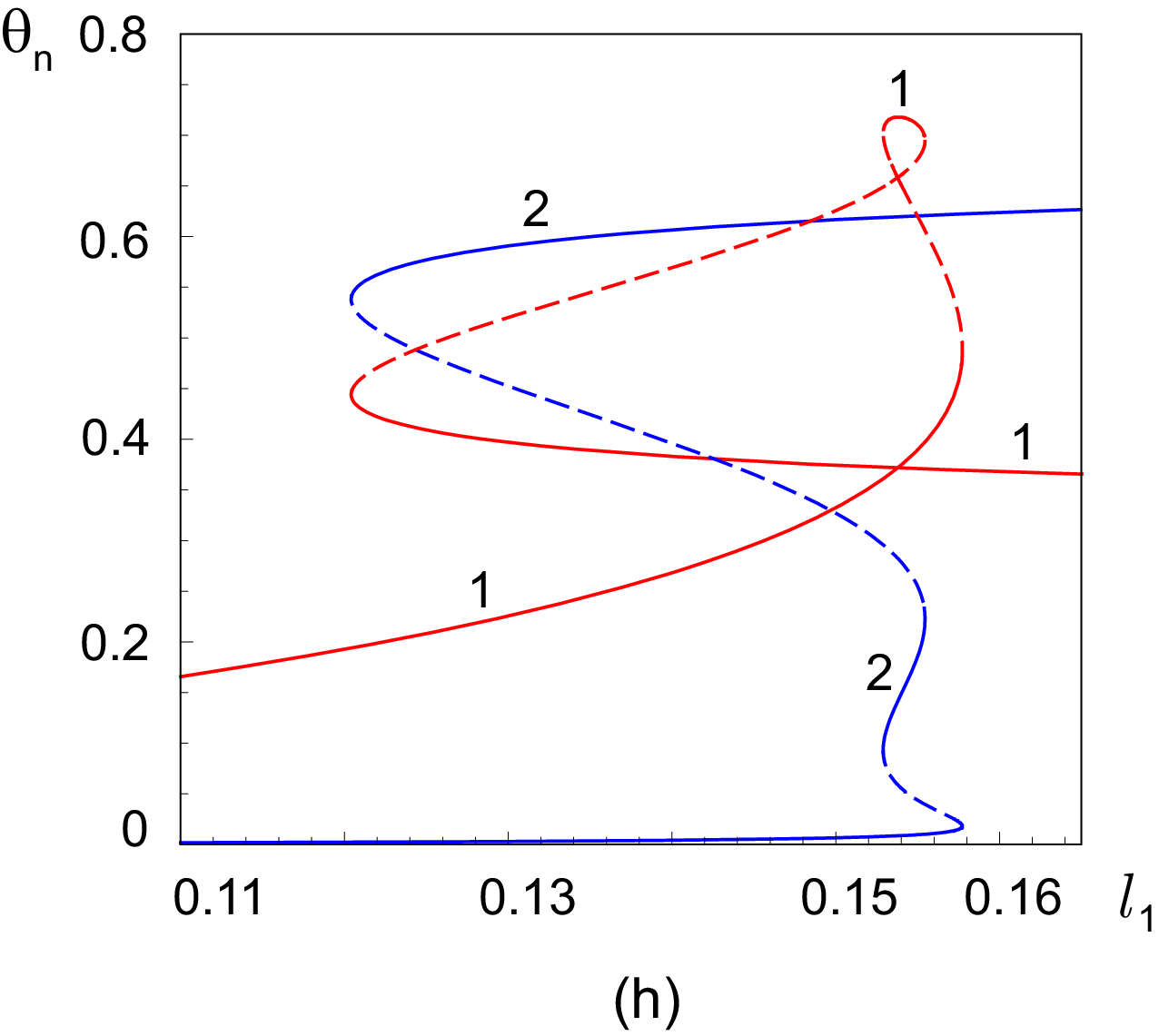}  \hfill
 \includegraphics[width=50mm, height=70mm]{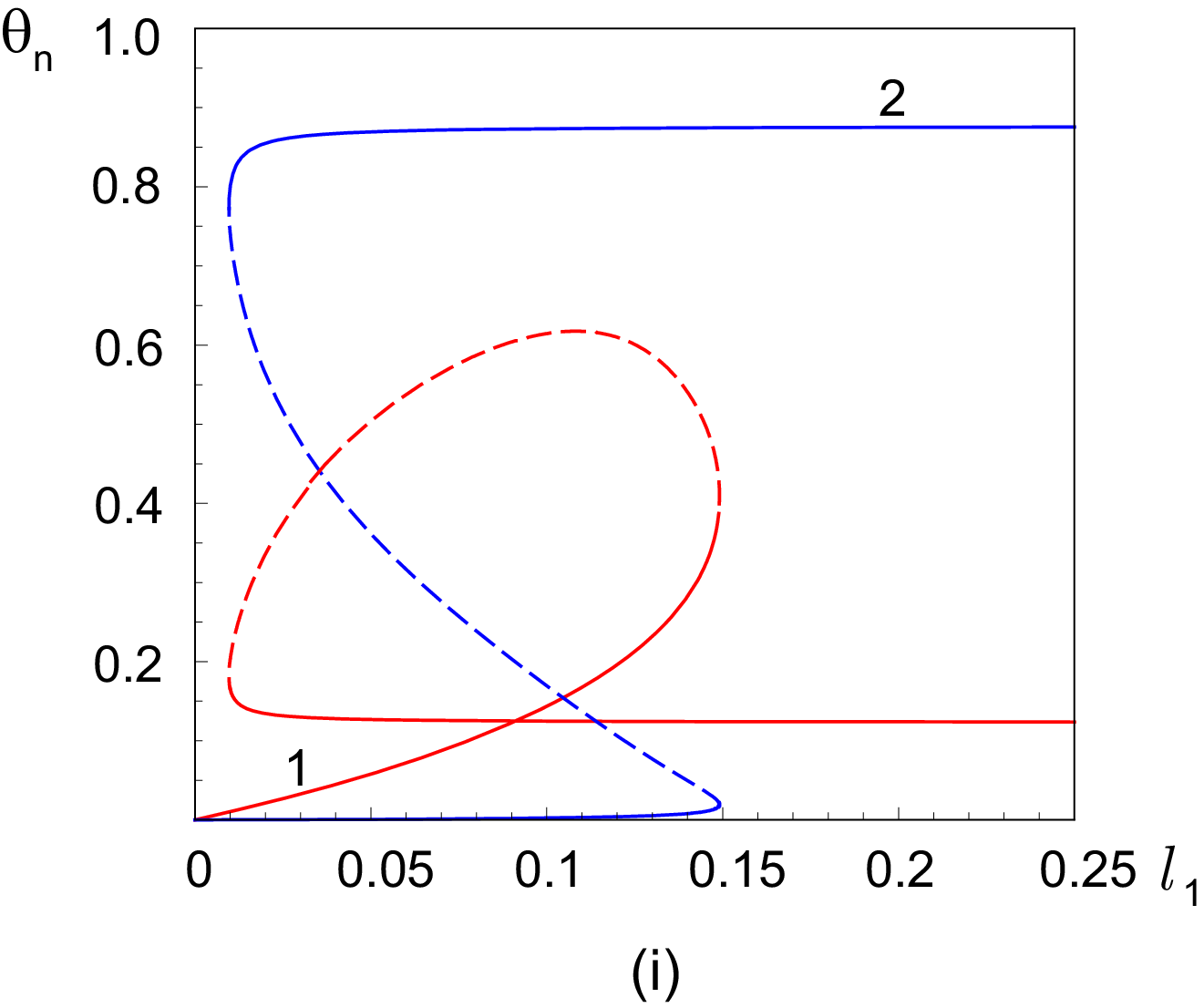}  \\
  \vfill
 \caption {Surface coverages $\theta_n$ by adparticles of species $n$
 ($n = 1$, curve 1; $n = 2$,  curve 2) vs the concentration $\ell_1$.
  The values of $S_0, G$, and $g_1$ are the same as in Fig.~6.}
 }
\label{pRGBfigc._7u_003}
\end{figure}

As the concentration  $\ell_1$ increases/decreases, the behavior of
$\xi(\ell_1)$ in Fig.~6d is similar to the behavior of this function in Fig.~6c
but the transitions of $\xi$ from the second (as $\ell_1$ increases) and third
(as $\ell_1$ decreases) stable branches to the second stable branch go along
the different vertical straight lines $\ell_1 = \ell^b_{1,2}$ and $\ell_1 =
\ell^b_{1,3}$ rather than the same one as in Fig.~6c. \looseness=-1

The behavior of the surface coverage $\theta_1(\ell_1)$ in Fig.~7d is similar
to its behavior in Fig.~7c but with the replacement of the self-tangency point
of $\theta_1(\ell_1)$ in Fig.~7c by two self-intersection points of
$\theta_1(\ell_1)$ in Fig.~7d one of which is the point of intersection of the
first and third stable branches and the second is the point of intersection of
the unstable branches. Note that the intersection of two stable branches of
$\theta_1(\ell_1)$ means only the same value of the surface coverage $\theta_1$
for two different displacements of adsorption sites $\xi$ for the corresponding
value of the concentration $\ell_1$, i.e., a partial degeneration of two
stationary three-component solutions of the problem with respect to one
component ($\theta_1$ in this case), rather than a continuous transition
between stable branches of  $\theta_1(\ell_1)$ at the point of their
intersection which is forbidden by the condition for transition between
stationary solutions of the system.

In the special case $S_0 = S_d$ where system (7) has two two-fold stationary
solutions, the behavior of the functions  $\xi(\ell_1)$ and $\theta_n(\ell_1)$
in Figs.~6e and 7e is similar to their behavior in Figs.~6d and 7d only for
increasing $\ell_1$. As $\ell_1$ decreases from a value greater than
$\ell^b_{1,4}$, these quantities vary along their third stable branches up to
their end at $\ell_1 = \ell^d_1$; then $\xi$ and $\theta_2$ successively jump
down, first, to the second stable branches and then to the first stable
branches, whereas $\theta_1$ successively, first, jumps up to the second stable
branch and then jumps down to the first stable branch. Then $\xi$ and
$\theta_n$ decrease along their first stable branches.

The curves in Figs.~6f and 7f distinctly illustrate discontinuous transitions
of $\xi$ and $\theta_n$ from the third stable branches directly to the first
stable branches at $\ell_1 = \ell^b_{1,3}$ for $S_0 \in (S_d, \, S_t)$ as
$\ell_1$ decreases, which implies that a stationary solution of system (7) on
the second stationary branch can be achieved only for increasing $\ell_1$.

In the special case  $S_0 = S_t$ where system (7) has two two-fold stationary
solutions (Figs.~6g and 7g), as $\ell_1$ increases from zero, $\xi$ and
$\theta_n$ increase along their first stable branches up to their end at
$\ell_1 = \ell^t_1$.  At this bifurcation concentration, $\xi$ and $\theta_2$
successively jump up, first, to their second stable branches and then to the
third stable branches, whereas $\theta_1$, first, jumps up to the second stable
branch and then jumps down to the third stable branch. Then $\xi$ and
$\theta_n$ vary along their third stable branches. As $\ell_1$ decreases from a
value greater than $\ell^t_1$, the behavior of $\xi(\ell_1)$ and
$\theta_n(\ell_1)$ (``disregard'' of the second stable branches) is similar to
their behavior in Figs.~6f and 7f.

The curves in Figs.~6h and 7h for $S_0 \in (S_t, \, S^c_1)$  illustrate that
$\xi(\ell_1)$ and $\theta_n(\ell_1)$  ``disregard'' the second stable branches
for both increasing (from a value lesser than $\ell^b_{1,1}$) and decreasing
(from a value greater than $\ell^b_{1,4}$) concentration $\ell_1$. Thus, a
stationary solution of system (7) on the second stable branch cannot be
achieved by transition from any other stable branch and, hence, a tristable
system behaves like a bistable one.

As is seen in Figs.~6c--h and 7c--h, the length of the second stable branch
decreases as $S_0 \in [S_u, \, S^c_1]$ increases (most clearly, it is
illustrated by the second stable branch of $\theta_1(\ell_1)$) and becomes
equal to zero for $S_0 = S^c_1$, which leads to the union of two unstable
branches. \looseness=-1

For $S_0 > S^c_1$, the coordinate  $\xi(\ell_1)$  (Fig.~6i) and the surface
coverage $\theta_2(\ell_1)$  (Fig.~7i) have a single hysteresis in the domain
$\ell \in [\ell^b_{1,3}, \, \ell^b_{1,2}]$, whereas the surface coverage
$\theta_1(\ell_1)$  has a loop: two intersecting stable branches connected by
the unstable branch (Fig.~7i). However, transitions between the stable branches
of $\theta_1(\ell_1)$ are discontinuous at $\ell = \ell^b_{1,2}$ (as $\ell_1$
increases) and $\ell = \ell^b_{1,3}$ (as $\ell_1$ decreases) rather than a
continuous transition at the point of their intersection.

The curves in Fig.~7i illustrate that the asymptotical value of the surfaces
coverage $\theta^a_2$ considerably exceeds the asymptotical value of the
surfaces coverage $\theta^a_1$ ($S(\xi^a) \approx 7.12$). Thus, due to a great
displacement of adsorption sites ($\xi^a \approx 1.88$), the adsorbent surface
is occupied mainly by adparticles of species 2 rather than adparticles of
species 1 as in the Langmuir case.  \looseness=-1



\subsection{Adsorption Isotherms with Several Asymptotes}  \label{Several asymptotes}

As has been shown in Sec.~3.1, the coordinate $\xi(\ell_1)$ and the surface
coverages $\theta_n(\ell_1)$ have three horizontal asymptotes (two stable and
one unstable) if $g_1 > g^a_c$ and, e.g., $S_0 \in (S^a_-, S^a_+)$ for $G
>1$. For $G = 2$, the value of $g^a_c$ coincides with the critical value of $g_1$ in
the case of one-component adsorption ($g^a_c = g_c = 4$) and  $S^a_c = \exp(-6)
\approx 0.00249$.

\begin{figure}[!ht]
 \centering{
 \includegraphics[width=50mm, height=80mm]{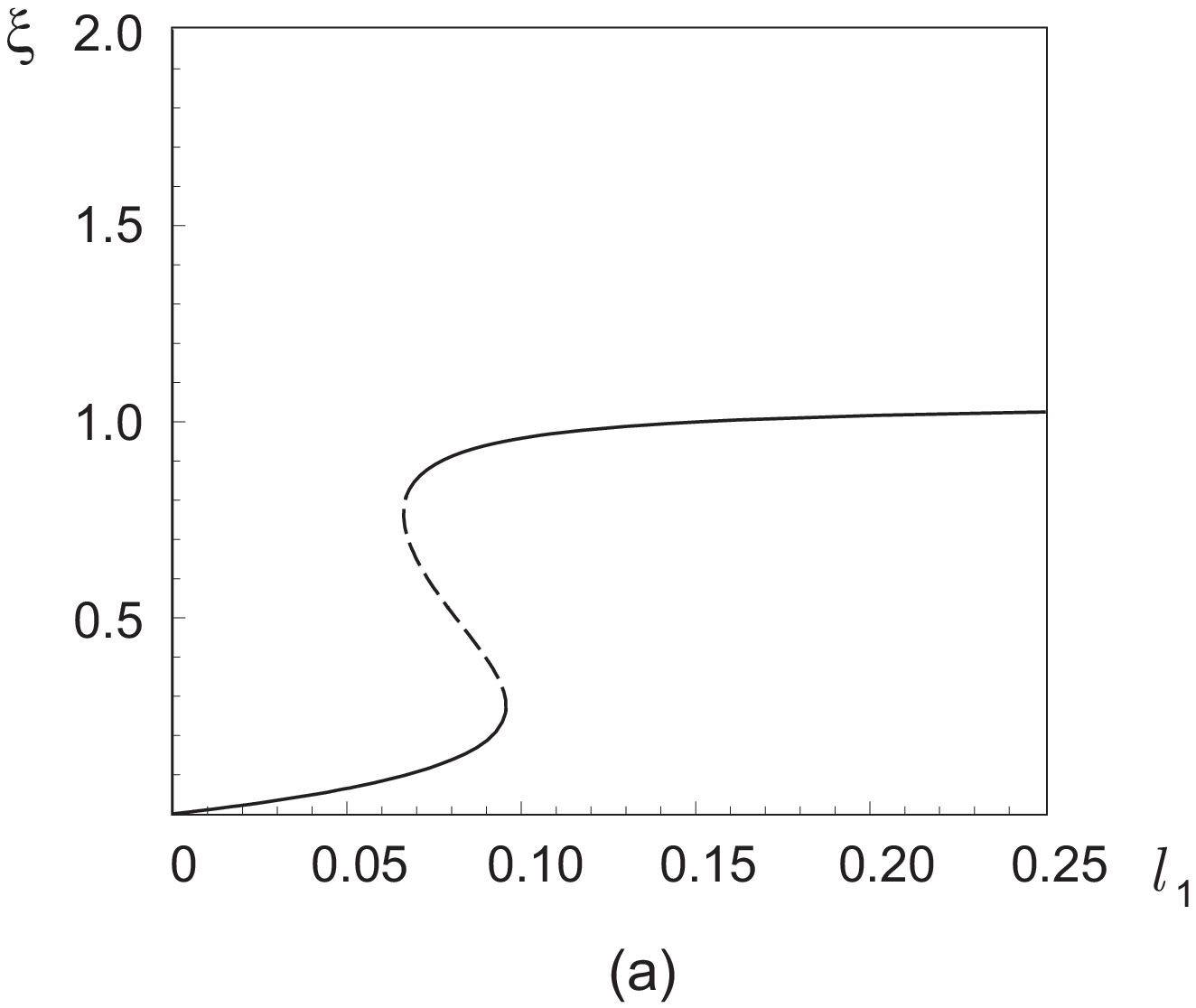} \hfill
 \includegraphics[width=50mm, height=80mm]{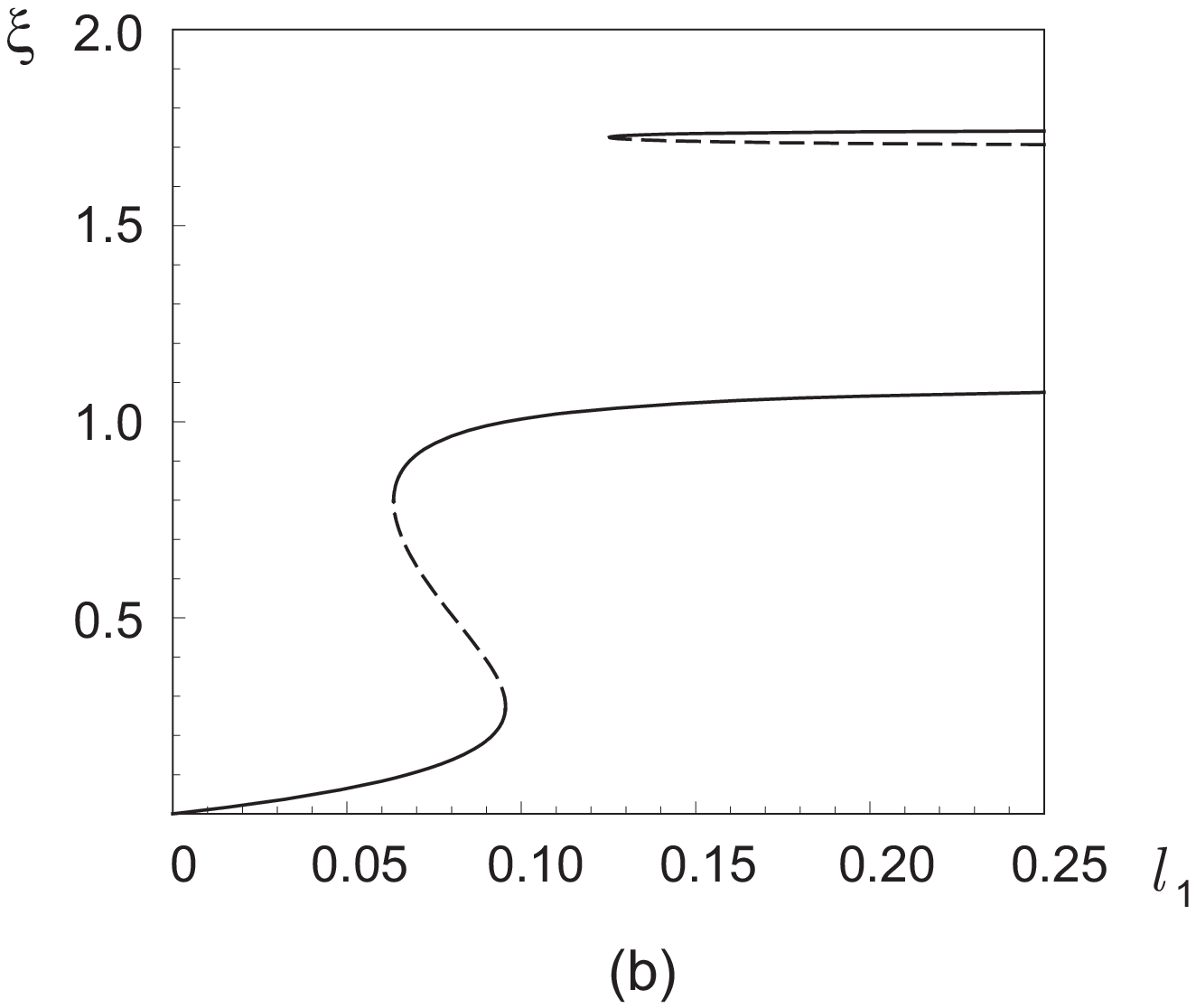} \hfill
 \includegraphics[width=50mm, height=80mm]{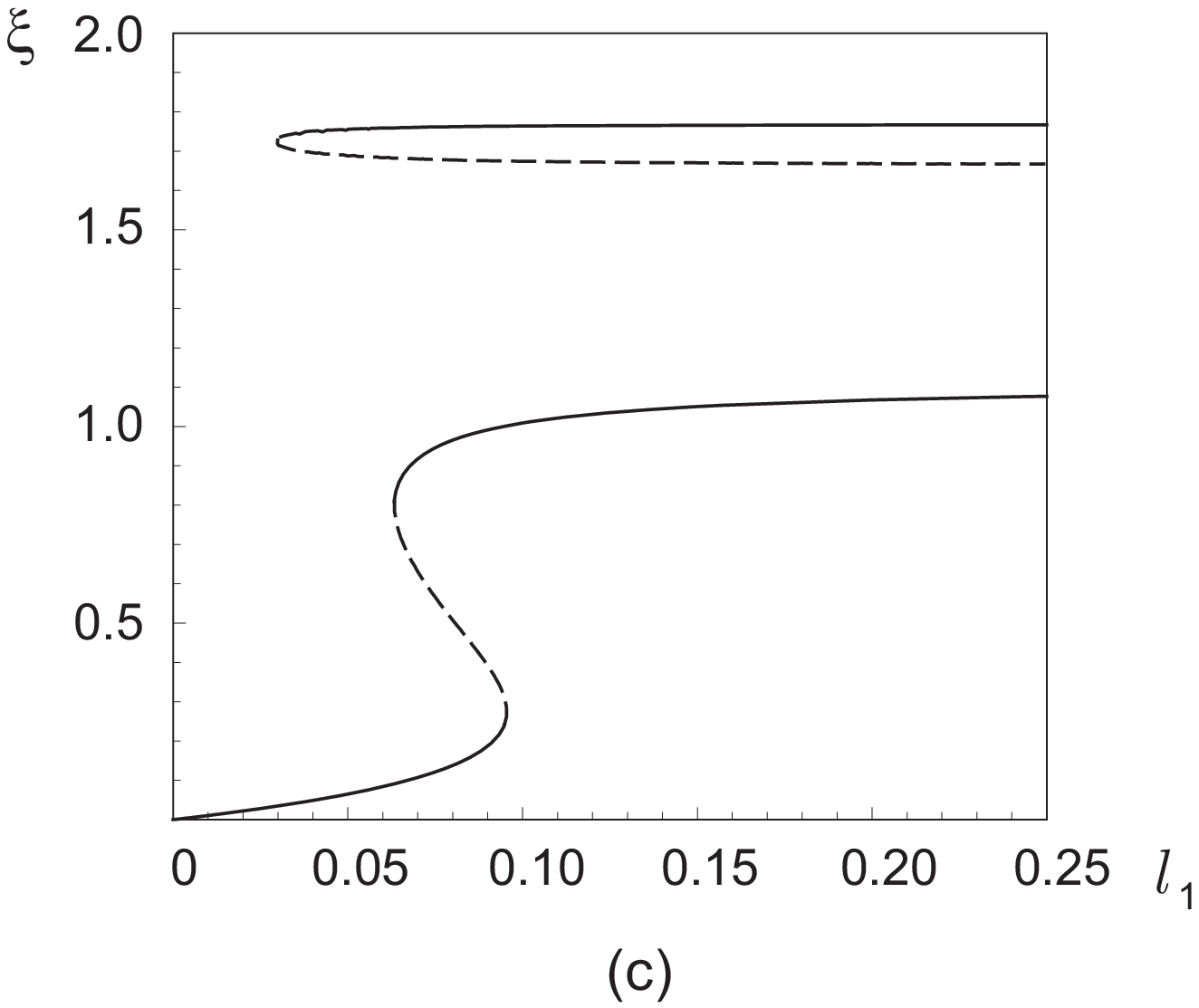}  \\
  \vspace{5mm}
 \includegraphics[width=50mm, height=80mm]{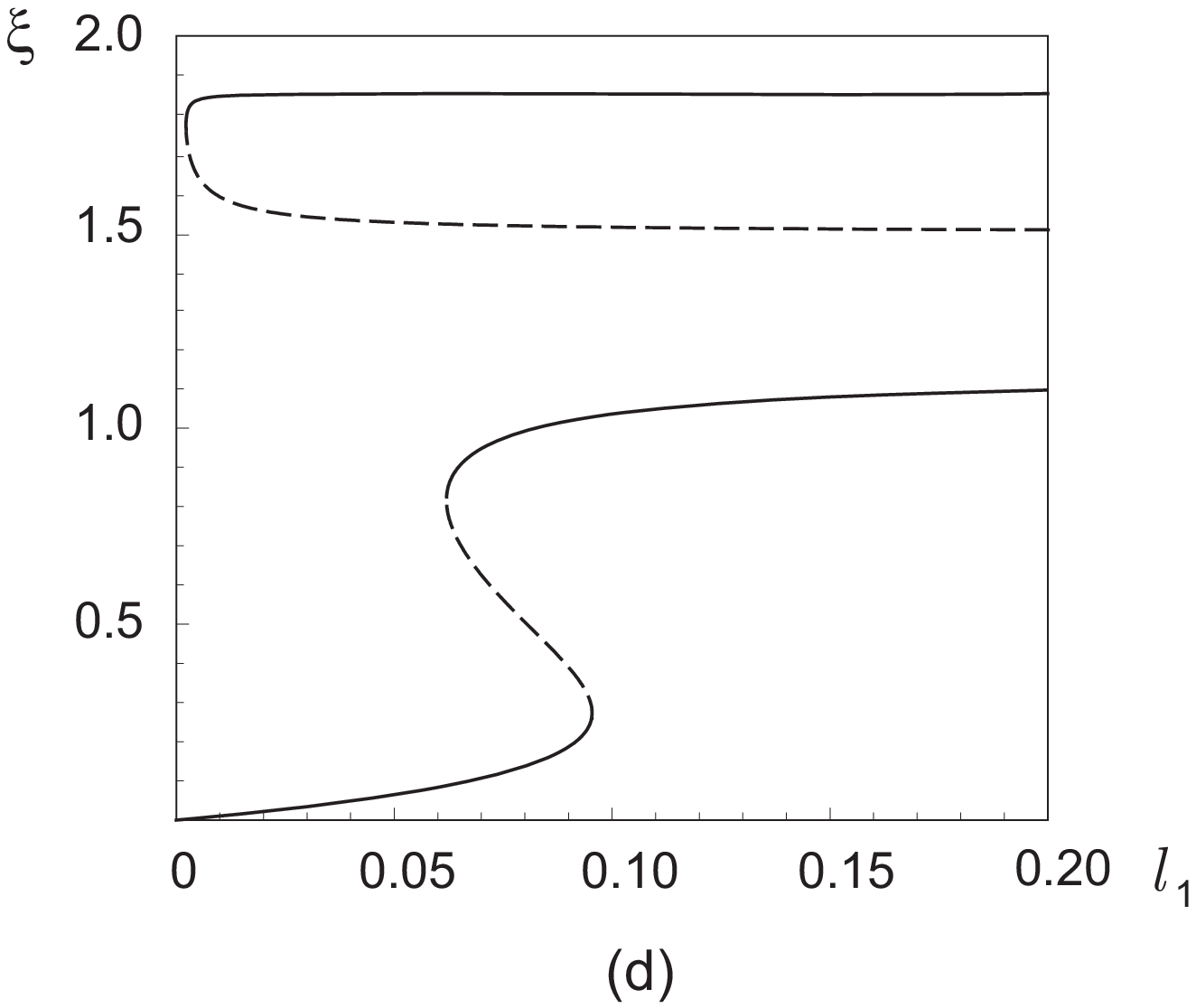} \hfill
 \includegraphics[width=50mm, height=80mm]{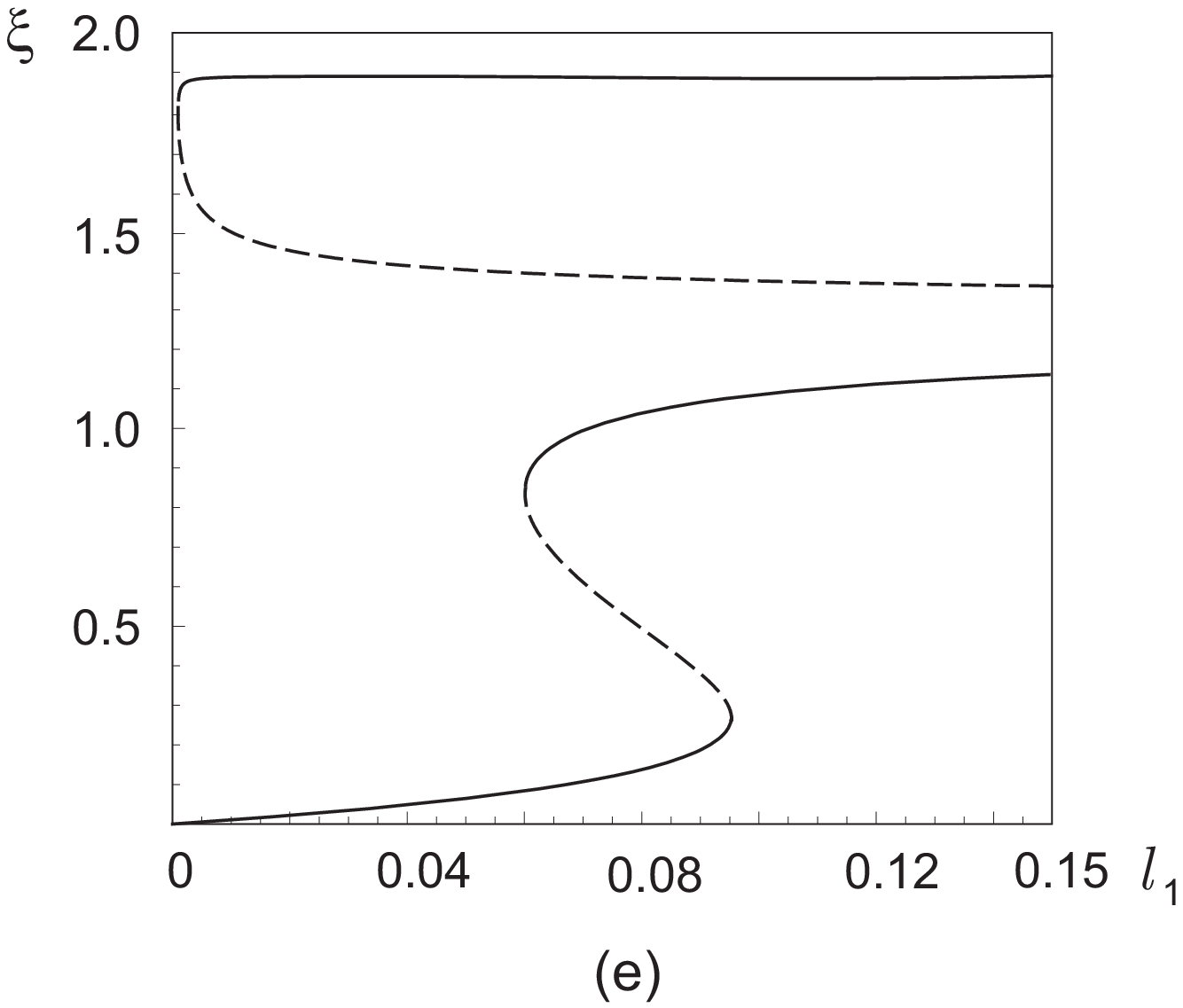} \hfill
 \includegraphics[width=50mm, height=80mm]{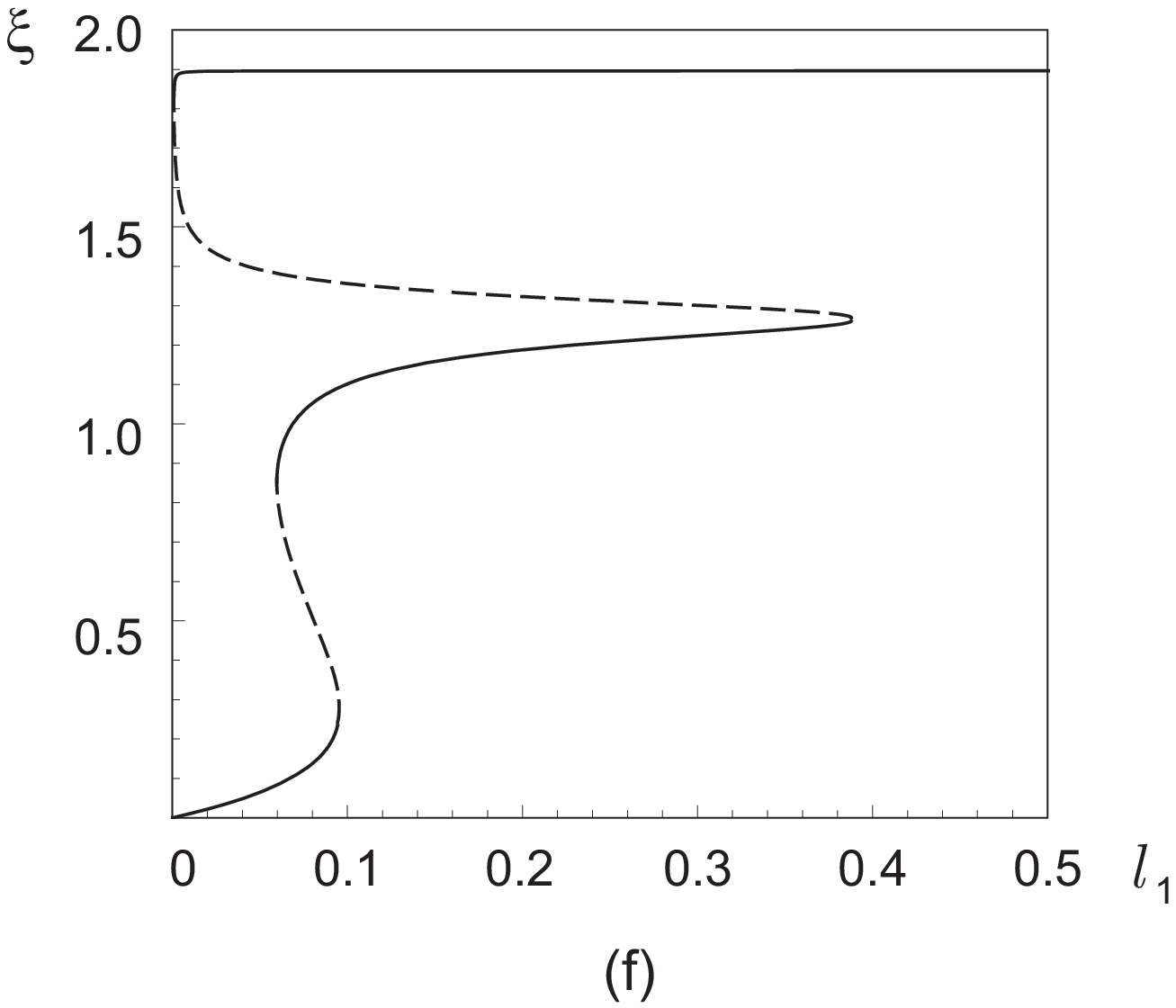}  \\
 \caption {Equilibrium position of oscillator  $\xi$  vs the concentration
  $\ell_1$ for different values of  $S_0$:
  $S_0$ = 0.0003~(a),   0.000475~(b),  0.00048~(c),
          0.00055~(d),  0.00064~(e),   0.00066~(f);
  $G = 2,  g_1 = 5$.}
 }
 \label{pfigc._8}
\end{figure}

\begin{figure}[!ht]
 \centering{
 \includegraphics[width=50mm, height=80mm]{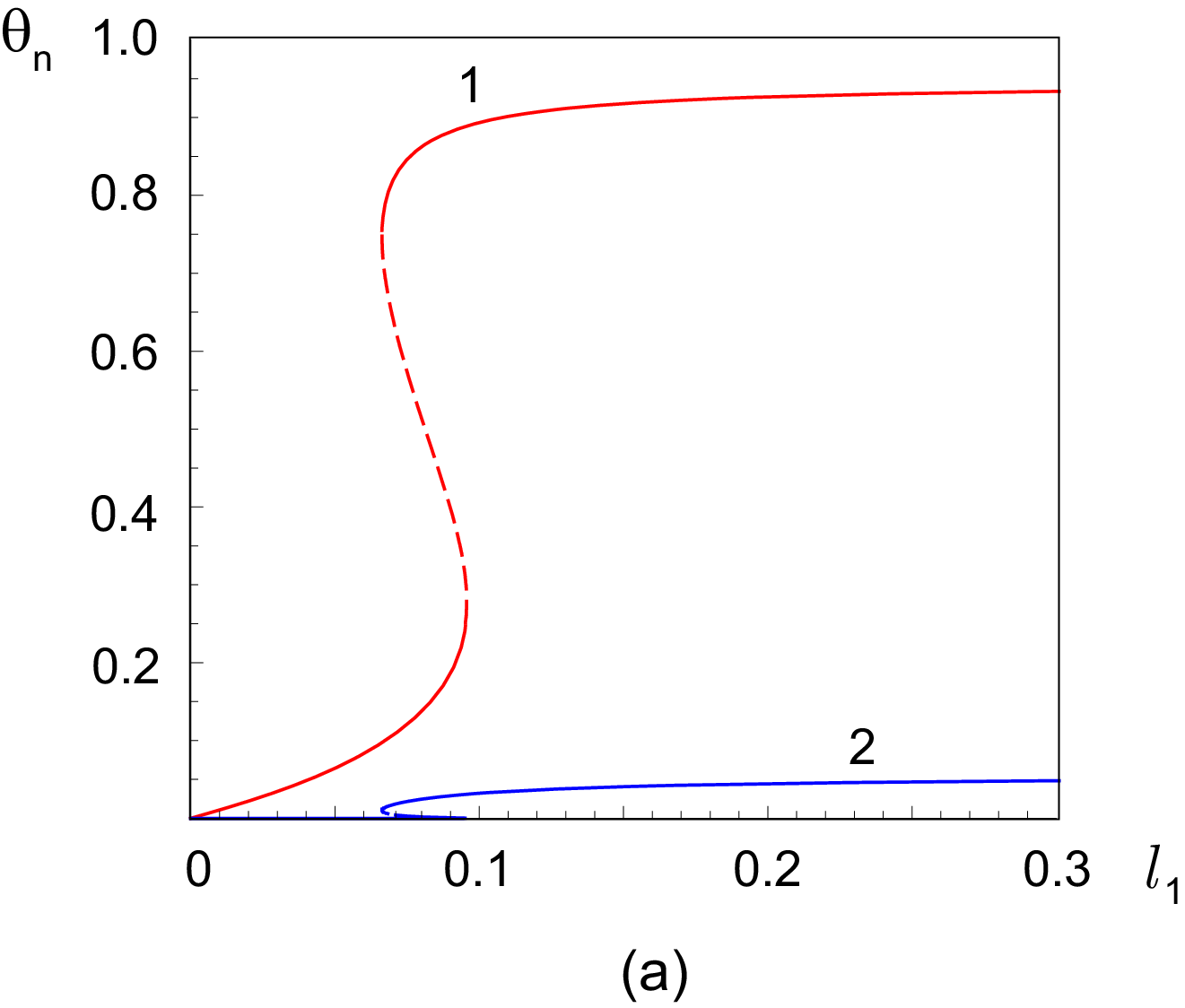} \hfill
 \includegraphics[width=50mm, height=80mm]{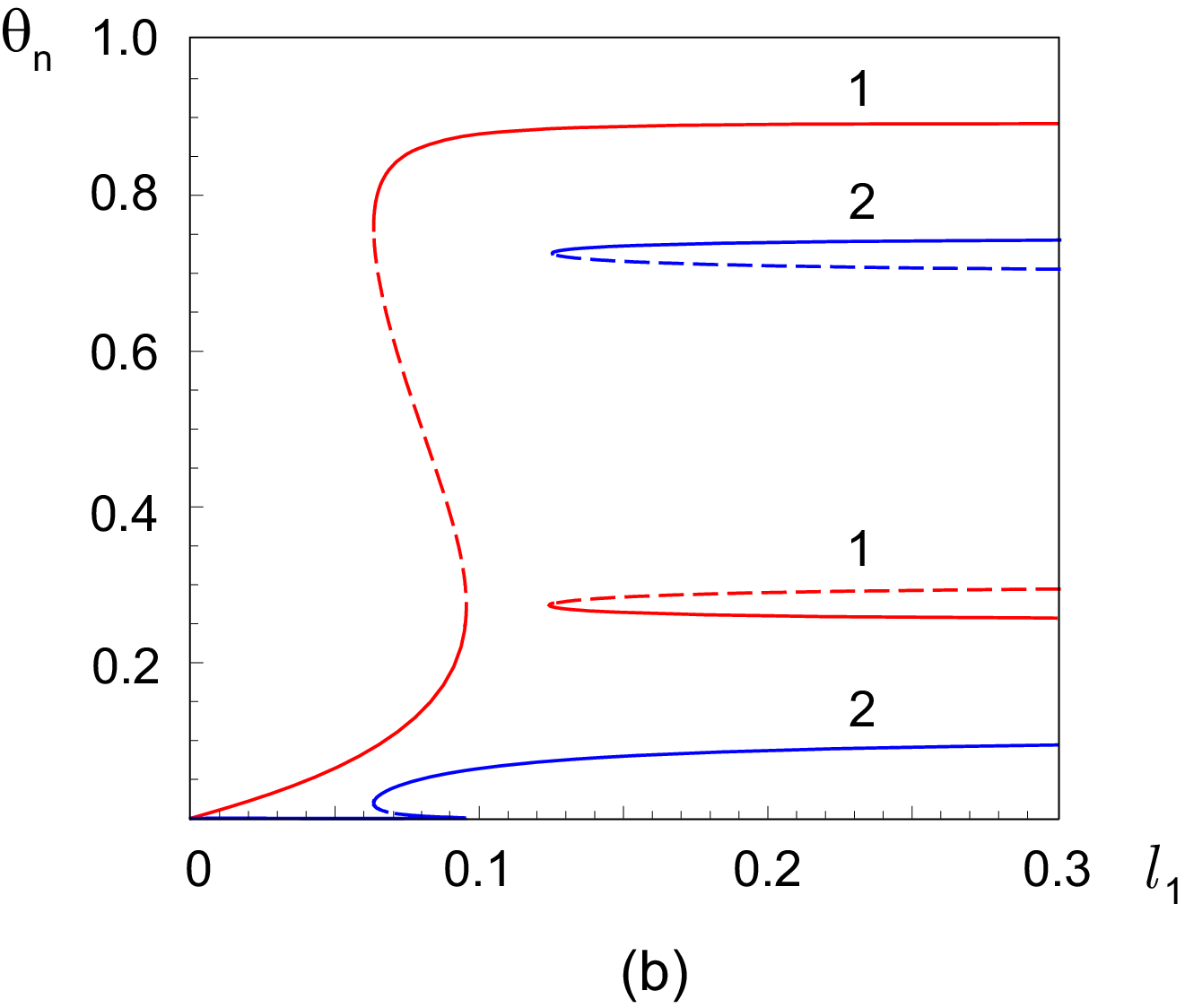} \hfill
 \includegraphics[width=50mm, height=80mm]{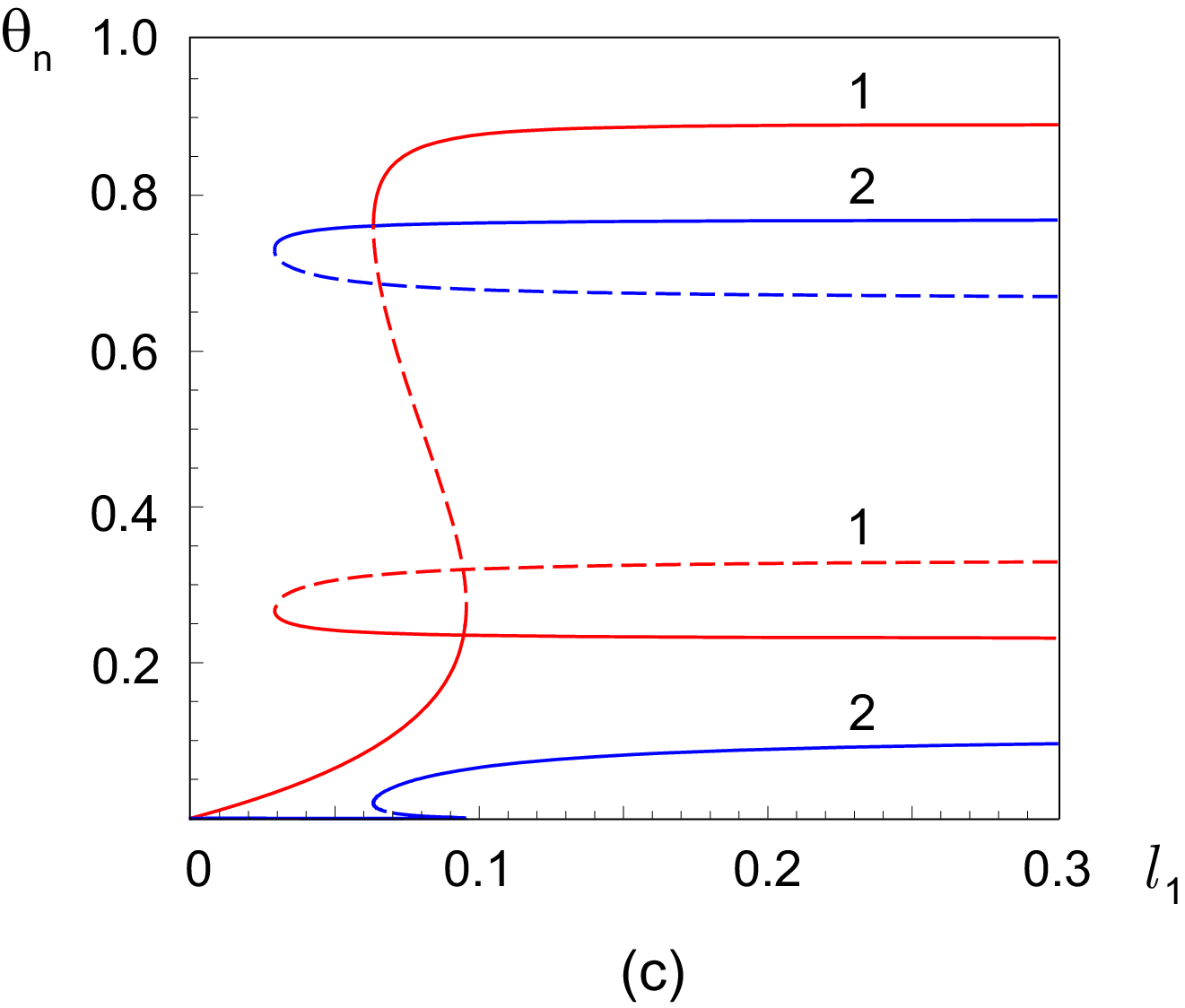}  \\
  \vspace{5mm}
 \includegraphics[width=50mm, height=80mm]{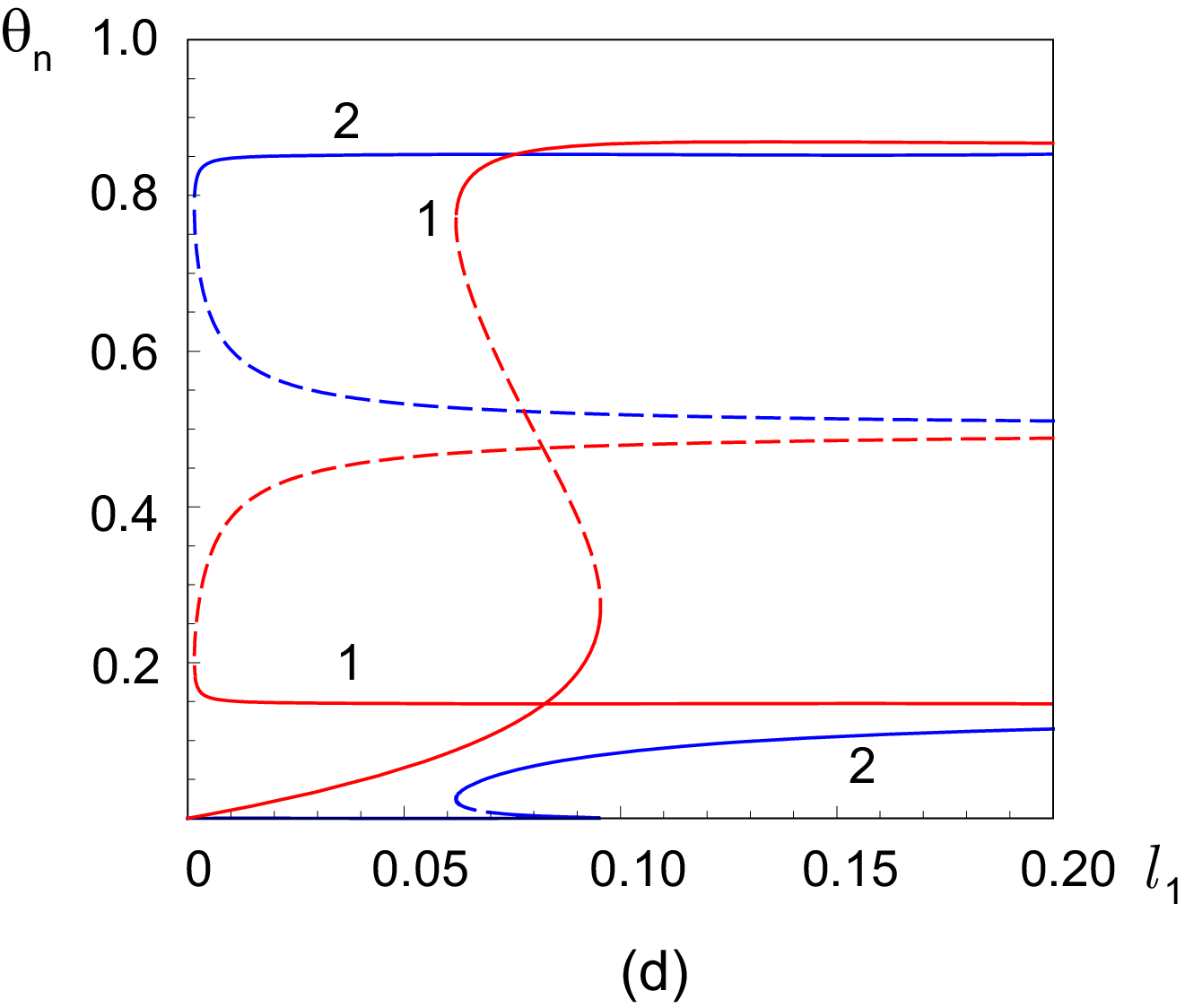} \hfill
 \includegraphics[width=50mm, height=80mm]{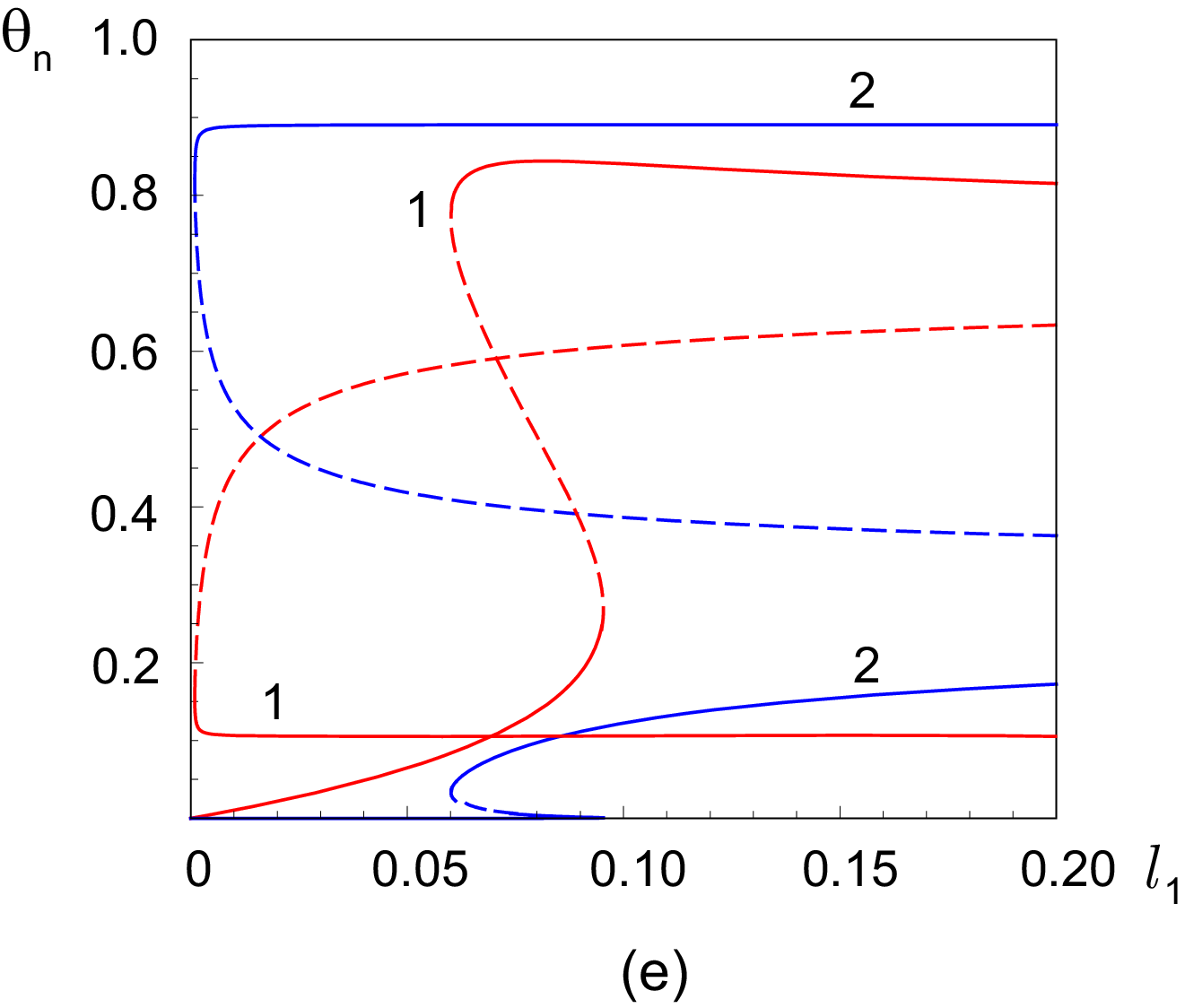} \hfill
 \includegraphics[width=50mm, height=80mm]{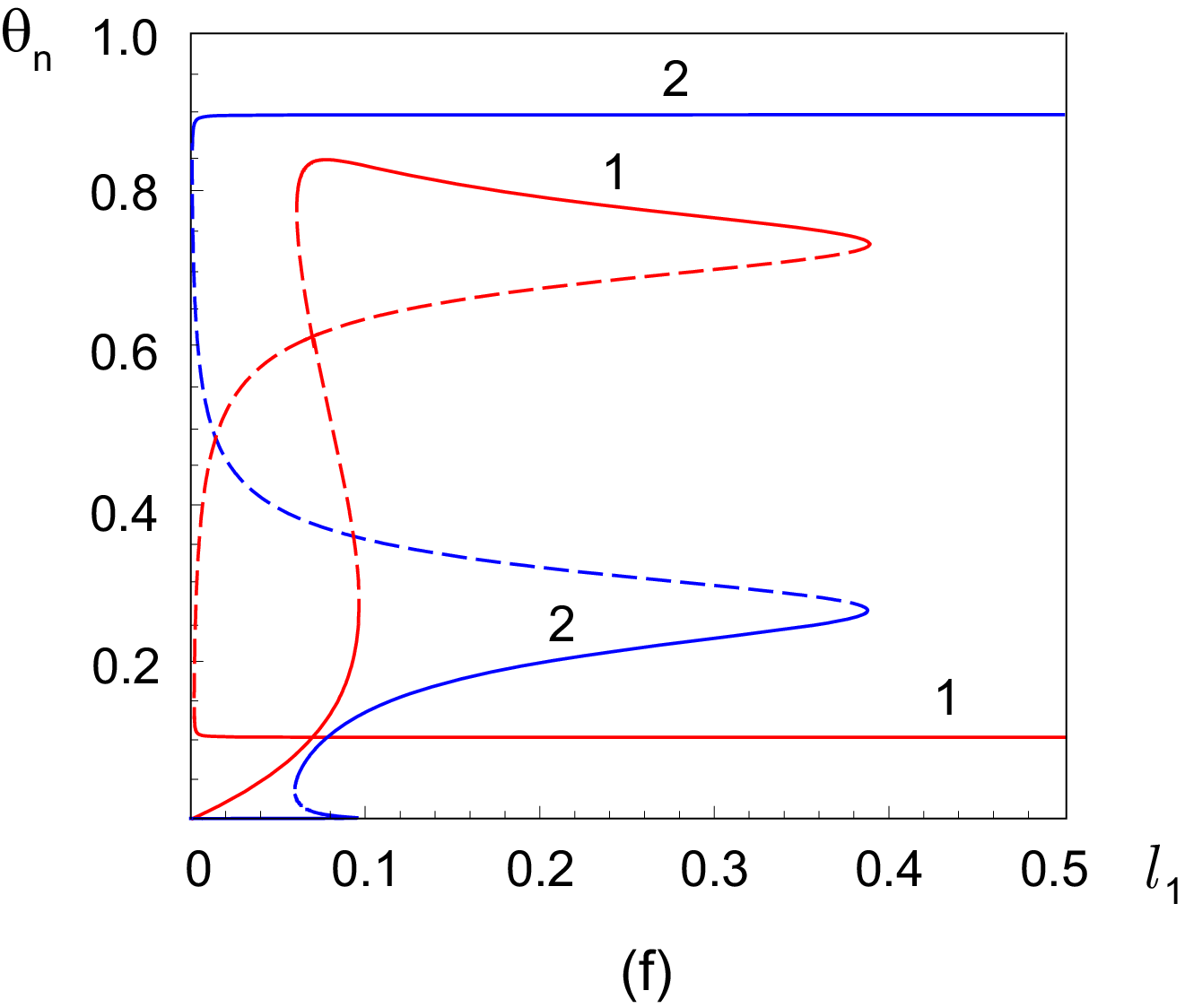}  \\
 \caption {Surface coverages  $\theta_n$  by adparticles of species $n$ ($n = 1$,
  curve 1; $n = 2$, curve 2) vs the concentration $\ell_1$.
  The values of $S_0, G$, and $g_1$ are the same as in Fig.~8.}
 }
 \label{RGBfigc._9}
\end{figure}

The graphs of the equilibrium position of oscillator $\xi(\ell_1)$ and the
surface coverages $\theta_n(\ell_1)$ for $g_1 = 5$ depicted in Figs.~8 and 9,
respectively, illustrate specific features of these functions in the case where
a stationary solution of system (7) can have several asymptotes. In this case,
$S^a_- \approx 0.0004734$ and $S^a_+ \approx 0.0006462$.

For $S_0 < S^a_-$, the coordinate  $\xi(\ell_1)$  (Fig.~8a) and the surface
coverages $\theta_n(\ell_1)$ (Fig.~9a) have a hysteresis typical of these
quantities in adsorption of a one-component gas for values of the coupling
parameter greater than critical \cite{ref.Usenko} or a two-component gas, e.g.,
for $g_1 \in (3, 4)$ and $S_0 \in (S^c_-,\, S^c_+)$ (see Figs.~6b and 7b). The
coordinate $\xi(\ell_1)$ consists of three branches: two stable branches (the
lower stable branch for $\ell_1 \in [0, \, \ell^b_{1,2}]$ and the upper stable
branch for $\ell_1 \in [\ell^b_{1,1},\, \infty)$ that approaches its horizontal
asymptote $\xi = \xi^a_1$ as $\ell_1$ increases) and one unstable branch
connecting them. The surface coverages $\theta_n(\ell_1)$ have a similar shape.
For all considered values of $S_0$, the lower stable branch of
$\theta_2(\ell_1)$ in Fig.~9 almost coincides with the abscissa axis.
\looseness=-1

For  $S_0 \in (S^a_-,\, S^a_+)$, the behavior of $\xi(\ell_1)$ and
$\theta_n(\ell_1)$ qualitatively differs from their behavior in Figs.~6 and 7.
For $S_0 >S^a_-$, there appears an isolated piece of $\xi(\ell_1)$  with
semiinfinite domain of definition $\ell_1 \in [\ell^b_{1,3},\, \infty)$
(Fig.~8b). This isolated piece consists of stable and unstable branches
starting at the bifurcation concentration $\ell^b_{1,3}$ and rapidly tending to
closely lying asymptotes $\xi = \xi^a_3$ and  $\xi = \xi^a_2$ ($\xi^a_3 >
\xi^a_2$), respectively, as $\ell_1$ increases. Thus, the range of values of
the positive-definite function $\xi(\ell_1)$ consists of two intervals ($\xi
\in [0,\, \xi^a_1)$  and $\xi \in (\xi^a_2,\, \xi^a_3)$) with gap $\xi \in
(\xi^a_1,\, \xi^a_2)$ between them. According to the principle of perfect delay
\cite{ref.PoS, ref.Gil}, the transition from the first piece ($\xi \in [0,\,
\xi^a_1)$) of $\xi(\ell_1)$ to the isolated piece ($\xi \in (\xi^a_2,\,
\xi^a_3)$) with variation in the concentration $\ell_1$ is impossible for any
initial value of  $\ell_1$. If the initial state of the system lies on the
stable branch of the isolated piece of $\xi(\ell_1)$, then, as $\ell_1$
decreases, the coordinate $\xi$ varies along this branch up to its end at
$\ell_1 = \ell^b_{1,3}$, then jumps down to the upper stable branch of the
first piece of  $\xi(\ell_1)$ and varies along it in the same way as in
Fig.~8a. \looseness=-1

Since the behavior of the surface coverage $\theta_2(\ell_1)$ in Fig.~9b is
similar to the behavior of the coordinate $\xi(\ell_1)$ in Fig.~8b, all
conclusions for $\xi(\ell_1)$ remain true for $\theta_2(\ell_1)$. Moreover,
this also holds for other values of $S_0$ (cf. curve 2 in Fig.~9c--f with curve
in Fig.~8c--f).

The surface coverage $\theta_1(\ell_1)$ in Fig.~9b also has the isolated piece.
However, unlike the isolated pieces of the surface coverage $\theta_2(\ell_1)$
and the coordinate $\xi(\ell_1)$ (Fig.~8b), it lies below the asymptote
$\theta_1 = \theta^a_{1,1}$ of the first piece of $\theta_1(\ell_1)$, $\
\theta^a_{1,1} > \theta^a_{1,2} > \theta^a_{1,3}$.

As $S_0 \in (S^a_-,\, S^a_+)$  increases, the isolated piece of $\xi(\ell_1)$
shifts to the ordinate axis (the bifurcation concentration $\ell^b_{1,3}$
decreases), its thickness increases, and, in a certain interval of $\ell_1$,
the system is tristable (Fig.~8c--e). As above, the transition from the first
piece of $\xi(\ell_1)$ to the isolated piece with variation in $\ell_1$ is
impossible.  However, as the concentration $\ell_1$ decreases, the transition
from the stable branch of the isolated piece of $\xi(\ell_1)$ to the lower
stable branch (rather than the upper stable branch as in Fig.~8b) of the first
piece of $\xi(\ell_1)$ occurs at $\ell_1 = \ell^b_{1,3}$. \looseness=-1

Unlike the surface coverage $\theta_2(\ell_1)$, the pieces of the surface
coverage $\theta_1(\ell_1)$ in Figs.~9c--e intersect one another. However, the
continuous transition between stable branches of the different pieces of
$\theta_1(\ell_1)$  at the point of their intersection is forbidden by the
condition for transition between stationary solutions of the system.
\looseness=-1

For $S_0 > S^a_+$, the gap between two pieces of  $\xi(\ell_1)$ disappears
($\ell^b_{1,4}$  is finite) and the function $\xi(\ell_1)$  is continuous and
has three stable and two unstable branches (Fig.~8f). There are two bistability
intervals $(\ell^b_{1,3}, \, \ell^b_{1,1})$ and $(\ell^b_{1,2}, \,
\ell^b_{1,4})$ and one tristability interval $(\ell^b_{1,1}, \, \ell^b_{1,2})$
between them. The graph of the surface coverage $\theta_1(\ell_1)$ in Fig.~9f
is also continuous and consists of three stable branches and two unstable
branches connecting them. However, the shapes of $\theta_1(\ell_1)$ and
$\theta_2(\ell_1)$ are essentially different. For $\ell_1 > \ell^b_{1,4}$, the
adsorption sites are considerably displaced from their nonperturbed equilibrium
position $\xi =0$ so that the coordinate $\xi(\ell_1)$ is, in fact, a constant
(Fig.~8f). In this case, an almost monolayer coverage of the surface mainly by
adparticles of species 2 occurs (cf. the flat regions of the curves in
Fig.~9f), whereas, in the classical case, the surface coverage by adparticles
of species 2 is less than 0.1\% of the total coverage. \looseness=-1



\section{Adiabatic Approximation}  \label{Adiabatic approximation}

The specific features of stationary solutions of system (7) investigated in
Sec.~3 can be explained with the use of a potential. To this end, we consider
the last equation of system (7) in the overdamped approximation where the
masses of an adsorption site and adparticles are low and the friction
coefficient is so large that the first term on the left-hand side of this
equation can be neglected as against the second. Using the well-known results
for a linear free oscillator of constant mass \cite{ref.AVKh}, this
approximation is correct if
\begin{equation}
 \tau^2_M  \ll \tau^2_r,
\end{equation}
\noindent where  $\tau_M = 1/\omega_M$, $\omega_M = \sqrt{\varkappa/M}$ is the
vibration frequency of an oscillator of mass $M$, and $\tau_r =
\alpha/\varkappa$ is the typical relaxation time of a massless oscillator.

Further, consider the case where the relaxation time of the coordinate $\xi(t)$
of a massless oscillator is much greater than the relaxation times of the
surface coverages $\theta_n(t),\ n = 1, 2$, in the linear case, i.e., the
variables $\xi$ and $\theta_n$ are slow and fast, respectively. In this case,
$\tau_r \gg \tau_{\theta}$, where  \looseness=-1

\begin{equation}
 \tau_{\theta} \approx 2\, \Biggl\{\biggl(\frac{1}{\tau^{ad}_1} + \frac{1}{\tau^{ad}_2} \biggr)
  - \sqrt{\biggl(\frac{1}{\tau^{ad}_1} - \frac{1}{\tau^{ad}_2} \biggr)^2
  + \frac{4}{\tau^{a}_1 \, \tau^{a}_2} \ } \, \Biggr\}^{-1},
\end{equation}
\noindent  $\tau^{ad}_n = \tau^a_n\, \tau^d_n/(\tau^a_n+\tau^d_n)$ is the time
taken for attaining the stationary value of the surface coverage $\theta_n$ in
the case of the Langmuir adsorption of a one-component gas particles of species
$n$ and $ \tau^a_n = 1/k^a_n C_n,\ n = 1, 2$, can be regarded as the typical
lifetime of a vacant adsorption site in this case. Using the principle of
adiabatic elimination of the fast variables $\theta_n(t)$ in (7)
\cite{ref.Hak}, we set $ d\theta_n/dt = 0, \ n = 1,2$, and express the surface
coverage $\theta_1$ vs the slow variable $\xi$ as follows: \looseness=-1
\begin{equation}
 \theta_1 = \frac{\ell_1}{\ell_1 \left(1 + S(\xi)\right) + \exp(-g_1\, \xi)}
 \,.
\end{equation}
The surface coverage $\theta_2$ is defined by relation (17) with $\theta_1$
given by relation (63). The coordinate  $\xi(t)$ is determined as a solution of
the nonlinear differential equation
\begin{equation}
 \alpha \frac{d\xi}{dt} = - \frac{dU(\xi)}{d\xi}
\end{equation}
\noindent  that describes the motion of a massless oscillator in the potential
\begin{equation}
 U(\xi) = \frac{\varkappa}{2} \, \biggl\{ \xi^2 - 2\,h(\xi)\biggr\},
 \end{equation}
\noindent  where the second term on the right-hand of (65) caused by the
adsorption-induced force acting on an adsorption site has the form
\begin{equation}
 h(\xi) = \ell_1 \int\limits_0^{\xi}\!\! dy\, \frac{1 + G\,S(y)}
  {\ell_1 \left(1 + S(y)\right) + \exp(-g_1\, y)}\, .
\end{equation}

Relations (17), (63), and (64) correctly describe the behavior of the dynamical
variables   $\xi(t)$ and $\theta_n(t)$ for times $t \gg \tau_{\theta}$ for
which the fast variables  $\theta_n(t)$  forget the initial data.

The shape of $U(\xi)$ essentially depends on the control parameters $\ell_1,
g_1, S_0$, and $G$. The stationary solutions $\xi_j$ of Eq.~(64), where the
subscript $j$ is the number of a stationary solution, are roots of Eq.~(18)
and, furthermore, the number of roots vary from 1 to 5 depending on the values
of the control parameters. Roots are enumerated so that $\xi_{j+1} > \xi_j$,
and $U_j \equiv U(\xi_j)$. In the case of simple roots, odd and even values of
$j$ correspond to stable (minima of $U(\xi)$) and unstable (maxima of $U(\xi)$)
stationary solutions of Eq.~(64), respectively. For a double root  $\xi_j$, the
potential has a horizontal point of inflection at  $\xi = \xi_j$ and Eq.~(64)
has a two-fold stationary solution. \looseness=-1

In the special case of the identical action of adparticles on the adsorbent ($G
= 1$), relation (65) is reduced to the potential in adsorption of a
one-component gas on a deformable adsorbent \cite{ref.Usenko} with  $\ell$
replaced by $\ell_+$
\begin{equation}
 U(\xi) = \frac{\varkappa}{2} \, \biggl\{ \xi^2 - 2\,\xi - \frac{2}{g}\,
   \ln{\frac{\ell_+ + \exp{\left(-g\, \xi\, \right)}}{\ell_+ + 1}} \biggr\}.
\end{equation}

Using results for one-component adsorption \cite{ref.Usenko}, we conclude that,
for $g > 4$ and $\ell_+ \in (\ell^b_{+,1}, \, \ell^b_{+,2})$, $U(\xi)$  is a
two-well potential with local minima at  $\xi = \xi_1$ and $\xi = \xi_3$
separated by a maximum at $\xi =\xi_2$, where $\xi_j, \ j = 1,2,3$, are the
coordinates determined from Eq.~(36) with regard for relation (35). Thus, in
this case, the system under study is bistable. For $g < 4$ and any
concentrations  $\ell_1$ and $\ell_2$ as well as for  $g > 4$ and $\ell_+
\notin [\ell^b_{+,1}, \, \ell^b_{+,2}]$, the potential $U(\xi)$ has one minimum
and, hence, the considered system is monostable.

In another special case where adparticles of species 2 do not affect the
adsorbent deformation ($G = 0$), the potential is also defined by relation (67)
with  $\ell_+$  replaced  by $\ell_1/(1+ \ell_2)$.

In what follows, we analyze the potential  $U(\xi)$ in the case $G = 2$ for
which the coordinate $\xi(\ell_1)$ and the surface coverages $\theta_n(\ell_1)$
have been investigated in Secs.~3.4, 3.5.

For $g_ 1\in(3, 4)$, $U(\xi)$ is a single-well potential if $S_0 < S^c_-$. For
the given values of the parameters $g_1$ and $S_0$, the depth of the well
$|U_1|$ and the position of its minimum $\xi_1$ increase with $\ell_1$.  As
$S_0$ increases, for $S_0 > S^c_-$, the situation cardinally changes and, for
the given value of $g_1$, the shape of the potential essentially depends on the
values of $\ell_1$ and $S_0$. If $S_0 \in (S^c_-,\, S^c_+)$, then the potential
has either two minima if $\ell_1 \in (\ell^b_{1,1}, \, \ell^b_{1, 2})$ or one
minimum if $\ell_1 \notin [\ell^b_{1,1}, \, \ell^b_{1, 2}]$. \looseness=-1

\begin{figure}[!ht]
 \centering{
 \includegraphics[width=75mm, height=90mm]{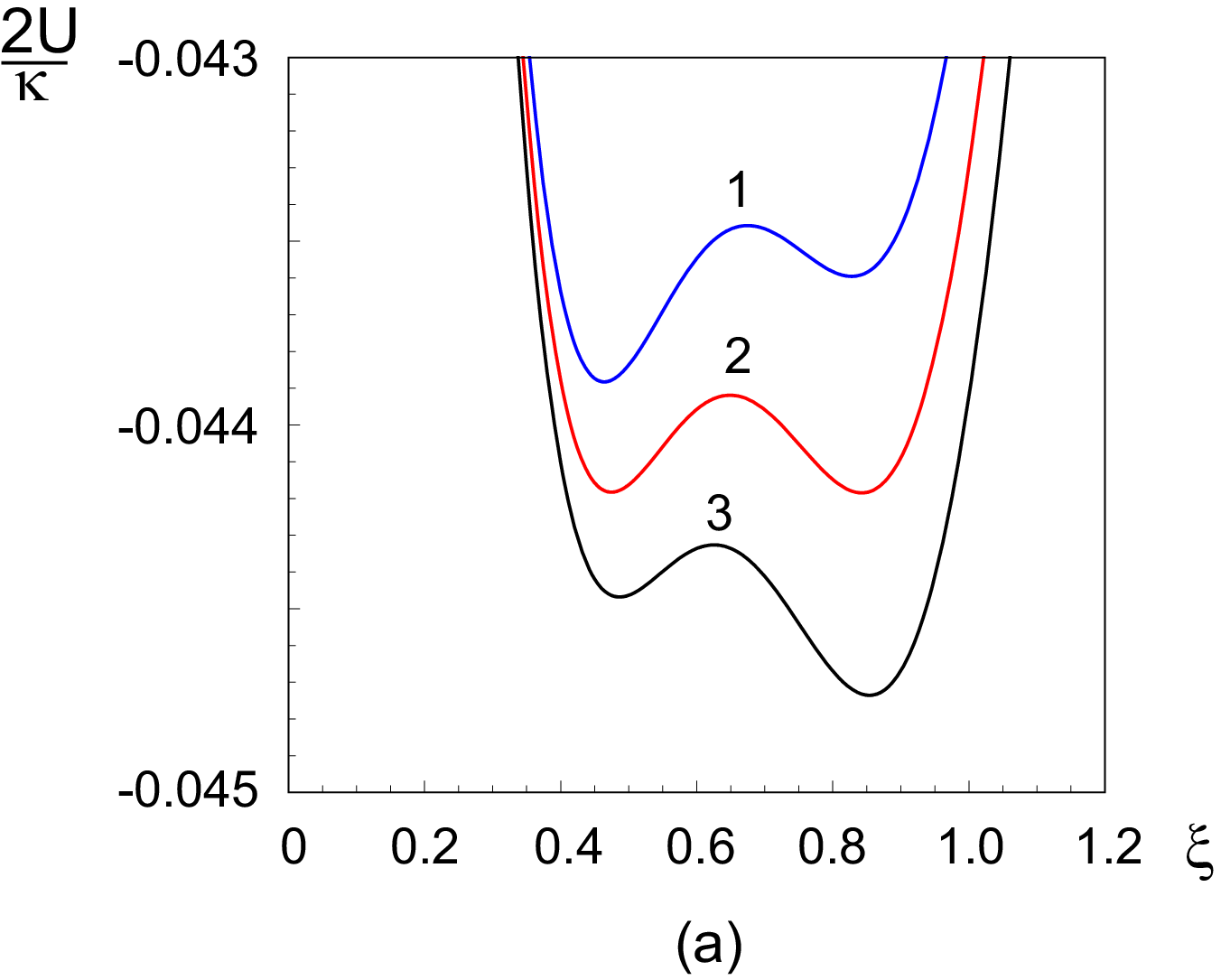} \hfill
 \includegraphics[width=75mm, height=90mm]{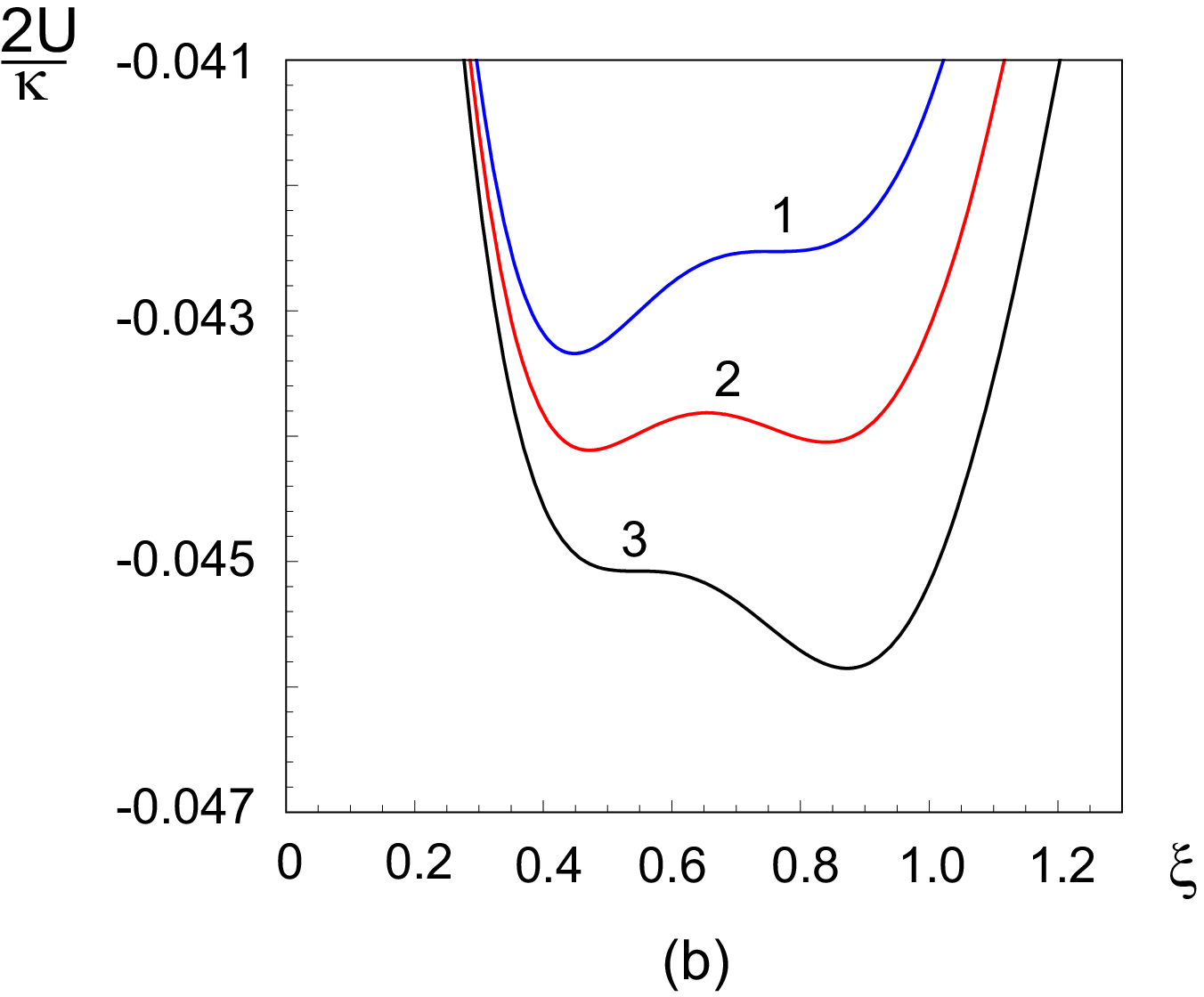}  \\
 \caption {Normalized potential for  $G = 2$, $g_1 = 3.5$, $S_0$ = 0.0055,
  and different values of the concentration $\ell_1$:
  $\ell_1 = $  0.158~(1),           0.15826~(2),  0.1585~(3)          (a);
               $\ell^b_{1,1}$~(1),  0.1582~(2),   $\ell^b_{1,2}$~(3)  (b).}
 }
 \label{pRGBfigc._10}
\end{figure}

For $S_0 = 0.0055 \in (S^c_-,\, S^c_+)$,  the curves in Fig.~10a illustrate the
two-well shape of the potential for $\ell_1 \in (\ell^b_{1,1}, \, \ell^b_{1,
2})$, where $\ell^b_{1,1} \approx 0.1575$ and $\ell^b_{1,2} \approx 0.159$. The
curves in Fig.~10b show essential changes in the shape of the potential for the
bifurcation concentrations $\ell^b_{1,1}$ (curve 1) and $\ell^b_{1,2}$ (curve
3), namely, as $\ell_1$  increases, the single-well potential for $\ell_1 <
\ell^b_{1,1}$ is transformed into a two-well one for $\ell_1 > \ell^b_{1,1}$
and then the two-well potential for $\ell_1 \in (\ell^b_{1,1}, \, \ell^b_{1,
2})$ is transformed into a single-well one for $\ell_1 > \ell^b_{1,2}$. Curve 1
in Fig.~10a shows the appearance of the second stationary (metastable because
$U_1 < U_3$) state of the system at the greater displacement  $\xi_3$ ($\xi_3 >
\xi_1$) of the oscillator from its nonperturbed equilibrium position $\xi = 0$.
An increase in $\ell_1$ is accompanied by an increase in the depths of both
wells and a decrease in the barrier $\delta_{2,1} = U_2 - U_1$ between the
wells. Since the increment of the depth of the second well with $\ell_1$ is
greater than that of the first well, for a certain value of $\ell_1$, the
depths of the wells become approximately equal (curve 2 in Fig.~10a) and, for
greater values of $\ell_1$, the second well is deeper than the first (curve 3
in Fig.~10a), i.e., the state of the system becomes metastable in the first
well and stable in the second. Nevertheless, within the framework of the
overdamped approximation, following the principle of perfect delay
\cite{ref.PoS,ref.Gil}, as $\ell_1$ increases, the oscillator remains in the
first well rather than moves to the second. For transition of the system from
the metastable state to the stable state according to the Maxwell principle of
the choice of the global minimum of the potential \cite{ref.PoS,ref.Gil},
thermal fluctuations or the inertia effect (as in \cite{ref.Usenko_Pr} in the
case of one-component adsorption) or both these factors should be taken into
account. This situation remains up to the bifurcation concentration
$\ell^b_{1,2}$ for which the barrier $\delta_{2,1} = 0$ (curve 3 in Fig.~10b).
A negligible excess of this bifurcation concentration leads to the
transformation of the potential into a single-well potential, which is
accompanied by the displacement of the oscillator to the unique equilibrium
position at the point $\xi_3$ or, in terms of $\xi(\ell_1)$, the transition of
the coordinate $\xi(\ell_1)$ from its lower stable branch to the upper one at
$\ell_1 = \ell^b_{1,2}$ (see Fig.~6b). In turn, this is accompanied by
transitions of the surface coverages $\theta_n(\ell_1)$ from their lower stable
branches to the upper ones (Fig.~7b).  \looseness=-1

\begin{figure}[!ht]
 \centering{
 \includegraphics[width=50mm, height=80mm]{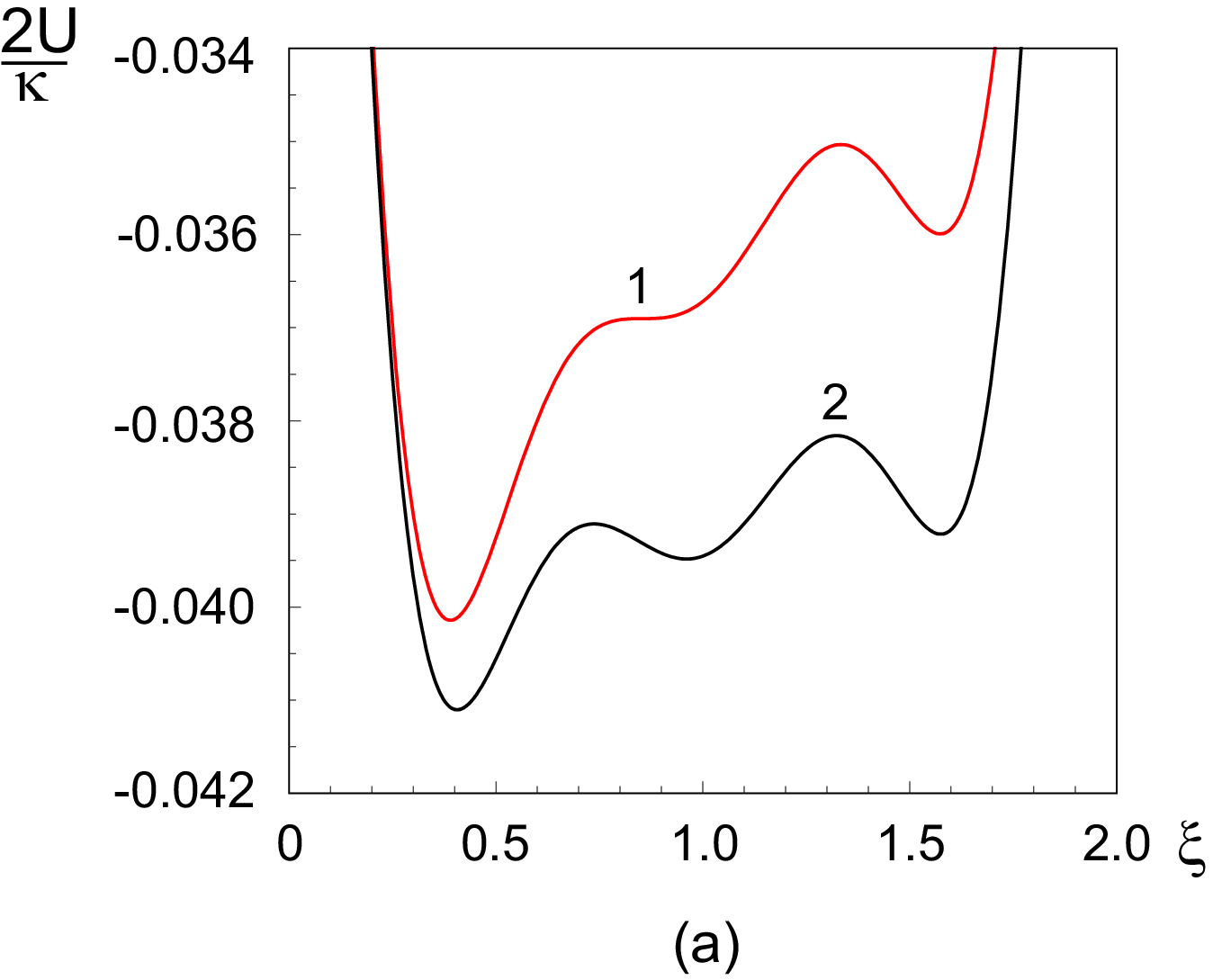} \hfill
 \includegraphics[width=50mm, height=80mm]{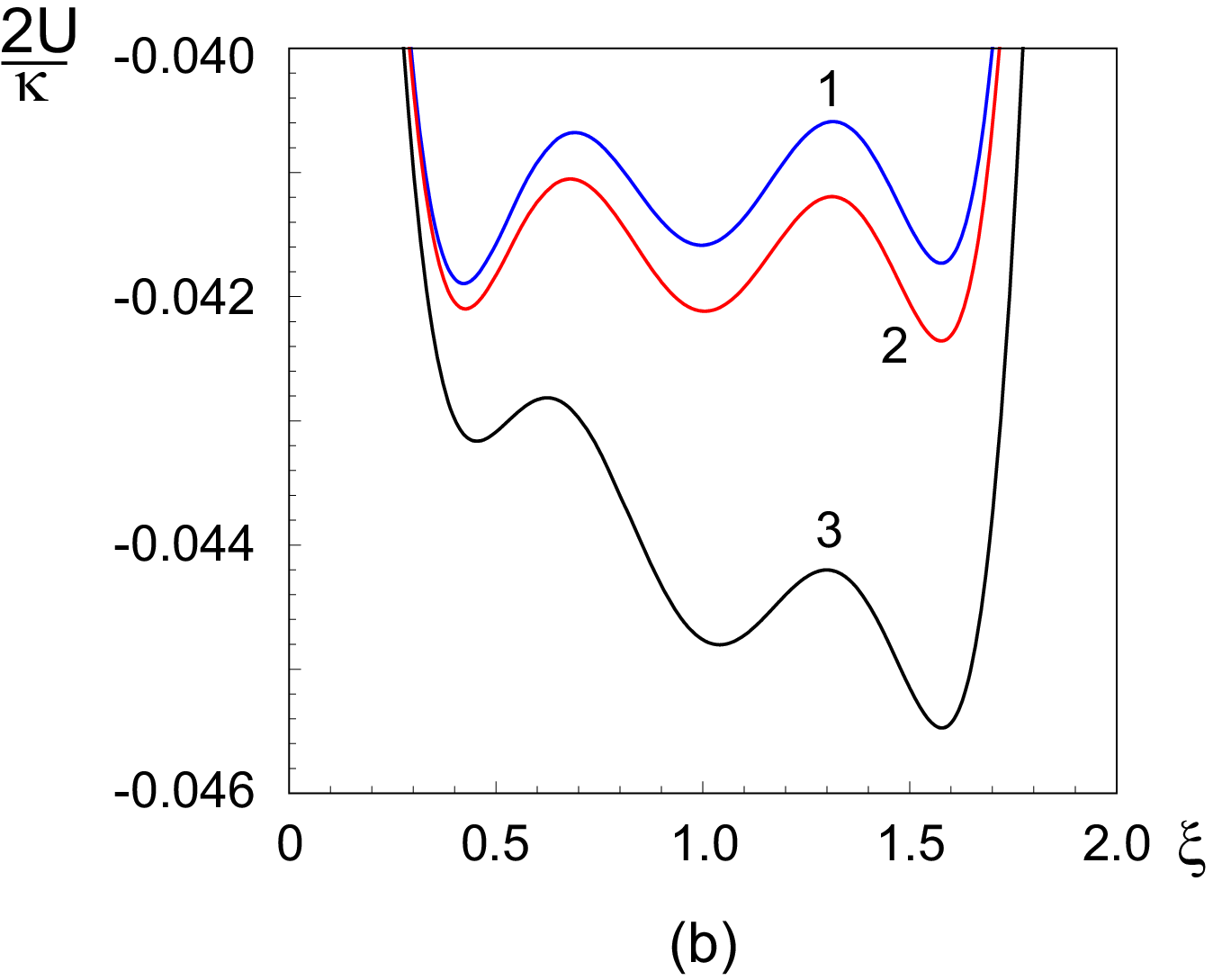} \hfill
 \includegraphics[width=50mm, height=80mm]{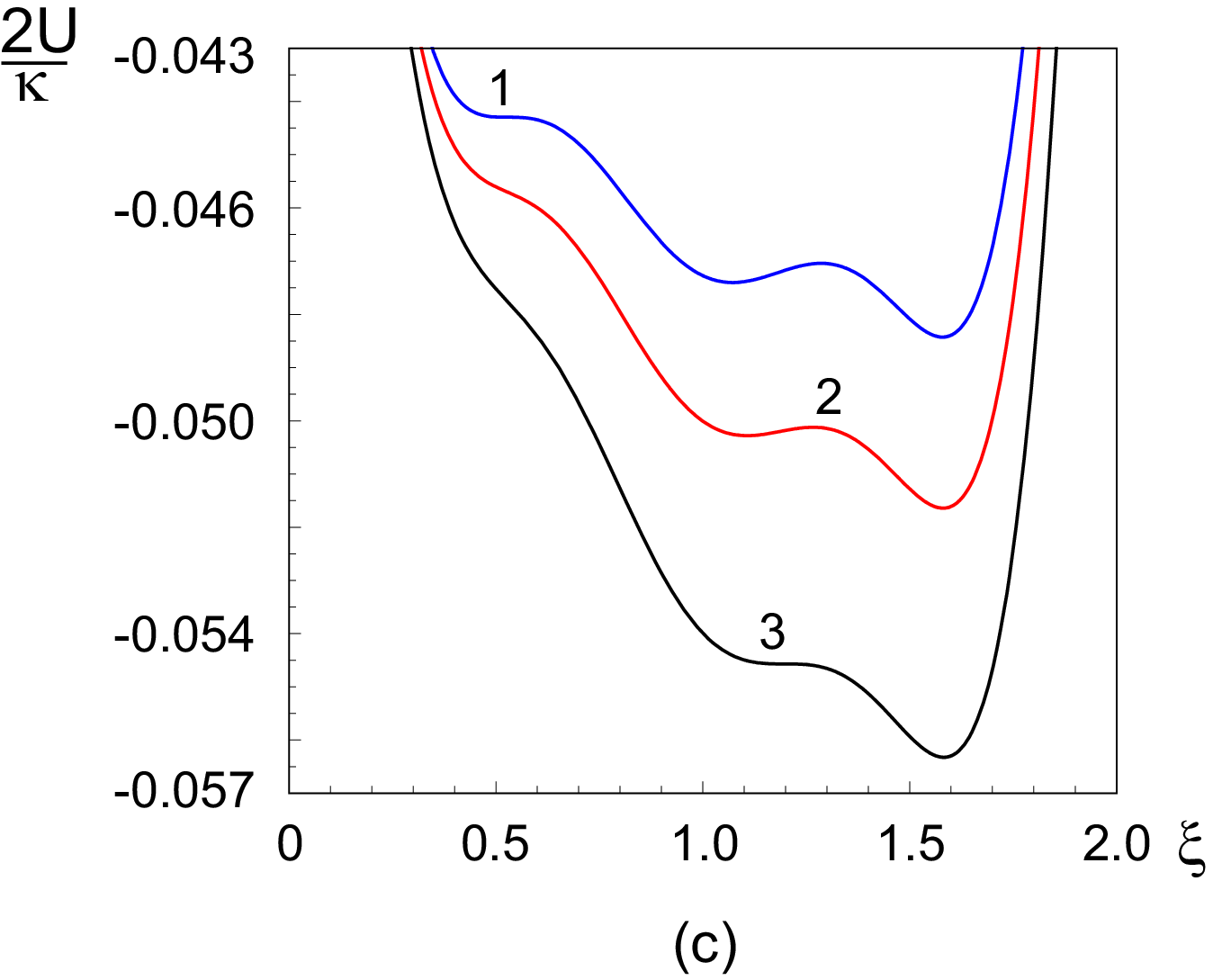}  \\
 \caption {Normalized potential for  $G = 2$, $g_1 = 3.5$, $S_0$ = 0.00585,
  and different values of the concentration $\ell_1$:
  $\ell_1 = $  $\ell^b_{1,1}$~(1),  0.155~(2)                       (a);
               0.1558~(1),          0.156~(2),  0.157~(3)           (b);
               $\ell^b_{1,2}$~(1),  0.159~(2),  $\ell^b_{1,4}$~(3)  (c).}
 }
 \label{pRGBfigc._11}
\end{figure}

The curves in Fig.~11 for $S_0 = 0.00585 \in (S^c_d,\, S^c_t)$ illustrate the
transformation of the two-well potential $U(\xi)$ for $\ell_1 < \ell^b_{1, 1}
\approx 0.154$ into a three-well one and vice versa for $\ell_1 > \ell^b_{1, 2}
\approx 0.158$ as $\ell_1$ increases. For the bifurcation concentration
$\ell^b_{1, 1}$, the potential has two wells and the horizontal point of
inflection between them (curve 1 in Fig.~11a). As the concentration $\ell_1$
increases, the potential is deformed in the neighborhood of this point of
inflection so that there appears one more well (curve 2 in Fig.~11a), which
corresponds to the case of five simple roots of Eq.~(18). The depths of the
wells at  $\xi = \xi_j$, where  $j = 1, 3, 5$, decrease with $\xi$: $|U_1| >
|U_3| > |U_5|$. As $\ell_1$ increases, this inequality is replaced, first, by
$|U_1| > |U_5| > |U_3|$ (curve 1 in Fig.~11b) and then by $|U_5| > |U_3| >
|U_1|$ (curves 2 and 3 in Fig.~11b), i.e., the deepest well successively moves
away from the nonperturbed equilibrium position $\xi =0$ with $\ell_1$. For the
bifurcation concentration $\ell^b_{1,2}$ the barrier between the first and
second wells disappears, $\delta_{2,1} = 0$ (curve 1 in Fig.~11c). A negligible
excess of this bifurcation concentration leads to the transformation of
$U(\xi)$ into a two-well potential and, as a result, the displacement of the
oscillator to the equilibrium position at the point $\xi_3$ or, in terms of
$\xi(\ell_1)$, the transition of the coordinate $\xi(\ell_1)$ from its first
stable branch to the second at $\ell_1 = \ell^b_{1,2}$ (Fig.~6f). In turn, this
is accompanied by transitions of the surface coverages $\theta_n(\ell_1)$ from
their first stable branches to the second (Fig.~7f). As $\ell_1$ increases, the
two-well potential is transformed so that the barrier between two remaining
wells $\delta_{4,3} = U_4 - U_3$ decreases and becomes equal to zero for the
bifurcation concentration $\ell^b_{1, 4} \approx 0.161$ (curve 3 in Fig.~11c).
For $\ell_1 > \ell^b_{1, 4}$, only one well of the potential $U(\xi)$  most
remote from the nonperturbed surface remains and the oscillator shifts to the
bottom of this well at $\xi = \xi_5$. \looseness=-1

Taking into account the different increase in the residence times of
adparticles of different species on the surface with displacement of adsorption
sites from the nonperturbed adsorbent surface [see (9)--(14)], we can draw a
conclusion on a considerable increase in the fraction of adparticles of species
2 in the total amount of adsorbed substance in transition of adsorption sites
to a more remote well.  This conclusion explains, in particular, the opposite
behavior of the surfaces coverages $\theta_1(\ell_1)$ and $\theta_2(\ell_1)$ in
Figs.~7c--f,~7i in passing through the bifurcation value $\ell^b_{1, 4}$: a
stepwise decrease in $\theta_1(\ell_1)$ and a stepwise increase in
$\theta_2(\ell_1)$ are caused by the displacement of the adsorption sites to
the most remote well.

As has been shown in Sec.~3.1, the specific feature of adsorption of a
two-component gas on a deformable adsorbent is two stable horizontal asymptotes
of the coordinate  $\xi(\ell_1)$ and the surface coverages $\theta_n(\ell_1)$
for certain values of control parameters. To explain this effect in terms of
the potential $U(\xi)$, we investigate its behavior in the limiting case of
infinitely large values of $\ell_1$. Passing in (66) to the limit $\ell_1
\rightarrow \infty$, we obtain  \looseness=-1

\begin{equation}
 U^a(\xi) \equiv \lim\limits_{\ell_1 \rightarrow \infty} U(\xi) =
  \frac{\varkappa}{2} \, \biggl\{ \xi^2 - 2\,\xi - \frac{2}{g_1}
   \ln{\frac{1 + S(\xi)}{1 + S_0}} \biggr\}.
\end{equation}

For a finite concentration of gas particles of species 2 (in the special case
of one-component adsorption,  $\ell_2 = 0$), relation (68) is reduced to the
parabolic potential
\begin{equation}
 U^a(\xi) = \varkappa\, \xi \biggl(\frac{\xi}{2} - 1 \biggr)
\end{equation}
with minimum at the point $\xi = 1$, which, according to (24), gives the unique
horizontal asymptotes  $\theta^a_1 = 1$ and   $\theta^a_2 = 0$  for the surface
coverages $\theta_1 (\ell_1)$ and $\theta_2 (\ell_1)$, respectively.

In the special case of adsorption of a two-component gas with identical action
of adparticles on the adsorbent ($G = 1$), relation (69) remains also true for
$\ell_2 = \infty$.

Taking into account that the functions $\xi(\ell_1)$ and $\theta_n(\ell_1)$
have two stable horizontal asymptotes if  $g_1 > g^a_c$ and, e.g., $S_0 \in
(S^a_-, S^a_+)$ for $G > 1$, we conclude that, for these values of the
parameters $G$, $g_1$, and $S_0$, $U^a(\xi)$ is a two-well potential and the
system under study is bistable in a semiinfinite interval of values of
$\ell_1$.

\begin{figure}[!htb]
 \centering{
 \includegraphics[width=140mm, height=100mm]{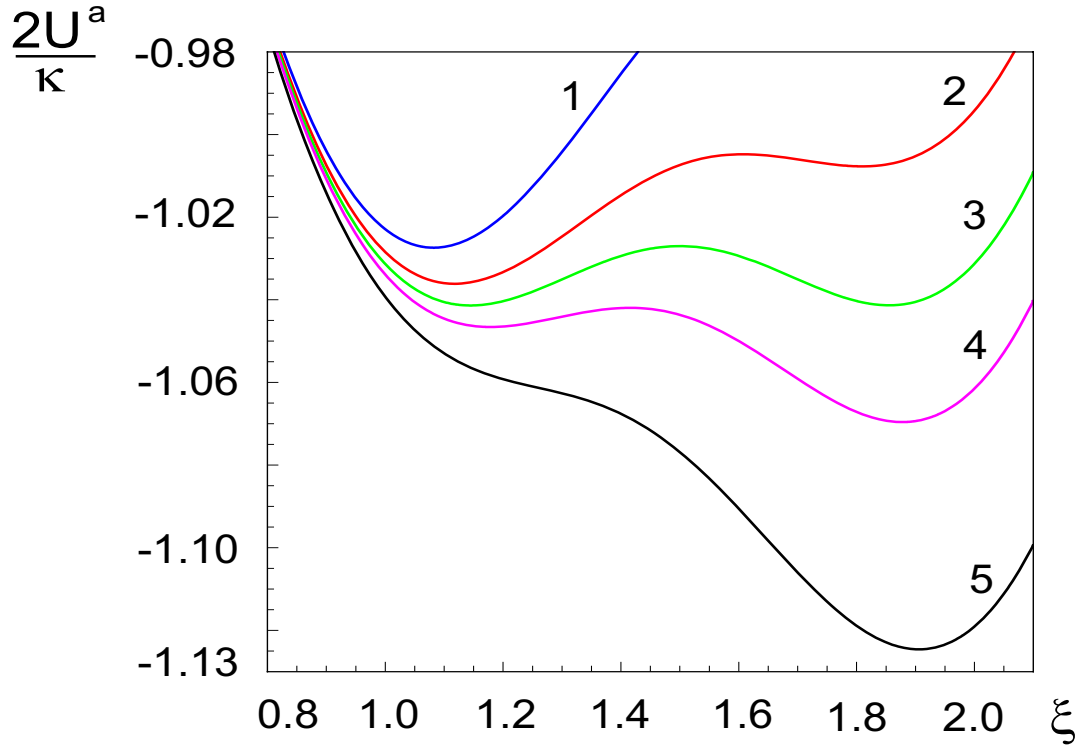}
 \caption {Normalized potential for  $\ell_1 = \infty$,
  $G = 2$, $g_1 = 5$,  and different values of $S_0$:
  $S_0$ = 0.0004~(1), 0.0005~(2), 0.000553~(3), 0.0006~(4), 0.0007~(5).}
 }
 \label{pRGBfigc._12}
\end{figure}

The curves in Fig.~12 clearly illustrate the essential dependence of the
potential  $U^a(\xi)$  on the value of $S_0$. For $S_0 < S^a_- \approx
0.000473$ (curve 1) and $S_0 > S^a_+ \approx 0.000646$ (curve 5), the potential
has one well; furthermore, in the second case, the well is deeper and
considerably more shifted from the nonperturbed adsorbent surface $\xi = 0$.
For $S_0 \in (S^a_-, S^a_+)$, the potential has two wells; moreover, if $S_0$
is close to $S^a_-$, then the first well is deeper than the second (curve 2)
and if $S_0$ is close to $S^a_+$, then the second well is deeper than the first
(curve 4). Curve 3 corresponds to the case of approximately equal depths of the
wells. The two-well potential $U^a(\xi)$ leads to two disconnected pieces of
the coordinate $\xi(\ell_1)$ (see Fig.~8b--e) and the corresponding specific
features of the surface coverages $\theta_1(\ell_1)$ and $\theta_2(\ell_1)$
(see Fig.~9b--e). \looseness=-1

\begin{figure}[!htb]
 \centering{
 \includegraphics[width=50mm, height=70mm]{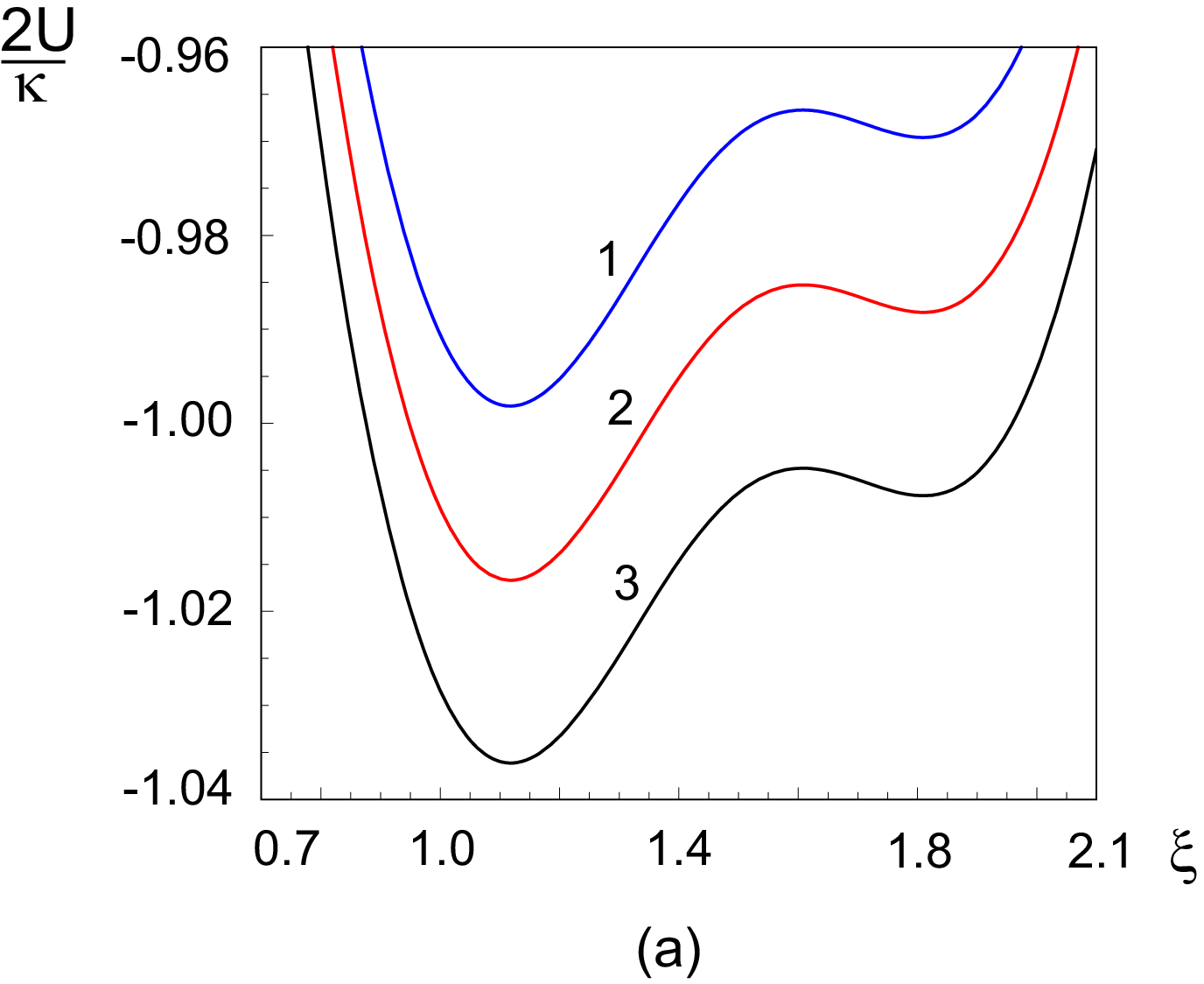} \hfill
 \includegraphics[width=50mm, height=70mm]{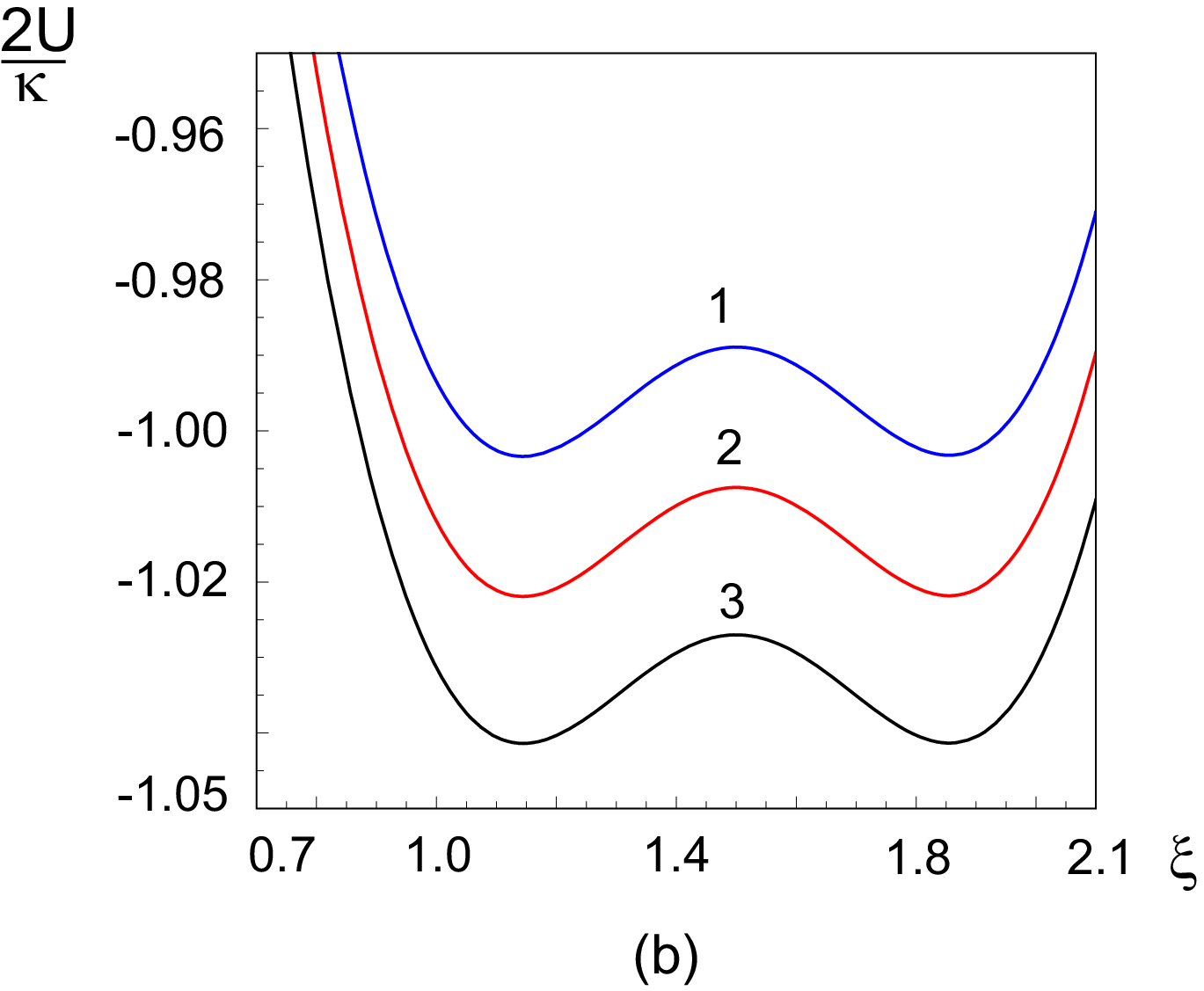} \hfill
 \includegraphics[width=50mm, height=70mm]{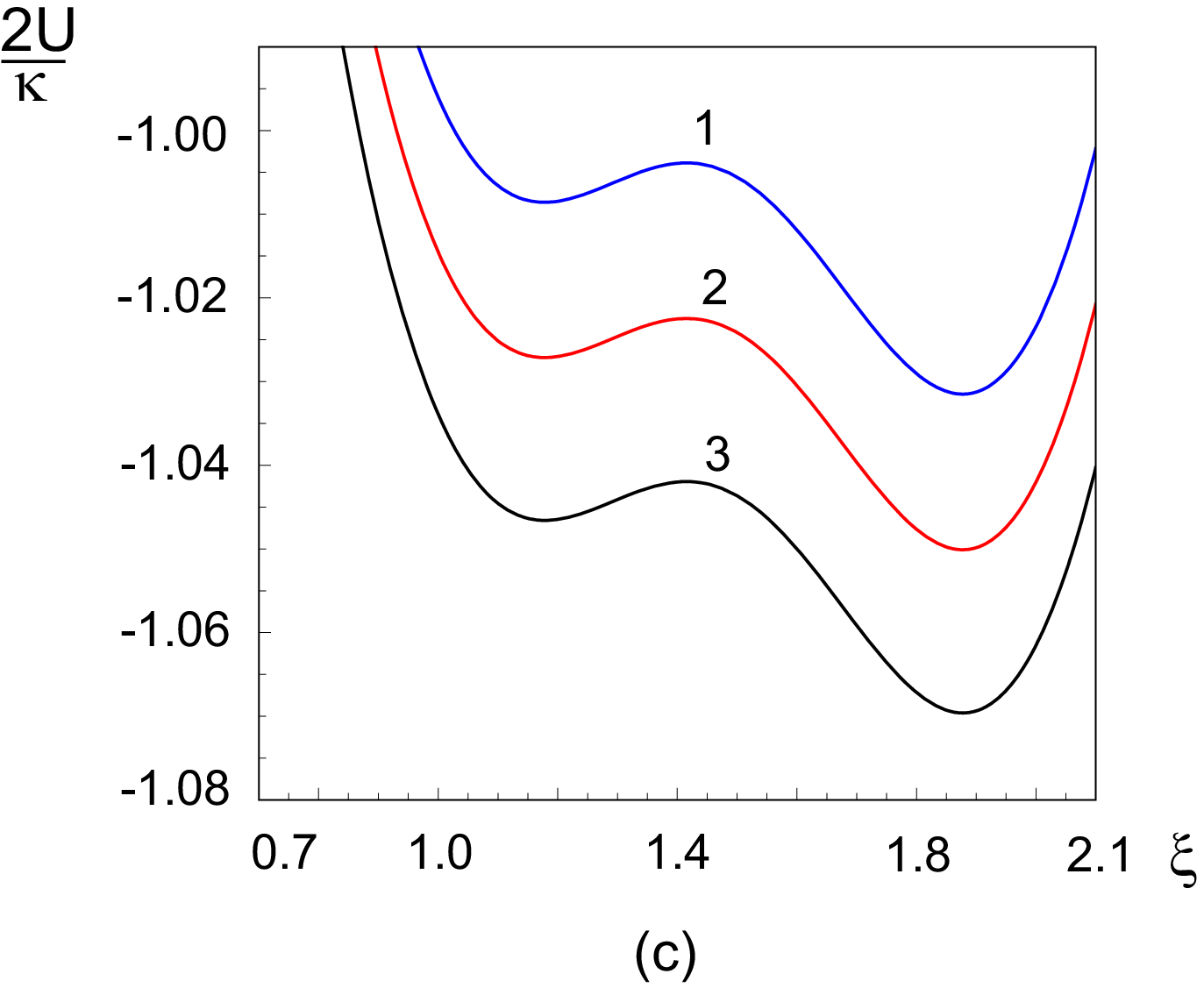}  \\
 \caption {Normalized potential for  $G = 2$, $g_1 = 5$,
  the concentration $\ell_1 = $  10~(1), 20~(2), $\infty$~(3),
  and $S_0$ = 0.0005~(a),  0.000553~(b),  0.0006~(c).}
 }
 \label{pRGBfigc._13}
\end{figure}

The curves in Fig.~13 illustrate the approach of the potential  $U(\xi)$  to
the two-well potential $U^a(\xi)$ as the concentration $\ell_1$ increases. For
large values of $\ell_1$, the behavior of the two-well potential $U(\xi)$ is
similar to $U^a(\xi)$: the first well is deeper if $S_0$ is close to $S^a_-$
(Fig.~13a), the second well is deeper if $S_0$ is close to $S^a_+$ (Fig.~13c),
the depths of two wells in figure~11(b) are approximately equal. Note that,
according to the principle of perfect delay \cite{ref.PoS,ref.Gil}, the
oscillator remains in the first well for arbitrarily large values of  $\ell_1$.




\section{Conclusions}  \label{Conclusions}

In the present paper, we have investigated isotherms of competitive adsorption
of a two-component gas on the surface of a solid adsorbent whose adsorption
properties vary in adsorption due to the adsorbent deformation. It has been
established that taking account of the adsorbent deformation in adsorption
essentially changes the shape of adsorption isotherms relative to the Langmuir
isotherms. The specific features of adsorption isotherms (bistability and
tristability of the system, two stable asymptotes of adsorption isotherms, an
essential redistribution of the quantities of adsorbed particles of different
species as compared with those in the classical case) depend on values of the
parameters expressed in terms of the phenomenological constant
adsorption-induced forces $\chi_1$ and $\chi_2$. The values of these forces can
be determined knowing experimental data of changes in the first interplanar
spacing (along the normal to the surface) $x_n^{max}$ due to the total
monolayer coverage of the adsorbent surface by adparticles of species $n$ in
adsorption of a one-component gas. In terms of the measured $x_n^{max}$, the
required forces and the coupling parameters are expressed as follows:
 \begin{equation}
 \chi_n = \varkappa\, x_n^{max},  \qquad
 g_n = \varkappa\, (x_n^{max})^2/k_B T, \quad n = 1,2, \qquad
 G = x_2^{max}/x_1^{max}.
\end{equation}

Based on the values of the parameters calculated by relations (70), one can
conclude whether the specific features of adsorption isotherms of a binary gas
mixture established in the paper are possible or not. According to (70), this
is more probable for adsorbents with relatively low elastic characteristics and
a surface layer susceptible to a change in the spatial distribution of the
charge density of adsorption sites in adsorption and resulting in not too low
absolute values of adsorption-induced forces. Possibly, in experiments aimed at
searching for these effects caused by the adsorbent deformation in adsorption,
it makes sense to use single crystal solid substrates of a ``soft'' material
admitting a considerable normal displacement of the adsorbent surface in
adsorption.

It is also worth noting that the used mean-field approximation requires that
the relaxation time of a bound adsorption site to a new equilibrium position
caused by adsorption be much greater than the average time between collisions
of gas particles with the adsorption site and the average residence time of an
adparticle on the surface. For this relaxation time to be much more greater
than the vibrational period of a vacant adsorption site, the friction
coefficient must be not negligible. This leads to a certain condition imposed
on its value, which depends on the concentration of particles in the gas phase,
so that many gas particles can successively take part in adsorption on the same
adsorption site before it reaches the equilibrium position.  \looseness=-1

The proposed model of competitive adsorption of a two-component gas on a
deformable adsorbent should be regarded as only the first step for describing
adsorption on a deformable adsorbent. The subsequent development of the model
requires taking account of various factors (lateral interactions between
adparticles, fluctuations, energy inhomogeneity of the adsorbent surface, etc.)
not considered here.



\section*{Acknowledgments}

The author expresses the deep gratitude to Prof. Yu.\,B.~Gaididei for the
valuable remarks and useful discussions of results.




\begin{thebibliography}{99}

\bibitem{ref.Mor}
S.\,R.~Morrison, \textit{The Chemical Physics of Surfaces} (Plenum, New York,
1977).

\bibitem{ref.RobMc}
M.\,W.~Roberts and C.\,S.~McKee, \textit{Chemistry of the Metal--Gas Interface}
(Oxford University Press, Oxford, 1978).

\bibitem{ref.Nau_78}
A.\,G.~Naumovets, \textit{Studies of surface structure by the low-energy
electron diffraction method: achievements and outlooks,} Ukr. Fiz. Zh.,
\textbf{23}, No.~10, 1585--1607 (1978).

\bibitem{ref.JaP}
M.\,J.~Jaycock and G.\,D.~Parfitt, \textit{Chemistry of Interfaces} (Wiley, New
York, 1981).

\bibitem{ref.KiK}
V.\,F.~Kiselev and  O.\,V.~Krylov, \textit{Adsorption Processes on
Semiconductor and Dielectric Surfaces} (Springer, Berlin, 1985).

\bibitem{ref.Vol}
F.\,F.~Vol'kenshtein, \textit{Electron Properties on Semiconductor Surfaces at
Chemisorption} (Nauka, Moscow, 1987).

\bibitem{ref.Zan}
A.~Zangwill, \textit{Physics at Surfaces} (Cambridge University Press,
Cambridge, 1988).

\bibitem{ref.Zhd}
V.\,D.~Zhdanov, \textit{Elementary Physicochemical Processes on Solid Surfaces}
(Plenum, New York, 1991).

\bibitem{ref.LNP}
I.\,F.~Lyuksyutov, A.\,G.~Naumovets, and V.\,L.~Pokrovsky,
\textit{Two-Dimensional Crystals} (Academic, Boston, 1992).

\bibitem{ref.AdC}
A.\,W.~Adamson and A.\,P.~Cast, \textit{Physical Chemistry of Surfaces} (Wiley,
New York, 1997).

\bibitem{ref.Nau_03}
A.\,G.~Naumovets, \textit{Use of surface phase transitions for control over
properties of surfaces,} in: I.\,K.~Pohodnya, A.\,H.~Kostornov, Yu.\,M.~Koval',
et al., \textit{Progressive Materials and Technologies,} Vol.~2
(Akademperiodyka, Kyiv, 2003), pp.~319--350.

\bibitem{ref.Bar}
P.~Barret, \textit{Cin\'{e}tique H\'{e}t\'{e}rog\`{e}ne} (Guathier-Villars,
Paris, 1973).

\bibitem{ref.Rog}
S.\,Z.~Roginskii, {\it Heterogeneous Catalysis. Some Problems of the Theory}
(Nauka, Moscow, 1979).

\bibitem{ref.Roz}
A.\,Ya.~Rozovskii, {\it Heterogeneous Chemical Reactions. Kinetics and
Macrokinetics} (Nauka, Moscow, 1980).

\bibitem{ref.Bor}
G.\,K.~Boreskov, \textit{Heterogeneous Catalysis} (Nauka, Moscow, 1986).

\bibitem{ref.KSh}
O.\,V.~Krylov and B.\,R.~Shub, \textit{Nonequilibrium Processes in Catalysis}
(CRC Press, Boca Raton, 1994).

\bibitem{ref.Kry}
O.\,V.~Krylov, \textit{Heterogeneous Catalysis} (Akademkniga, Moscow, 2004).

\bibitem{ref.FrK}
D.\,A.~Frank-Kamenetskii, \textit{Diffusion and Heat Transfer in Chemical
Kinetics} (Plenum, New York, 1969).

\bibitem{ref.Rut}
D.\,M.~Ruthven, \textit{Principles of Adsorption and Adsorption Processes}
(Willey, Chichester, 1984).

\bibitem{ref.Tov}
Yu.\,L.~Tovbin, \textit{Theory of Physical Chemistry Processes at a Gas--Solid
Interface} (CRC Press, Boca Raton, 1991).

\bibitem{ref.Do}
D.\,D.~Do, \textit{Adsorption Analysis: Equilibria and Kinetics} (Imperial
College Press, London 1998).

\bibitem{ref.KSt}
J.~Keller and R.~Staudt, \textit{Gas Adsorption Equilibria: Experimental
Methods and Adsorption Isotherms} (Springer, 2005).

\bibitem{ref.IEr}
R.~Imbihl and G.~Ertl, \textit{Oscillatory kinetics in heterogeneous
catalysis,} Chem. Rev., \textbf{95}, No.~3, 697--733 (1995).

\bibitem{ref.Imb_1}
R.~Imbihl, \textit{Nonlinear dynamics on catalytic surfaces,} Catal. Today,
\textbf{105}, 206--222 (2005).

\bibitem{ref.Imb_2}
R.~Imbihl, \textit{Nonlinear dynamics on catalytic surfaces: The contribution
of surface science,} Surf. Sci., \textbf{603}, 1671--1679 (2009).

\bibitem{ref.Ert}
G.~Ertl, \textit{Reactions at Solid Surfaces} (Wiley, Hoboken, 2009).

\bibitem{ref.Usenko}
A.\,S.~Usenko, \textit{Adsorption on a surface with varying properties,} Phys.
Scr., \textbf{85}, 015601 (2012).

\bibitem{ref.Zeld}
Ya.\,B.~Zeldovich, \textit{Adsorption on a uniform surface,} Acta Physicochim.
URSS, \textbf{8}, No.~5, 527--530 (1938).

\bibitem{ref.AFY}
T.~Ala-Nissila, R.~Ferrando, and S.\,G.~Ying, \textit{Collective and single
particle diffusion on surfaces,}  Adv. Phys., \textbf{51}, No.~3, 949--1078
(2002).

\bibitem{ref.TZHZJ}
Z.-J.~Tan, X.-W.~Zou, S.-Y.~Huang, W.~Zhang, and Z.-Z.~Jin, \textit{Patterns of
particle distribution in multiparticle systems by random walks with memory
enhancement and decay,} Phys. Rev.~B, \textbf{66}, 011101 (2002).

\bibitem{ref.HZJ}
S.-Y.~Huang, X.-W.~Zou, and Z.-Z.~Jin, \textit{Multiparticle random walks on a
deformable medium,} Phys. Rev.~B, \textbf{66}, 041112 (2002).

\bibitem{ref.Yang}
R.\,T.~Yang, \textit{Adsorbents: Foundations and Applications} (Wiley, Hoboken,
2003).

\bibitem{ref.Gun}
V.\,M.~Gun'ko, \textit{Competitive adsorption,} Theor. Experim. Chem.,
\textbf{43}, No.~3, 133--182 (2007).

\bibitem{ref.Nau_94}
A.\,G.~Naumovets, \textit{Adsorption on metals: a look from the not-too-far
East,} Surf. Sci., \textbf{299/300}, 706--721 (1994).

\bibitem{ref.Chr}
L.\,N.~Christophorov, A.\,R.~Holzwarth, V.\,N.~Kharkyanen, and F.~van~Mourik,
\textit{Structure-function self-organization in nonequilibrium macromolecular
systems,} Chem. Phys., \textbf{256}, 45--60 (2000).

\bibitem{ref.PoS}
T.~Poston and I.~Stewart, \textit{Catastrophe Theory and Its Applications}
(Pitman, London 1978).

\bibitem{ref.Gil}
R.~Gilmore, \textit{Catastrophe Theory for Scientists and Engineers} (Wiley,
New York, 1981).

\bibitem{ref.AVKh}
A.\,A.~Andronov, A.\,A.~Vitt, and S.\,\'{E}.~Khaikin, \textit{Theory of
Oscillators} (Pergamon, New York, 1966).

\bibitem{ref.Hak}
H.~Haken, \textit{Synergetics} (Springer, Berlin, 1978).

\bibitem{ref.Usenko_Pr}
A.\,S.~Usenko, \textit{Adsorption on a Surface with Varying Properties,} arXiv:
0907.5569v2 (2009).

\end{thebibliography}
\end{document}